\documentclass[a4paper,12pt]{article}
\usepackage[onehalfspacing]{setspace}
\usepackage[flushmargin]{footmisc}
\usepackage{amsmath,amsfonts,amssymb,bbm}
\usepackage{rotating,graphics,eurosym}
\usepackage{longtable,lscape}
\usepackage[latin1]{inputenc}
\usepackage{enumitem}
\usepackage{fullpage}
\usepackage{hyperref}
\usepackage{harvard}
\usepackage[english]{babel}
\usepackage{float}
\usepackage[dvips]{epsfig}
\usepackage[dvips]{graphicx}
\usepackage[T1]{fontenc}
\usepackage{tabulary}
\usepackage{tabularx}
\usepackage{subfigure}
\usepackage{lscape}
\usepackage{multirow}
\usepackage{color}
\usepackage{calc}
\usepackage{extarrows}
\usepackage{stackrel}
\usepackage{booktabs,caption}
\usepackage[flushleft]{threeparttable}
\usepackage{eqnarray}
\usepackage{float}
\usepackage{amsthm}
\usepackage{longtable}
\usepackage{adjustbox}
\usepackage{tablefootnote}

\theoremstyle{plain}
\newtheorem{assumption}{Assumption}

\DeclareGraphicsExtensions{.eps .jpg} 
\graphicspath{{graphics/}}

 \paperheight29,7cm \paperwidth21cm
\begin{document}
\begin{center}
{\LARGE The finite sample performance of instrumental variable-based estimators of the Local Average Treatment Effect when controlling for covariates \\}
{\LARGE \vspace{1.3cm}}
{\large Hugo Bodory+, Martin Huber++, Michael Lechner*,** }
\vspace{0.3 cm}\\
{\small {+ University of St.Gallen, Vice-President's Board (Research \& Faculty)}}\\
{\small {++ University of Fribourg, Dept.\ of Economics}}\\
{\small {* University of St.Gallen, Dept.\ of Economics}}\\
\medskip
\vspace{0.5cm}
\end{center}

\begin{abstract}
\noindent This paper investigates the finite sample performance of a range of parametric, semi-parametric, and non-parametric instrumental variable estimators when controlling for a fixed set of covariates to evaluate the local average treatment effect. Our simulation designs are based on empirical labor market data from the US and vary in several dimensions, including effect heterogeneity, instrument selectivity, instrument strength, outcome distribution, and sample size. Among the estimators and simulations considered, non-parametric estimation based on the random forest (a machine learner controlling for covariates in a data-driven way) performs competitive in terms of the average coverage rates of the (bootstrap-based) 95\% confidence intervals, while also being relatively precise. Non-parametric kernel regression as well as certain versions of semi-parametric radius matching on the propensity score, pair matching on the covariates, and inverse probability weighting also have a decent coverage, but are less precise than the random forest-based method. In terms of the average root mean squared error of LATE estimation, kernel regression performs best, closely followed by the random forest method, which has the lowest average absolute bias.

\end{abstract}

\vfill

\small{
  \par
\vspace{0.3cm}
 \par
\vspace{0.3cm}
\hrule
\vspace{0.3cm}

Correspondence to: Hugo Bodory, University of St.\ Gallen, Varnb\"{u}elstrasse 14, CH-9000 St.\ Gallen, hugo.bodory@unisg.ch; Martin Huber, University of Fribourg, Bd.\ de P\'{e}rolles 90, CH-1700 Fribourg, martin.huber@unifr.ch; Michael Lechner, University of St.\ Gallen, Varnb\"{u}elstrasse 14, CH-9000 St.\ Gallen, michael.lechner@unisg.ch. **: Michael Lechner is also affiliated with CEPR and PSI, London, CESIfo, Munich, IAB, Nuremberg, and IZA, Bonn. \thispagestyle{empty}

\setcounter{page}{0}
\newpage

\section{Introduction}\setcounter{page}{1}

The evaluation of the causal effect of a treatment (e.g.,\ fertility)  on an outcome (e.g.,\ labor supply) is frequently complicated by endogeneity, implying that the treatment is associated with unobserved characteristics affecting the outcome (e.g.\ personality traits, preferences, and values concerning family and working life). One may nevertheless assess treatment effects in the presence of an instrumental variable (IV) which affects the treatment of (at least) some subjects in a monotonic way, does not directly affect the outcome (other than through treatment) and is as good as randomly assigned. Under these conditions, the local average treatment effect (LATE) on the compliers, the subpopulation whose treatment state reacts positively to the instrument, is identified, as discussed in \citeasnoun{Imbens+94} and \citeasnoun{Angrist+96}. In many empirical contexts, it may seem unlikely that the IV assumptions hold unconditionally, in particular when the treatment evaluation relies on observational data in which the instrument is not explicitly randomized like in an experiment. Depending on the application, it might, however, appear plausible that the IV assumptions hold conditional on covariates observed in the data. In this case, the LATE is identified and can be consistently estimated under certain conditions, see the discussions in  \citeasnoun{Abadie00}, \citeasnoun{Tan2006}, and \citeasnoun{Froel2007}.

This paper assesses the finite sample performance of various parametric, semi-parametric, and non-parametric IV estimators when controlling for a fixed (i.e., pre-defined and low-dimensional) set of covariates by Monte Carlo simulations that are based on empirical labor market data from \citeasnoun{Angrist+98}. The latter study assesses the effect of fertility, defined as having at least three vs.\ two children, on mother's labor supply (for instance, a binary employment status or weeks employed per year), using twins at the second birth as instrument. The intuition for this IV strategy is that if a mother with one child get twins at the second birth, then fertility immediately increases to three rather than two children, implying a first stage effect of the twins instrument on the treatment. In the spirit of \citeasnoun{HuLeWu13}, our empirical Monte Carlo simulation makes to a certain extent use of the empirical associations in the labour market data when assessing the various IV estimators, with the aim that our analysis is more closely linked to real world applications.

We vary the simulation designs in several dimensions, including treatment effect heterogeneity, instrument selectivity across observed covariates (namely age, race, and quarter of birth), instrument strength, the outcome distribution, and sample size. We analyse the performance of a range of estimators commonly considered in treatment and policy evaluation, including two stage least squares, inverse probability weighting, matching,  doubly robust estimation, and nonparametric regression. We find that overall, non-parametric estimation based on the random forest, a machine learning algorithm controlling for covariates in a data-driven way, performs best in terms of coverage rates. The latter are defined as the share of simulations in which the true LATE is included in the 95\% confidence interval of a LATE estimator, where the standard error required for the construction of the confidence interval is obtained by the non-parametric bootstrap. Furthermore, the random forest-based estimator is relatively precise, implying that the confidence interval is comparably short, which (conditional on having a decent coverage) appears desirable from the perspective of statistical power. Non-parametric kernel regression as well as certain versions of semi-parametric radius matching on the propensity score, pair matching on the covariates, and inverse probability weighting also have a decent coverage, but are less precise than the random forest-based method. Concerning the average root mean squared error of LATE estimation, kernel regression performs best (and also has the smallest average standard deviations), closely followed by the random forest method, which has the lowest average absolute bias.

The remainder of this paper is organized as follows. Section \ref{id} discusses the identifying assumptions for IV-based LATE evaluation in the presence of covariates. Section \ref{iv_estimators} introduces various parametric, semi-parametric, and non-parametric LATE estimators, as well as a bootstrap procedure for computing standard errors. Section \ref{datasim} presents our empirical Monte Carlo simulation approach, namely the empirical data and the simulation designs. Section \ref{results} presents the results on the finite sample performance of the LATE estimators. Section \ref{conclusion} concludes.

\section{Identification of the LATE}\label{id}

In this section, we present the assumptions underlying the identification of the Local Average Treatment Effect (LATE) when controlling for covariates. To formalize the discussion, let us denote by $D_i$ a possibly endogenous treatment received by unit $i$, and by $Y_i$ the outcome variable based on which the treatment effect is to be evaluated. In their seminal paper, \citeasnoun{Imbens+94} define the LATE as the mean effect of $Y_i$ in response to a change in $D_i$ among the compliers, a subgroup whose $D_i$ reacts to an exogenous shift in the instrumental variable, which is denoted by $Z_i$. To discuss the identification of the LATE, we make use of the potential outcomes framework introduced by \citeasnoun{Rubin74}, which expresses causal effects as differences between potential outcomes under treatment and non-treatment. We adapt this concept to our instrumental variable setting with binary indicators  $D_i$ and $Z_i$, and define potential outcome and treatment variables for unit $i$ in the following way:

\begin{eqnarray}\label{obs_pot}
Y_{i,z}^d &=& Y_i(D_i=d_i,Z_i=z_i) \\
D_{i,z}   &=& D_i(Z_i=z_i),
\end{eqnarray}

with $d_i,z_i\in\{0,1\}$. Using this framework, \citeasnoun{Angrist+96} show that units can be divided into two subgroups, compliers and noncompliers. Compliers are those induced to take the treatment when being assigned to it. Formally, this type of units is characterized by $D_{i,1}-D_{i,0}=1$. The subgroup of noncompliers may consist of three further types, namely always-takers with $D_{i,1}=D_{i,0}=1$, never-takers with $D_{i,1}=D_{i,0}=0$, and defiers with $D_{i,1}-D_{i,0}=-1$. Note that the type of a single unit cannot be identified because the counterfactual potential treatment (that would have occurred under the alternative, rather than the factual instrument assignment) is not observed.

\citeasnoun{Abadie00}, \citeasnoun{Tan2006}, and \citeasnoun{Froel2007} consider non-parametric LATE identification and estimation when controlling for observed covariates, denoted by $X_i$. We subsequently present the identifying assumptions in this context, which consist of (i) a monotonicity restriction on the treatment, (ii) the existence of compliers, (iii)  conditional independence of the instrument and the share of compliance types, (iv) conditional mean independence of the outcome and the instrument, and (v) common support.

\begin{assumption}[Monotonicity]\label{as1} ~\\ \indent $P(D_{i,0}>D_{i,1})=0$.
\end{assumption}
\begin{assumption}[Existence of compliers]\label{as2} ~\\ \indent $P(D_{i,0}<D_{i,1})>0$.
\end{assumption}
\begin{assumption}[Unconfounded type]\label{as3} ~\\ \indent $P(\tau_i=t \vert X_i=x_i,Z_i=0)=P(\tau_i=t \vert X_i=x_i,Z_i=1)$ for $t \in \{a,n,c\}$. ~\\
	\indent The types $\tau$ include always-takers $a$, never-takers $n$, and compliers $c$.
\end{assumption}
\begin{assumption}[Conditional mean independence of the outcome]\label{as4} ~\\
  \indent $E[Y_{i,Z_i}^0 \vert X_i=x_i,Z_i=0,\tau_i=t]=E[Y_{i,Z_i}^0 \vert X_i=x_i,Z_i=1,\tau_i=t]$ for $t \in \{n,c\}$, ~\\
	\indent $E[Y_{i,Z_i}^1 \vert X_i=x_i,Z_i=0,\tau_i=t]=E[Y_{i,Z_i}^1 \vert X_i=x_i,Z_i=1,\tau_i=t]$ for $t \in \{a,c\}$.
\end{assumption}
\begin{assumption}[Common support]\label{as5} ~\\ \indent $Supp(X_i \vert Z_i=1)=Supp(X_i \vert Z_i=0)$.
\end{assumption}

Assumption~\ref{as1} rules out the presence of defiers, a type whose treatment never complies with the instrument. Assumption~\ref{as2} implies that the subgroup of compliers exists. Due to the conditional independence of the instrument and the shares of compliers, always-takers, and never-takers stated in Assumption~\ref{as3},  the first stage effect of the instrument on the treatment is identified conditional on covariates, such that any variables affecting both the instrument and the treatment are controlled for. The conditional mean independence in Assumption~\ref{as4} rules out a direct average effect of the instrument on the outcome (exclusion restriction) and unobservables that jointly affect the instrument and the outcome when controlling for covariates. Finally, Assumption~\ref{as5} ensures that for all covariate values occurring in the population, either instrument value $Z_i\in\{0,1\}$ exists such that the instrument is not deterministic in the covariates.

Under Assumptions~\ref{as1} to \ref{as5}, the LATE, denoted as $\theta=E[Y_{i,Z_i}^1-Y_{i,Z_i}^0\vert D_{i,1}-D_{i,0}=1]$, is identified by

\begin{equation}\label{eq:late_x}
	\theta=\frac{E_X[E[Y_i \vert Z_i=1,X_i]-E[Y_i \vert Z_i=0,X_i]]}{E_X[E[D_i \vert Z_i=1,X_i]-E[D_i \vert Z_i=0,X_i]]}.
\end{equation}

Based on the insights of \citeasnoun{rosenbaum1983}, \citeasnoun{Froel2007} shows that identification is also obtained by conditioning on the instrument propensity score $p(x):=P(Z_i=1 \vert X_i=x)$ rather than the covariates, because it possesses the so-called `balancing property'. That is, conditioning on the one-dimensional propensity score balances the distribution of the covariates across the states of the instrument. For this reason, the LATE is alternatively identified by

\begin{equation}\label{eq:late_px}
	\theta=\frac{E_X[E[Y_i \vert Z_i=1,p(X_i)]-E[Y_i \vert Z_i=0,p(X_i)]]}{E_X[E[D_i \vert Z_i=1,p(X_i)]-E[D_i \vert Z_i=0,p(X_i)]]}.
\end{equation}

\section{Estimation and inference}\label{iv_estimators}

In this section, we present parametric, semi-parametric, and non-parametric methods for estimating the LATE parameter $\theta$ introduced in  Section \ref{id}. We also discuss a trimming rule that tackles limited common support in covariate values across instrument states, based on dropping observations which would obtain large weights in the estimator because their covariate values occur (almost) exclusively in only one of the instrument states. Finally, we provide an bootstrap procedure for estimating the standard errors of the LATE estimators.

\subsection{Estimation}\label{estim}

One method for the estimation of $\theta$ frequently applied in empirical work is two-stage least-squares (2SLS), which is easy to implement and computationally fast. However, the linearity assumption of the 2SLS estimator implies effect homogeneity, a restriction that may not hold in empirical studies. We consider 2SLS as a benchmark method, but also include more general LATE estimators that allow for effect heterogeneity of the LATE across values of the covariates.

Equations~\ref{eq:late_x} and \ref{eq:late_px} imply that $\theta$ can be expressed as the ratio of two treatment effect estimators that account for covariate differences in the presence and absence of the instrument. The numerator gives the reduced form effect of $Z_i$ on $Y_i$ and the denominator the first stage effect of $Z_i$ on $D_i$. Thus, a natural choice for the construction of estimators for $\theta$ is to substitute the expressions in the numerators and denominators of Equations~\ref{eq:late_x} and \ref{eq:late_px} by estimators standardly applied in treatment or policy evaluation, see for instance the surveys by \citeasnoun{Imbens03} and \citeasnoun{ImWo08}.

Many treatment effect estimators are semi-parametric in the sense that (parametric) propensity score estimation is combined with non-parametric treatment effect estimation, using weighting, matching, or doubly robust methods. A growing number of simulation studies has investigated the finite sample behavior of such treatment effect estimators when the treatment is exogenous conditional on covariates, see for instance \citeasnoun{Froe00a}, \citeasnoun{Zh04}, \citeasnoun{LuncefordDavidian2004}, \citeasnoun{BuDNMC09}, \citeasnoun{HuLeWu13}, and \citeasnoun{FrHuWi14}. We consider such methods to estimate the LATE based on estimates of the instrument propensity score. We also vary the degree of flexibility of the estimators and implement parametric, semi-parametric, and non-parametric approaches to compute the reduced form and first stage effects in the numerators and denominators of Equations~\ref{eq:late_x} and \ref{eq:late_px}.

\citeasnoun{SmithTodd00}, among others, regard treatment effect estimators as weighted differences in outcomes. We apply this definition to the Wald formula and express the LATE as:

\begin{equation}\label{est}
\hat{\theta}=\frac{\frac{1}{n_1}\sum_{i=1}^n z_i\hat{w}_iy_i-\frac{1}{n_0}\sum_{j=1}^n(1-z_j)\hat{w}_jy_j}
									{\frac{1}{n_1}\sum_{i=1}^n z_i\hat{w}_id_i-\frac{1}{n_0}\sum_{j=1}^n(1-z_j)\hat{w}_jd_j}.
\end{equation}

$n$ denotes the size of an i.i.d.\ sample of realizations of $\{Y_i,D_i,Z_i,X_i\}$ with observation $i \in {1,...,n}$. $n_1=\sum_{i=1}^n Z_i$ is the size of the subsample of those with $Z_i=1$, $n_0=n-n_1$, and $\hat{w}_i$ are weights that may depend on $X_i$ or $\hat{p}(x)$, an estimate of the propensity score $p(x)$. Next, we discuss different methods of estimating $\hat{p}(x)$ and $\hat{w}_i$.

\subsection{Instrument propensity scores}\label{cond}

We consider two different approaches to balance the covariates across groups for units with $Z_i=0$ and $Z_i=1$. One is to directly control for covariates $X_i$, but some LATE estimators alternatively control for estimates of $p(x)$, which is motivated by the propensity score's balancing properties discussed in \citeasnoun{rosenbaum1983}. Their results imply that $p(x)$ is capable of equalizing the covariate distributions across instrument states, such that the instrument is conditionally independent of potential outcomes and treatments given the propensity score whenever independence holds conditional on the covariates. A practical advantage of controlling for the propensity score (rather than a vector of covariates) is that it is one-dimensional and thus, avoids the curse of dimensionality.

We compute $\hat{p}(x)$ in three different ways. Firstly, we specify a probit model to estimate the conditional probability $P(Z_i=1 \vert X_i=x_i)$ by

\begin{equation}\label{p1}
\hat{p}(x)^{probit}=\Phi(x_i^T\tilde{\beta}_{ML}),
\end{equation}

where $\tilde{\beta}_{ML}$ denotes the estimated probit coefficients based on maximum likelihood and $\Phi(x_i^T\tilde{\beta}_{ML})$ is the cumulative distribution function of the standard normal distribution evaluated at $X_i^T\tilde{\beta}_{ML}$.

Secondly, we apply the covariate balancing propensity score (CBPS) method by \citeasnoun{ImaiRatkovic2014} to compute $\hat{p}(x)$.  This methodology maximizes covariate balancing when predicting treatment assignment using the generalized method-of-moments (GMM) framework. \citeasnoun{ImaiRatkovic2014} show that the CBPS method is robust to mild misspecifications of the propensity score model, which is estimated by the following expression:

\begin{equation}\label{p2}
\hat{p}(x)^{CBPS}=\Lambda(x_i^T\tilde{\beta}_{GMM}),
\end{equation}

where $\tilde{\beta}_{GMM}$ are coefficients estimated by GMM and $\Lambda(x_i^T\tilde{\beta}_{GMM})$ is the cumulative distribution function of the standard logistic distribution evaluated at $x_i^T\tilde{\beta}_{GMM}$. We use the overidentified version of CBPS, with more moment conditions (based on the covariate balancing condition and the score of a logit model) than coefficients $\beta_{GMM}$, which are estimated by continuously updated GMM estimation:

\begin{equation}\label{bGMM}
\tilde{\beta}_{GMM}=\arg\min_{\beta}\bar{g}_{\beta}(Z,X)^T\Sigma_{\beta}(Z,X)^{-1}\bar{g}_{\beta}(Z,X).
\end{equation}

$\bar{g}_{\beta}(Z,X)$ is the sample mean of the moment conditions and $\Sigma_{\beta}(Z,X)$ is a consistent variance estimator, described in more detail in Chapter 2.2 of \citeasnoun{ImaiRatkovic2014}.

Our third estimator of the instrument propensity score is fully non-parametric and based on kernel regression:

\begin{equation}\label{p3}
\hat{p}(x)^{lc}=\frac{\sum_{i=1}^n z_i K\Big(\frac{x_i-x}{h}\Big)}{\sum_{i=1}^n K\Big(\frac{x_i-x}{h}\Big)}.
\end{equation}

Equation~\ref{p3} corresponds to the Nadaraya-Watson (local constant) kernel estimator, where $K$ denotes the Epanechnikov kernel and bandwidth $h$ is chosen by least-squares cross-validation, i.e., by minimizing the least squares cross validation error w.r.t.\ $h$, see \citeasnoun{LiRacine06}. As an alternative to using $\hat{p}(x)^{lc}$ as weighting function, we also apply the Nadaraya-Watson estimator for estimating the outcome and treatment models in Equation~\ref{eq:late_x}, see our discussion on non-parametric estimation methods in Chapter~\ref{nonparam}. 

A practically relevant issue of treatment effect methods is thin or lacking common support (or overlap) in the propensity score, which may compromise estimation due to a non-comparability across groups, see the discussions in \citeasnoun{Imbens03}, \citeasnoun{ImWo08}, and \citeasnoun{LeSt19}. If specific propensity score values among one group are either very rare (thin common support) or absent (lack of common support) among the opposite group, as it may occur close to the boundaries of the propensity score, some units may receive a very large weight $\hat{w}_i$ in LATE estimation as provided in Equation~\ref{est}. In the case of thin common support, these observations could dominate the estimator of the LATE which may potentially entail an explosion of the variance. In the case of lacking common support, this even introduces asymptotic bias by giving a large weight to observations that are not comparable to observations in the opposite group in terms of the propensity score.

\citeasnoun{HuLeWu13} and \citeasnoun{Bodory20} consider a trimming procedure to tackle common support issues in the sample also discussed in \citeasnoun{Imbens03}, which is asymptotically unbiased if common support holds asymptotically. It is based on setting the weights of those observations to zero whose relative share of all weights within either instrument state in Equation~\ref{est} exceeds a particular threshold value in \% (denoted by $t$):

\begin{equation}\label{trim}
\hat{w}_{i}=\hat{w}_i \mathbb{I}\left\{\frac{z_i/\hat{p}(x)^{lc}}{\sum_{j=1}^{n}z_j/\hat{p}(x)^{lc}}+
                                       \frac{(1-z_i)/(1-\hat{p}(x)^{lc})}{\sum_{j=1}^{n}(1-z_j)/(1-\hat{p}(x)^{lc})}\leq t\%\right\}.
\end{equation}

We set the threshold $t$ to 5\% and trim observations based on the weights of normalized IPW, see (\ref{ipw}), irrespective of the LATE estimator considered. This changes (in finite samples) the target parameter due to discarding observations with extreme weights, but ensures common support prior to estimation. Note that our bootstrap variance estimators discussed in Section~\ref{inf} account for the stochastic nature of trimming.

\subsection{Inverse probability weighting (IPW)}\label{ipw}

Inverse probability weighting (IPW) reweighs (instrument) group-specific outcomes such that the distribution of the covariates in the total population is matched, see \citeasnoun{Hirano+00} for a more detailed discussion. We consider a normalized IPW estimator in our simulations, which performed well in several simulation studies on conditionally exogenous treatments, see for instance \citeasnoun{HuLeWu13} and \citeasnoun{BuDNMC09}. The IPW-based LATE estimator corresponds to

\begin{equation}\label{late_ipw}
\hat{\theta}_{\text{IPW}}=
\frac{
\sum_{i=1}^{n}   z_{i} y_{i}\left\{\frac{\frac{1}{  \hat{p}(x_{i})}}{\sum_{j=1}^{n}\frac{  z_j}{  \hat{p}(x_{j})}} \right\}-
\sum_{i=1}^{n}(1-z_{i})y_{i}\left\{\frac{\frac{1}{1-\hat{p}(x_{i})}}{\sum_{j=1}^{n}\frac{1-z_j}{1-\hat{p}(x_{j})}} \right\}	
}{
\sum_{i=1}^{n}   z_{i} d_{i}\left\{\frac{\frac{1}{  \hat{p}(x_{i})}}{\sum_{j=1}^{n}\frac{  z_j}{  \hat{p}(x_{j})}} \right\}-
\sum_{i=1}^{n}(1-z_{i})d_{i}\left\{\frac{\frac{1}{1-\hat{p}(x_{i})}}{\sum_{j=1}^{n}\frac{1-z_j}{1-\hat{p}(x_{j})}} \right\}	
}.
\end{equation}
															
\sloppy
The normalizations $\sum_{j=1}^{n}\frac{  z_j}{  \hat{p}(x_{j})}$ and $\sum_{j=1}^{n}\frac{1-z_j}{1-\hat{p}(x_{j})}$ ensure that the weights in curly brackets add up to one. It is easy to see that (\ref{late_ipw}) corresponds to (\ref{est}) when setting $\hat{w}_i$ in the latter to
$z_i n_1 \left\{\frac{\frac{1}{  \hat{p}(x_{i})}}{\sum_{j=1}^{n}\frac{  z_j}{  \hat{p}(x_{j})}} \right\}+
(1-z_i) n_0 \left\{\frac{\frac{1}{1-\hat{p}(x_{i})}}{\sum_{j=1}^{n}\frac{1-z_j}{1-\hat{p}(x_{j})}} \right\}$.
IPW possesses the desirable property that it can attain the semiparametric efficiency bound (implying the smallest possible asymptotic variance) derived by \citeasnoun{Ha98}, if the propensity score is estimated non-parametrically (while this is generally not the case for parametric propensity scores). Furthermore, it is computationally inexpensive and easy to implement. However, evidence in the treatment effect literature suggests that IPW also has an important drawback: at the boundaries of the support of the propensity score, estimation may be unstable and the variance may explode in finite samples, see \citeasnoun{Froe00a} and \citeasnoun{KhTa07}.

\subsection{Doubly robust estimation}\label{dr}

Doubly robust (DR) estimation combines IPW with outcome regression. It reweighs outcome models for different instrument states by the inverse of the propensity scores. Denoting the conditional mean outcomes in the presence and absence of the instrument by $\mu_z^y(x):=E[Y_i \vert Z=z_i,X_i=x_i]$ and $\mu_z^d(x):=E[D_i \vert Z_i=z_i,X_i=x_i]$, the DR LATE estimator corresponds to

\begin{equation}\label{late_dr}
\hat{\theta}_{\text{DR}}=
\frac{
\frac{1}{n}\sum_{i=1}^{n}\left(\hat{\mu}_1^y(x)+\frac{z_i(y_i-\hat{\mu}_1^y(x))}{\hat{p}(x)}-\hat{\mu}_0^y(x)-\frac{(1-z_i)(y_i-\hat{\mu}_0^y(x))}{1-\hat{p}(x)}\right)
}{
\frac{1}{n}\sum_{i=1}^{n}\left(\hat{\mu}_1^d(x)+\frac{z_i(d_i-\hat{\mu}_1^d(x))}{\hat{p}(x)}-\hat{\mu}_0^d(x)-\frac{(1-z_i)(d_i-\hat{\mu}_0^d(x))}{1-\hat{p}(x)}\right)
}.
\end{equation}

For non-binary outcomes, we run OLS regression to compute $\hat{\mu}_z^y(x)=x_i^T\hat{\beta}_{z,OLS}$. For binary outcome and treatment variables, we apply probit regression to compute $\hat{\mu}_z(x)=\Phi(x_i^T\hat{\beta}_{z,ML})$. The coefficients $\beta_z$ are estimated in the subgroups with $Z_i\in\{0,1\}$. Differently to IPW, which exclusively relies on reweighing by the propensity score, the DR estimator remains consistent even if either $\hat{p}(x)$ or $\hat{\mu}_z(x)$ is misspecified, as it makes use of both, the treatment and outcome models. If both are correctly specified, the DR estimator is semi-parametrically efficient, as discussed in \citeasnoun{RobinsRotnitzkyZhao1994}.

\subsection{Matching}

Matching is based on assigning (matching) to each observation in one instrument state one or more units in the other instrument state with comparable covariates, in order to estimate the LATE based on the ratio of average differences in the outcome and the treatment across units with and without instrument in the matched sample. We implement multiple variants of two types of matching methods, pair and radius matching, to estimate $\theta$.

Pair (or one-to-one) matching with replacement (implying that an observation may be matched several times) as discussed in \citeasnoun{Ru73a} matches to each reference observation exactly the observation with the most similar covariates in the opposite instrument state. This implies the following weights in  Equation~\ref{est}:

\begin{equation}\label{pairm}
\varpi_{i,j}=\mathbb{I}\left\{ \ |\hat{f}(x_{i},x_{j})|=\underset{k:Z_{k}\in\{0,1\}}{\min }%
|\hat{f}(x_{i},x_{k})|\right\}.
\end{equation}

$\varpi_{i,j}$ is the weight of the outcome (or treatment) of observation $j$ in one instrument group (e.g., $Z_j=0$) when matched to unit $i$ in the opposite group (e.g., $Z_i=1$), with $Z_k=1-Z_i$. $\mathbb{I}\{\cdot \}$ is the indicator function, which is one if its argument is true and zero otherwise. $\hat{f}(\cdot)$ is a function of the difference in covariates between observations $i$ and $j$. For example, the function could be defined as the difference in propensity score estimates of observations $i$ and $j$ in the case of propensity score matching or as a distance metric w.r.t.\ the covariate values of $i$ and $j$ like the Euclidean distance in the case of matching directly on the covariates. In pair matching, all weights are zero except for the observation $j$ with the smallest difference with reference unit $i$, which receives a weight of one. For propensity score matching, we base the weights on the distance of the one-dimensional propensity score, while for direct matching, we use a normalized Euclidean distance metric, where differences in the covariates are weighed by the inverse of the variances of $X_i$. Because only one observation is matched to each unit irrespective of the sample size and the potential availability of several suitable matches with similar covariates, pair matching is not efficient (i.e., does not attain the smallest possible variance asymptotically). On the other hand, it is likely more robust to propensity score misspecification than IPW, in particular if the misspecified propensity score model is only a monotone transformation of the true model, see for instance \citeasnoun{Zh08}, \citeasnoun{MiTc09}, \citeasnoun{Waernbaum2012}, and \citeasnoun{HuLeWu13}.

Radius matching as discussed in \citeasnoun{RosenbaumRubin1985} and \citeasnoun{DehejiaWahba99} uses \emph{all} matches with propensity scores within a predefined radius around the reference unit, which trades off some bias in order to increase efficiency (or precision). This approach expectedly works relatively well if several comparable potential matches are available for a reference unit. In the simulations, we consider the radius matching algorithm of \citeasnoun{LeMiWu11}, which performed well in \citeasnoun{HuLeWu13}, who also provide details on the radius matching-related weighting function $\hat{w_i}$ in Equation~\ref{est}. The estimator combines distance-weighted radius matching, where units within the radius are weighted proportionally to the inverse of their distance to the reference unit, with a regression-based bias correction, see \citeasnoun{Ru79} and \citeasnoun{AbIm11}. For the bias correction, we apply an OLS regression adjustment for $Y$ and a probit regression adjustment for $D$ to remove small and large sample bias due to mismatches. \citeasnoun{HuLeSt2014} provide a detailed description of the estimator. As in \citeasnoun{LeMiWu11}, the radius size in our simulations is defined as a function of the distribution of distances between reference units and matches in pair matching. Namely, it is set to 3 times the maximum pair matching distance. Note that we include radius matching both with and without conditioning on the covariate `age at first birth' in addition to the propensity score to account for this influential confounder.

\subsection{Parametric regression estimators}\label{param}

In our simulations, IPW, DR estimation, and matching are implemented with various degrees of flexibility in terms of parametric assumptions. We consider both semi-parametric versions based on parametric propensity score models, $\hat{p}(x)^{probit}$ and $\hat{p}(x)^{CBPS}$, as well as fully non-parametric estimators using the non-parametric propensity scores $\hat{p}(x)^{lc}$ (based on a local constant kernel regression) or when directly conditioning on $X_i$. For non-parametric DR estimation, also the conditional means of the binary treatment and binary (or non-binary) outcome $\hat{\mu}_z(x)$ are estimated by local constant (or local linear) kernel regressions. 

In addition, we also consider several parametric treatment effect estimators. The first parametric approach computes the LATE by differences in the conditional mean functions $\hat{\mu}_z(x)$, which are estimated by OLS regressions for non-binary outcomes and by probit regressions for the treatment and binary outcome variables (see Section~\ref{dr}). Formally, this regression-based LATE estimator corresponds to the following expression:

\begin{equation}\label{late_reg_param}
\hat{\theta}_{\text{REGR}}=
\frac{
\frac{1}{n} \sum_{i=1}^n (\hat{\mu}_1^y(x)-\hat{\mu}_0^y(x))
}{
\frac{1}{n} \sum_{i=1}^n (\hat{\mu}_1^d(x)-\hat{\mu}_0^d(x))
}.
\end{equation}

Furthermore, we apply two-stage least-squares (2SLS) estimation, which was also applied by \citeasnoun{Angrist+98} for analysing the data our simulations are based on. 2SLS may be regarded as a benchmark method for instrumental variable estimation under the assumption of homogeneous treatment effects. Formally, the 2SLS estimator is given by

\begin{eqnarray}\label{2sls}
\hat{\theta}_{\text{2SLS}}=\left[
\left(\frac{1}{n}\sum_{i=1}^n \tilde{x}_i^T \tilde{z}_i\right) \left(\frac{1}{n}\sum_{i=1}^n \tilde{z}_i^T \tilde{z}_i\right)^{-1} \left(\frac{1}{n}\sum_{i=1}^n \tilde{z}_i^T \tilde{x}_i\right)
\right]^{-1} \notag\\
\left(\frac{1}{n}\sum_{i=1}^n \tilde{x}_i^T \tilde{z}_i\right) \left(\frac{1}{n}\sum_{i=1}^n \tilde{z}_i^T \tilde{z}_i\right)^{-1} \left(\frac{1}{n}\sum_{i=1}^n \tilde{z}_i^T y_i\right),
\end{eqnarray}

where $\tilde{x}_i:=(1,x_{i,1},\cdots,x_{i,K})$, $\tilde{z}_i:=(\tilde{x}_i,z_i)$, and $K$ denotes the number of covariates $X_i$. Note that in our just-identified settings with one treatment and one instrumental variable, the 2SLS estimator is numerically identical to the limited information maximum likelihood (LIML) estimator.

\subsection{Further non-parametric estimators}\label{nonparam}

We analyze the performance of three further non-parametric estimation methods that do not impose any functional form assumptions on the regression functions of the outcome or the treatment.

Firstly, we apply the generalized random forest (GRF) method, a non-parametric estimator introduced by \citeasnoun{ATW18}. GRF is a variant of random forest algorithms, a machine learning approach, see for instance the discussion in \citeasnoun{LUW20} and citations therein. As described in \citeasnoun{Breiman2001}, random forests consist of averaging the predictions of many decision trees applied to different subsamples that are repeatedly drawn from the original data. In each of these samples, a decision tree partitions the space of $X_i$ into a set of rectangles and computes the fitted value of $Y_i$ as the average outcome in each of the rectangles. The partitions are chosen in a data-driven way such that the predictive performance is maximized (e.g.\ by minimizing the squared residuals based on the fitted values in each rectangle). A popular estimation algorithm for decision trees is CART (classification and regression tree), see for instance Chapter 9.2 in the textbook of \citeasnoun{hastie01statisticallearning}. 

GRF shares the core features of `traditional' random forest algorithms like recursive partitioning, subsampling from the original data, and the random selection of a subset of covariates at each partitioning step. However, as a methodological twist, GRF uses a gradient-based partitioning scheme and a particular (so-called `honest') sample splitting technique (within any of the drawn sub-samples) that avoids overfitting the predictive models to the specificities of the data, see \citeasnoun{WagerAthey2018}. Using the conditional expectation function $\mu_z(X_i)$ in Section~\ref{dr} and applying the GRF to estimate the latter for the outcome and the treatment to obtain $\hat{\mu}_{z,RF}^Y(x)$ and $\hat{\mu}_{z,RF}^D(x)$ for $z\in\{0,1\}$ (where the subscript RF indicates the random forest approach), we compute the LATE as follows:

\begin{equation}\label{late_rf}
\hat{\theta}_{\text{RF}}=
\frac{
\frac{1}{n} \sum_{i=1}^n (\hat{\mu}_{1,RF}^Y(x)-\hat{\mu}_{0,RF}^Y(x))
}{
\frac{1}{n} \sum_{i=1}^n (\hat{\mu}_{1,RF}^D(x)-\hat{\mu}_{0,RF}^D(x))
}.
\end{equation}

Algorithm~1 in \citeasnoun{ATW18} provides more details on the GRF method. We estimate the conditional expectations in Equation~\ref{late_rf} using the default options of the \textit{causal\_forest} function of the \textit{grf} package for the statistical software \textit{R}, see \citeasnoun{grf2020}.

Alternatively, we could have estimated the predictions $\hat{\mu}_{z,RF}(x)$ by standard Breiman-type random forests \cite{Breiman2001}, or considered double/debiased machine learning estimators based on Neyman orthogonal scores \cite{DML18} or alternative causal forest algorithms \cite{MCF19} for estimation. Such methods would also be appropriate to evaluate the finite sample performance of LATE estimators in high-dimensional settings with many potential covariates, as they are capable of selecting control variables in a data-driven way, an interesting topic that we leave for future research.

Secondly, we use non-parametric kernel regression to estimate the conditional mean functions $\hat{\mu}_z(x)$ defined in Section~\ref{dr}, see the subscript NP in the respective estimates in Equation~\ref{late_np}. For non-binary outcomes, $\hat{\mu}_{z,NP}^y(x)$ is estimated by local linear kernel regression, for the binary outcome and treatment variables, $\hat{\mu}_{z,NP}^y(x)$ and $\hat{\mu}_{z,NP}^d(x)$ are estimated by local constant kernel regression.

\begin{equation}\label{late_np}
\hat{\theta}_{\text{NP}}=
\frac{
\frac{1}{n}\sum_{i=1}^n(\hat{\mu}_{1,NP}^y(x)-\hat{\mu}_{0,NP}^y(x))
}{
\frac{1}{n}\sum_{i=1}^n(\hat{\mu}_{1,NP}^d(x)-\hat{\mu}_{0,NP}^d(x))
}.
\end{equation}

Finally, we consider a (naive) LATE estimator that is based on the mean differences of the outcome and treatment variables, respectively, across instrument states, which in contrast to the other methods does not control for the covariates. Therefore, the consistency of this approach provided in Equation~\ref{late_mean} generally requires that the IV assumptions hold unconditionally, i.e., without conditioning on $X$.

\begin{equation}\label{late_mean}
\hat{\theta}_{\text{MEANS}}=
\frac{
\frac{1}{n_1}\sum_{i=1}^{n} y_i z_i-\frac{1}{n_0}\sum_{i=1}^{n} y_i(1-z_i)
}{
\frac{1}{n_1}\sum_{i=1}^{n} d_i z_i-\frac{1}{n_0}\sum_{i=1}^{n} d_i(1-z_i)
}.
\end{equation}

Table~\ref{T1} summarizes the LATE estimators analysed in our simulation study along with the corresponding conditioning sets.

\begin{table}[htbp]
\centering\footnotesize
\caption{Point estimators}
\label{T1}
\begin{adjustbox}{max width=\textwidth}
\begin{tabular}{l|ccccc}
\hline
Estimators & \multicolumn{5}{c}{Conditioning sets} \\	
\hline
& \multicolumn{3}{c}{\underline{propensity scores}} & \underline{covariates} & \underline{none} \\	
&        &      & local    & & 	\\				
& probit & cbps & constant & & \\				
\hline
Inverse probability weighting     & $ipw^{probit}$       & $ipw^{cbps}$       & $ipw^{lc}$       &                 &         \\
Doubly robust                     & $dr^{probit}$        & $dr^{cbps}$        & $dr^{lc}$        &                 &         \\
Pair matching                     & $pairmatch^{probit}$ & $pairmatch^{cbps}$ & $pairmatch^{lc}$ & $pairmatch^{x}$ &         \\
Radius matching on propensity score                   & $radmatch^{probit}$  & $radmatch^{cbps}$  & $radmatch^{lc}$  &                 &         \\
Radius matching on propensity score + covariate & $radmatchx^{probit}$ & $radmatchx^{cbps}$ & $radmatchx^{lc}$ &                 &         \\
Parametric regressions            &                      &                    &                  & $reg$           &         \\
2SLS                              &                      &                    &                  & $tsls$          &         \\
Random forests                    &                      &                    &                  & $randforest$    &         \\
Non-parametric regressions        &                      &                    &                  & $reg^{kernel}$  &         \\
Mean differences (ignoring covariates)           &                      &                    &                  &                 & $means$ \\
\hline	
\end{tabular}
\end{adjustbox}
\end{table}

\subsection{Inference}\label{inf}

Treatment effect estimation frequently relies on the non-parametric bootstrap for statistical inference \cite{Ef79,Horowitz2001}. In an extensive simulation study with a conditionally exogenous treatment, \citeasnoun{Bodory20} find evidence that variance estimation of treatment effect estimators based on bootstrap procedures outperforms asymptotic variance approximations in terms of rejection and coverage probabilities in finite samples. These results even hold for matching estimators in small samples, despite the inconsistency of the non-parametric bootstrap for the (non-smooth) pair matching estimator, see the discussion in \citeasnoun{AbadieImbens06}.

For this reason, we apply the non-parametric bootstrap to estimate the standard errors of all LATE estimators. This algorithm randomly draws $B$ bootstrap samples of size $n$ (the size of a simulation sample) with replacement out of each simulation sample and estimates the LATE in every draw. Denoting the $B$ bootstrapped LATE estimators by $\hat{\theta}^b$, with $b\in\{1,2,\dots,B\}$, we estimate the standard error $\sigma$ of a LATE estimator by

\begin{equation}\label{sigma}
\hat{\sigma}=\sqrt{\frac{1}{B-1}\sum_{b=1}^B\left(\hat{\theta}^b-\frac{1}{B}\sum_{b=1}^B\hat{\theta}^b\right)^2}.
\end{equation}

In line with \citeasnoun{Bodory20}, we set $B=199$. Bootstrapping naturally accounts for heteroscedasticity as well as uncertainty due to trimming of influential observations and propensity score estimation.

\section{Simulation design with empirical data}\label{datasim}

Simulations often rely on randomly generated data drawn from a probability distribution that is selected by the researcher. However, the data generating processes (DGPs) of such simulations may appear somewhat arbitrary in the sense that they might be far from reflecting typical associations between variables in empirical data. To improve upon this caveat, \citeasnoun{HuLeWu13} suggest a simulation design based on empirical data, also called Empirical Monte Carlo Study (EMCS), an idea that has been subsequently applied in several papers, see for instance \citeasnoun{FrHuWi14}, \citeasnoun{HuLeMe2016}, and \citeasnoun{Bodory20}, among others. Briefly, the idea of an EMCS is to randomly draw small samples from large real data sets while relying as much as possible on the empirical associations between the variables when generating the simulation designs.

Our study follows this EMCS approach to evaluate the properties of various IV estimators of the LATE, with the aim that the simulation designs are more closely linked to real world data. However, we point out that also in an EMCS, several important choices about the simulation features are to be made by the researcher such that the DGPs are not fully determined by the data, see the caveats raised by \citeasnoun{AdSl2013}. The remainder of this section describes the implementation of our EMCS. We first present the empirical labor market data underlying our simulations and then provide the steps for generating the various simulation designs.

\subsection{Database}\label{database}

Our simulations are based on empirical data analysed in \citeasnoun{Angrist+98}, who aim at exploiting exogenous variation in family size to evaluate the treatment effect of fertility, defined as having at least three vs. two children, on female labor supply. This database is well suited to analyze the finite sample properties of IV estimators by means of an EMCS for several reasons. First, the data set is large, as it comprises 394840 observations and therefore easily allows one to draw many different random subsamples. Furthermore, the data contains a strong instrument that importantly affects fertility, namely twins at second birth.\footnote{There may be cases where the randomness of twin births is violated, see \citeasnoun{FGV18} for a discussion on dizygotic twinning. In our simulation study, we artificially generate random and non-random instrument assignments.} Finally, it provides demographic information on the mothers, which may be used as covariates to control for potential confounders of the instrument and the outcome.

Coming from the 1980 Census Public Use Micro Samples (PUMS), the data set contains information on young mothers aged 21 to 35, all of which gave birth to at least two children. Our analysis considers two different outcomes, the number of weeks worked within one year (with 43\% zeros) and an indicator for being employed at all in that year. The binary treatment variable indicates if a mother has more than two kids (treatment is one) or two kids (treatment is zero). The binary instrumental variable is one if a mother gave birth to twins at second birth and zero otherwise. The covariates considered in our simulation include mother's age, mother's age at first birth, race, and quarter of birth.

\begin{table}[H]
\centering
\caption{Descriptive statistics of the full sample}
\label{tabdescr}
\begin{adjustbox}{max width=\textwidth}
\begin{tabular}{l|cccccccccc}
\hline
Variables & \multicolumn{5}{c}{More than two kids} & \multicolumn{5}{c}{Twins at second birth} \\
\hline
& \multicolumn{2}{c}{No} & \multicolumn{2}{c}{Yes} & & \multicolumn{2}{c}{No} & \multicolumn{2}{c}{Yes} \\
\cmidrule{2-5} \cmidrule{7-10}       	      	
& mean & std & mean & std & st.diff &  mean & std & mean & std & st.diff \\
\hline
\textit{Outcomes} \\
Weeks worked / 10                    & 2.32 & 2.26 & 1.73 & 2.13 & 19 & 2.09 & 2.23 & 1.89 & 2.19 & 6 \\
Worked for pay (binary)              & 0.61 &      & 0.49 &      & 17 & 0.57 &      & 0.52 &      & 7 \\
\hline
\textit{Treatment} \\
More than two kids (binary)          &      &      &      &      &      & 0.40 &      & 1    &      & 123 \\
\hline
\textit{Instrument} \\
Twins at second birth (binary)       & 0    &      & 0.02 &      & 15  \\
\hline
\textit{Covariates: Mothers' characteristics} \\
Age / 10                             & 2.98 & 0.36 & 3.05 & 0.34 & 14 & 3.01 & 0.35 & 3.04 & 0.34 & 5 \\
Age at first birth /10               & 2.06 & 0.30 & 1.94 & 0.27 & 29 & 2.01 & 0.29 & 2.04 & 0.30 & 5 \\
African American (binary)            & 0.10 &      & 0.15 &      & 11 & 0.12 &      & 0.15 &      & 6 \\
Other race (binary)                  & 0.18 &      & 0.18 &      & 1  & 0.18 &      & 0.20 &      & 3 \\
First quarter of birth (binary)      & 0.24 &      & 0.24 &      & 1  & 0.24 &      & 0.24 &      & 1 \\
Second quarter of birth (binary)     & 0.24 &      & 0.25 &      & 1  & 0.24 &      & 0.24 &      & 0 \\
Third quarter of birth (binary)      & 0.27 &      & 0.27 &      & 0  & 0.27 &      & 0.27 &      & 0 \\
Fourth quarter of birth (binary)     & 0.25 & 0.43 & 0.25 & 0.43 & 0  & 0.25 & 0.43 & 0.26 & 0.44 & 1 \\
\hline
$X_i\tilde{\beta}$                   &-0.35 & 0.36 &-0.14 & 0.36 & 42 &-2.39 & 0.05 &-2.38 & 0.05 &10 \\
$\Phi(X_i\tilde{\beta})$             & 0.37 & 0.13 & 0.45 & 0.14 & 42 & 0.01 & 0    & 0.01 & 0    &10 \\
\hline
Number of obs., Pseudo-R$^2$ in \% & \multicolumn{2}{c}{236089} & \multicolumn{2}{c}{158751} & 10.5 &\multicolumn{2}{c}{391460} & \multicolumn{2}{c}{3380} & 0.2   \\
\hline
\end{tabular}
\end{adjustbox}
\caption*{\scriptsize Notes: The statistics mean, std, and st.diff stand for mean, standard deviation, and standardized difference in percent, respectively. The standardized difference is defined as the absolute difference of means normalized by the square root of the sum of estimated variances of the particular variables in both subsamples (see e.g. Imbens and Wooldridge (2009),p. 24). $\tilde{\beta}$ denotes the estimated probit coeffcients and $\Phi(X_i\tilde{\beta})$ is the cumulative distribution function of the standard normal distribution evaluated at $X_i\tilde{\beta}$. Pseudo-R$^2$ is the so-called Nagelkerke's R$^2$: $(1-exp(-(-2(l_0-l_1))/n)/(1-exp(-(-2l_0)/n)$, where $l_0$ and $l_1$ are the log likelihoods for the null and full model, respectively, and $n$ denotes the number of observations.}
\end{table}

Table~\ref{tabdescr} reports descriptive statistics of the database, by treatment indicator (more than two kids) and the instrument (twins at second birth). The upper part presents descriptives for the two labor market outcomes `weeks worked' (in weeks) and `worked for pay' (binary). There are large differences between the outcomes of the treated and non-treated in terms of the standardized difference statistic as suggested by \citeasnoun{RosenbaumRubin1985}  (the literature considers values around 20 and above as severely unbalanced). The line underneath the outcomes in Table~\ref{tabdescr} gives details on the treatment variable. Not surprisingly, the treatment fully complies with the instrument if the latter equals one, because all mothers with twins at second birth ($Z_i=1$) necessarily have more than two children ($D_i=1$). The subsequent row of Table~\ref{tabdescr} provides information on the instrument. It reveals that 2\% of women with at least three children have twins at their second birth. Considering the covariates, the standardized differences show that mothers' characteristics are partly unbalanced across treatment states, whereas they are well balanced across instrument states, in line with a randomly assigned instrument. The randomness of the instrument is also supported by the pseudo-R2 statistic with a value of 0.2\% when regressing the instrument on the covariates.

\subsection{Simulation designs}\label{simdesign}

Data generating processes (DGPs) may differ in (infinitely) many dimensions. We select ten practically relevant dimensions for varying the specifications of our simulation models. These dimensions include: effect homogeneity vs.\ heterogeneity, randomness vs.\ non-randomness of the instrument, varying levels of instrument strength, binary vs.\ non-binary outcome distributions, and different sample sizes. Summary statistics of all DPGs are presented in Table~\ref{sum_dgps}.

We start by assuming homogeneous treatment effects with a randomly assigned instrument and the empirically observed instrument strength. To evaluate the performance of the estimators under these conditions, we define a new \textit{population} for which the true LATE is equal to zero. To this end, we drop all 3380 observations from the database who receive the instrument ($Z_i=1$). Among the remaining 391460 observations with instrument state $Z_i=0$ (no twins at second birth), there is no reduced form effect of the instrument on the outcome or first stage effect of the instrument on the treatment, such that there exists no LATE. After that, we create a pseudo-instrument and artificially assign $Z_i=1$ to those who are similar to the 3380 discarded observations in terms of observed characteristics. This similarity is determined by $1:M$ matching on the covariates without replacement. By setting $M=58$, we assign $Z_i=1$ to approximately half of the observations, see column 4 of Table~\ref{sum_dgps}. In addition, we set the treatment state of everyone with $Z_i=1$ to $D_i=1$ (as in the original database) to maintain the empirically observed instrument strength. Finally, we draw small samples from our new \textit{population} to compare the finite sample properties of alternative LATE estimators.

To simulate specifications with a weaker instrument, we reduce the first stage effect by lowering the impact of $Z_i$ on $D_i$. Instead of setting all observations with $Z_i=1$ to $D_i=1$, we change the treatment status from zero to one only for those with $Z_i=1$ for which the condition $D_i=\mathbbm{1}(u_i>1.25 )$ holds. $\mathbbm{1}(\cdot)$ denotes the indicator function which is one if its argument is true, otherwise it is zero, and $u_i$ is a standard normally distributed random variable. Column 7 of Table~\ref{sum_dgps} displays the first stage coefficients for the different DGPs.

The randomness of the instrument implies that the covariates are balanced across groups. To mimic a non-random assignment of the instrument, we increase the magnitude of instrument selectivity in the following way. We first estimate the propensity score $\hat{p}(1.5X_i)^{probit}$ (see Equation~\ref{p1}) using the original database. Then, we change the instrument status $Z_i$ from zero to one for observations with characteristics similar to the 3380 observations dropped from the original database (with $Z_i=1$). We obtain such similar matches by $1:M$ matching on the estimated propensity score $\hat{p}(1.5X_i)^{probit}$, with $M=22$. Next, we assign $D_i=1$ to all observations with $Z_i=1$. Based on this modified data set, we increase the selection into the instrument by discarding the best matches for the newly created observations with $Z_i=1$ among observation with $Z_i=0$. To find the best matches to be discarded, we apply $1:M$ matching on a newly estimated propensity score $\hat{p}(X_i)^{probit}$ (with the modified instrument assignments), where $M=3$. The selectivity of the instrument is provided in columns 5 and 6 in Table~\ref{sum_dgps}.

To model a scenario with non-constant treatment effects, we introduce effect heterogeneity with respect to age and race as follows. We add to the existing control variables squared and cubic terms of both age variables (`age' and `age at first birth') and interact the unmodified age variables with the indicator variable for African Americans. This new set of control variables for settings with effect heterogeneity is denoted by $X_i^{het}$ for each unit $i$. We generate $Y_i$ and $D_i$ in each simulation sample according to the rules $Y_i=Y_{i,1}^d Z_i+Y_{i,0}^d (1-Z_i)$ and $D_i=D_{i,1}Z_i+D_{i,1}(1-Z_i)$. To this end, we compute the non-binary potential outcomes based on the equation $Y_{i,z}^d=X_i^{het}\hat{\beta}_{OLS}+\hat{\sigma}v_i$, where $v_i$ is a standard normally distributed random variable.  $\hat{\beta}_{OLS}$ and $\hat{\sigma}$ are the coefficients and residual standard deviation of OLS regressions in subsamples by instrument state $Z_i\in\{0,1\}$ of our new \textit{population}. The binary potential outcomes are computed based on $Y_{i,z}^d=\mathbbm{1}(X_i^{het}\hat{\beta}_{probit}+v_i>0)$, where $\mathbbm{1}(\cdot)$ is the indicator function and $\hat{\beta}_{probit}$ are the coefficients estimated from probit models in subsamples by instrument state of our new \textit{population}. The potential treatments $D_{i,1}$ are set to one, whereas $D_{i,0}$ is computed analogously to $Y_{i,z}^d$ in the binary outcome case.

We combine these variations in the DGPs with respect to effect heterogeneity, instrument strength, and instrument selectivity with smaller and larger sample sizes of 1000 and 2000, respectively, and with binary and non-binary outcome distributions. We run 2000 simulations for the smaller and 1000 simulations for the larger samples. Table~\ref{sum_dgps} presents summary statistics of the DGPs considered in our simulation study.

\begin{table}[!t]
\centering
\caption{Summary statistics (DGPs)}
\label{sum_dgps}
\begin{adjustbox}{max width=\textwidth}
\begin{tabular}{llrrrrrrrrr}
\hline
Random &Empiri-    &Share     &Share     &St.diff.     &Pseudo-      &First  &\multicolumn{2}{c}{LATE}&\multicolumn{2}{c}{Trimming} \\
selec- &cally      &(\%)      &(\%)      &(\%)         &R$^2$ (\%)   &stage  &  1000 & 2000           &\multicolumn{2}{c}{(\%)}     \\
tion   &observed   &More      &Twins-2   &Twins-2      &Twins-2      &(\%)   &       &                & 1000 & 2000                 \\
Twins-2&strength   &than 2    &          &$P(Z\vert X)$&$P(Z\vert X)$&Twins-2&       &                &      &                      \\
       &Twins-2    &children  &          &             &             &       &       &                &      &                      \\
\hline
\multicolumn{11}{c}{Effect homogeneity (Weeks worked)} \\
\hline
Yes & Yes & 70 & 50 & 23 & 4  & 60 & 0    & 0    & 0.03 & 0.02 \\  
Yes & No  & 43 & 50 & 23 & 4  & 6  & 0    & 0    & 0.03 &	0.02 \\  
No  & Yes  & 70 & 50 & 51 & 16 & 61 & 0    & 0    & 0.06 &	0.03 \\  
No  & No   & 43 & 50 & 51 & 16 & 7  & 0    & 0    & 0.06 &	0.03 \\  
\hline
\multicolumn{11}{c}{Effect homogeneity (Worked for pay)} \\
\hline
Yes & Yes  &&&&&                 & 0    & 0    &&	\\ 
Yes & No   &&&&&                 & 0    & 0    && \\ 
No  & Yes  &&&&&                 & 0    & 0    &&	\\ 
No  & No   &&&&&                 & 0    & 0    &&	\\ 
\hline
\multicolumn{11}{c}{Effect heterogeneity (Weeks worked)} \\
\hline
Yes & Yes& 71 & 50 & 23  & 4  & 57 &-0.91 & 0.27 & 0.03 & 0.02 \\  
Yes & No & 85 & 50 & 23  & 4  & 30 & 0.52 & 0.51 & 0.03 & 0.02 \\  
No  & Yes& 71 & 50 & 51 & 16 & 58 &-0.09 &-0.09 & 0.06 &	0.05 \\  
No  & No & 82 & 50 & 51 & 16 & 34 &-0.21 &-0.21 & 0.06 &	0.05 \\  
\hline
\multicolumn{11}{c}{Effect heterogeneity (Worked for pay)} \\
\hline
Yes & Yes &&&&&                  & 0.00 & 0.00 &&	\\  
Yes & No  &&&&&                  & 0.01 & 0.01 &&	\\  
No  & Yes &&&&&                  &-0.01 &-0.01 &&	\\  
No  & No  &&&&&                  &-0.02 &-0.02 &&	\\  
\hline
\end{tabular}
\end{adjustbox}
\caption*{\scriptsize Note: The numbers in columns 3-9 are averages over all simulation replications with samples of sizes 1000 and 2000. 'St.diff': Standardized difference defined as absolute mean difference normalized by the square root of the sum of the estimated variances of the particular variables in both subsamples \cite*[p.~24]{ImWo08}.
'std': standard deviation. Pseudo-R$^2$ is the so-called Nagelkerke's R$^2$: $(1-exp(-(-2(l_0-l_1))/n)/(1-exp(-(-2l_0)/n)$, where $l_0$ and $l_1$ are the log likelihoods for the null and full model, respectively, and $n$ is the sample size. Trimming: Share of dropped units due to the trimming rule in Equation~\ref{trim} or non-convergence of propensity scores. All statistics that do not dependend on the outcome variables are only presented for Weeks worked.}
\end{table}

\clearpage

\section{Results}\label{results}

This section presents results about the finite sample performance of various LATE estimators across different DGPs. We rank the estimators by their coverage rates, which are defined as the share of simulations in which the true LATE is included in the 95\% confidence interval of the respective LATE estimator. We recall that the standard errors for computing those confidence intervals come from the non-parametric bootstrap, as discussed in Section \ref{inf}.  For the sake of brevity, we subsequently only discuss a selection of our results, which conveys the main message of our findings. In the Appendix, we include more detailed results.

\begin{table}[H]
\centering
\begin{threeparttable}
\caption{Coverage rates and intervals}
\label{results1}
\begin{tabular}{lcccc}
\toprule
point      & coverage & diff & confidence & diff \\
estimators & rates    & (pp) & intervals  & (\%) \\
\midrule
$randforest$        & 95.0 & 0.0 & 228.8 & 8.0  \\
$radmatch^{probit}$  & 95.1 & 0.1 & 356.5 & 68.3 \\
$reg^{kernel}$   & 94.8 & 0.2 & 329.4 & 55.5 \\
$pairmatch^{x}$   & 94.6 & 0.4 & 265.6 & 25.4 \\
$ipw^{cbps}$   & 95.5 & 0.5 & 290.4 & 37.1 \\
$radmatchx^{cbps}$ & 94.4 & 0.6 & 284.5 & 34.3 \\
$radmatchx^{probit}$ & 94.1 & 0.9 & 265.6 & 25.4 \\
$reg$  & 96.0 & 1.0 & 211.8 & 0.0  \\
$tsls$      & 96.0 & 1.0 & 260.9 & 23.2 \\
$dr^{probit}$   & 96.0 & 1.0 & 257.5 & 21.6 \\
$means$   & 90.5 & 4.5 & 234.5 & 10.7 \\
\bottomrule
\end{tabular}
\begin{tablenotes}
\footnotesize
\item Notes: `diff' indicates the difference to the left best performer in percentage points (pp) or in percent (\%).
\end{tablenotes}
\end{threeparttable}
\end{table}

Table~\ref{results1} provides the average coverage rates and lengths of confidence intervals across all DGPs of any parametric, semi-parametric, or non-parametric LATE estimator which performs best (in terms of coverage) in at least one of the ten DGPs discussed in Section~\ref{simdesign}.
We find that only the non-parametric random forest-based LATE estimator described in Equation~\ref{late_rf} attains exactly the nominal coverage size of 95\% on average. Furthermore, its average length of confidence intervals is the second shortest among the estimators analyzed in Table~\ref{results1}, 8\% larger than the average interval of the parametric regression estimator, the nominal size of which is 96\%. Conditional on obtaining a decent coverage, a short confidence interval is desirable in terms of precision, as it implies a lower estimation uncertainty. Three out of the four LATE estimators whose average coverage rates come closest to 95\% are non-parametric, with those of  non-parametric kernel regression (94.8\%) and pair matching on the covariates (94.6\%) having a minor under-coverage. Also semi-parametric radius matching on the propensity score performs decent in terms of coverage rates, with the probit-based version attaining an average rate of 95.1\%, and two further versions achieving 94.4\% and 94.1\%, respectively. Furthermore, also IPW using the CBPS method for propensity score estimation reaches a satisfactory average coverage rate of 95.5\%. However, the average length of the confidence intervals of radius matching, kernel regression, pair matching, and IPW is substantially larger than that of parametric regression or of the random forest.

\begin{table}[H]
\centering
\begin{threeparttable}
\caption{Average absolute biases, standard deviations, and root mean square errors}
\label{results2}
\begin{tabular}{lcccccccc}
\toprule
point      & bias   & diff & sd  & diff & rmse & diff & bias & diff \\
estimators & effect & (\%) &     & (\%) &      & (\%) & se   & (\%) \\
\midrule
$randforest$        & 0.6 & 0.0   &  8.0 & 14.1 & 8.0  & 11.9 &  1.7 & 0.0   \\
$radmatch^{probit}$  & 1.3 & 100.1 & 12.0 & 70.0 & 12.0 & 67.3 & 10.8 & 540.7 \\
$reg^{kernel}$   & 1.4 & 124.8 &  7.0 & 0.0  & 7.2  & 0.0  &  4.2 & 149.3 \\
$pairmatch^{x}$  & 0.8 & 21.4  &  9.9 & 40.3 & 9.9  & 37.6 &  5.8 & 247.4 \\
$ipw^{cbps}$   & 1.2 & 85.4  &  7.4 & 5.4  & 7.5  & 4.7  &  4.4 & 162.4 \\
$radmatchx^{cbps}$ & 1.1 & 76.9  &  8.6 & 22.9 & 8.7  & 21.5 &  4.1 & 142.8 \\
$radmatchx^{probit}$ & 1.2 & 86.4  &  9.7 & 38.1 & 9.8  & 36.4 &  3.4 & 100.1 \\
$reg$   & 0.7 & 10.1  &  7.3 & 3.7  & 7.3  & 1.9  &  4.0 & 140.0 \\
$tsls$    & 0.7 & 11.3  &  7.9 & 12.7 & 7.9  & 10.6 &  3.8 & 124.0 \\
$dr^{probit}$    & 0.8 & 31.9  &  7.6 & 8.6  & 7.7  & 6.9  &  3.9 & 133.3 \\
$means$    & 3.4 & 440.1 &  7.9 & 11.7 & 8.8  & 21.9 &  5.6 & 235.3 \\
\bottomrule
\end{tabular}
\begin{tablenotes}
\footnotesize
\item Notes: `bias effect' denotes the absolute bias from the true treatment effect, `sd' is the standard deviation of the estimator, `rmse' stands for root mean squared error, and `bias se' indicates the median bias of the estimated bootstrap standard error. `diff' indicates the difference to the left best performer in percent (\%).
\end{tablenotes}
\end{threeparttable}
\end{table}

The coverage accuracy of the different estimation methods is related to the bias and variance of the LATE estimators, as well as the bias of the bootstrap-based standard error. Table~\ref{results2} provides details on these statistics. We find that two non-parametric methods perform best when considering the bias of LATE estimation, the standard deviation, and the mean squared error (i.e., the sum of the squared bias and the variance), as well as the bias of the standard error (relative to a LATE estimator's true standard deviation). The random forest-based LATE estimator has on average the smallest deviation from the true LATE, with its absolute bias amounting to 0.6. The non-parametric kernel regression estimator has the smallest average standard deviation among the estimators in Table~\ref{results2}, amounting to 7.0. It also performs best in terms of root mean squared errors across DGPs with an average value of 7.2. When considering the median bias of the bootstrap standard errors relative to the true standard deviations of the respective LATE estimators, the inference method of the random forest-based LATE estimator performs best, with an average median bias of 1.7. The averages of the median biases of its competitors are on average at least 100\% larger.

Our findings suggest that the coverage accuracy is mainly driven by a LATE estimator's bias. This is for instance the reason why the mean differences estimator (which ignores covariates), the bias of which exceeds the bias of the random forest-based LATE estimator by 440.1\%, shows a poor coverage in Table~\ref{results1}. Also the OLS estimator performs poorly in terms of coverage, due to its high bias, while its variance is small (results not presented but available on request). The performance of all LATE estimators by DGP is presented in Section~\ref{appendix_dgps} of the Appendix. Tables~\ref{DGP 1}-\ref{DGP 32} provide details on the coverage rates, biases, standard deviations, and root mean squared errors.

\begin{table}[H]
\centering
\begin{threeparttable}
\caption{Best performing estimators in terms of coverage rates}
\label{results3}
\begin{tabular}{lccc}
\toprule
DGP feature & \multicolumn{3}{c}{point estimator} \\
              & \multicolumn{1}{l}{parametric} & \multicolumn{1}{l}{semi-parametric} & \multicolumn{1}{l}{non-parametric} \\
\midrule
effect homogeneity    & $tsls $          & \boldmath{$radmatchx^{probit}$}  & $means$ \\
effect heterogeneity  & \boldmath{$tsls$}  & $dr^{probit}$                  & $randforest$ \\
standard selection    & $reg$            & $radmatchx^{cbps}$             & \boldmath{$randforest$} \\
strong selection      & $tsls$           & $ipw^{cbps}$                   & \boldmath{$randforest$} \\
weaker first stage    & $reg$            & \boldmath{$dr^{probit}$}         & $randforest$ \\
strong first stage    & $reg$            & \boldmath{$radmatchx^{probit}$}  & $pairmatch^{x}$ \\
non-binary outcome      & $reg$            & \boldmath{$radmatch^{probit}$}   & $randforest$ \\
binary outcome        & $tsls$           & \boldmath{$radmatch^{probit}$}   & $randforest$ \\
small sample size     & $reg$            & $radmatch^{probit}$            & \boldmath{$reg^{kernel}$} \\
larger sample size    & $tsls$           & \boldmath{$radmatch^{probit}$}   & $reg^{kernel}$ \\
\bottomrule
\end{tabular}
\begin{tablenotes}
\footnotesize
\item Notes: The best performing LATE estimator under each DGP feature is shown in bold print.
\end{tablenotes}
\end{threeparttable}
\end{table}

Table~\ref{results3} lists the best performing LATE estimators in terms of average coverage, separately for each of ten DGP features (see the rows) as well as for parametric, semi-parametric, and non-parametric methods (see the columns). The results suggest that semi-parametric and non-parametric estimators come closest to the nominal coverage rate of 95\%. The radius matching algorithm of \citeasnoun{LeMiWu11} (with or without controlling for a covariate in addition to the propensity score) most frequently performs best both among the semi-parametric LATE estimators (in 70\% of cases) and overall (in 50\% of cases). Radius matching achieves the best average coverage in settings with effect homogeneity, a strong instrument, non-binary and binary outcomes, and under a larger sample size. For specifications with standard and strong selection into the instrument, the non-parametric random forest-based estimator is closest to the nominal size. Considering scenarios with effect heterogeneity, a weaker instrument, and a small sample size, the best performers are LATE estimators based on 2SLS, DR, and non-parametric regression, respectively.

The performance of the best performing LATE estimators across the ten DGP features is presented in Section~\ref{appendix_specifications} of the Appendix. Tables~\ref{Effect homogeneity}-\ref{Larger sample size} provide information on the average coverage rates, biases, standard deviations, and root mean squared errors.

\section{Conclusion}\label{conclusion}

This paper presented a simulation study based on empirical labor market data to investigate the finite sample properties of a range of point estimators of the local average treatment effect (LATE) when controlling for a fixed (and low-dimensional) set of covariates. The structure of these estimators is inspired by the Wald estimator, consisting of the ratio of the estimated reduced form effect of the instrument on the outcome and the estimated first stage effect of the instrument on the treatment. Furthermore, we applied the non-parametric bootstrap to estimate the standard errors and the 95\% confidence intervals of the LATE estimators. We find that among the LATE estimators considered, non-parametric kernel regression has the smallest average root mean squared error across the different simulations, closely followed by the random forest-based approach, which has the lowest average absolute bias. The random forest method also performs very competitive in terms of average coverage rates, while at the same time having relatively narrow confidence intervals, which is attractive in terms of precision. Specific versions of semi-parametric radius matching on the propensity score, nonparametric kernel regression, inverse probability weighting, and pair matching on the covariates perform decently in terms of coverage, too, but have substantially wider confidence intervals.

\pagebreak

\bibliographystyle{econometrica}
\bibliography{research}

\pagebreak



\pagebreak
{\large \renewcommand{\theequation}{A-\arabic{equation}} %
\setcounter{equation}{0} \appendix
}
\appendix \numberwithin{equation}{section}
\section{Appendix}\label{further}

\subsection{DGPs}

Table~\ref{dgps} summarizes the 32 data generating processes of our Empirical Monte Carlo study. They differ in several dimensions, namely with respect to effect heterogeneity, instrument selectivity, instrument strength, outcome distribution, and sample size.

\begin{table}[H]\footnotesize
\centering
\caption{Numbered list of DGPs}
\label{dgps}
\footnotesize
\begin{adjustbox}{max width=\textwidth}
\begin{tabular}{l|ccccc}
\hline
& \multicolumn{1}{l}{Effect} & \multicolumn{1}{l}{Strong}    & \multicolumn{1}{l}{Empirically}&\multicolumn{1}{l}{Binary}  &\multicolumn{1}{l}{Sample} \\
& \multicolumn{1}{l}{hetero-}& \multicolumn{1}{l}{instrument}& \multicolumn{1}{l}{observed}   & \multicolumn{1}{l}{outcome}& \multicolumn{1}{l}{size}  \\
& \multicolumn{1}{l}{geneity}& \multicolumn{1}{l}{selection} & \multicolumn{1}{l}{instrument} &                            &                           \\
DGP&                         &                               & \multicolumn{1}{l}{strength}   &                            &                           \\
\hline
    1     & 0     & 0     & 1     & 0     & 1000   \\
    2     & 1     & 0     & 0     & 0     & 1000   \\
    3     & 0     & 1     & 1     & 0     & 1000   \\
    4     & 0     & 0     & 0     & 0     & 1000   \\
    5     & 1     & 1     & 0     & 0     & 1000   \\
    6     & 1     & 0     & 1     & 0     & 1000   \\
    7     & 0     & 1     & 0     & 0     & 1000   \\
    8     & 1     & 1     & 1     & 0     & 1000   \\
    9     & 0     & 0     & 1     & 1     & 1000   \\
    10    & 1     & 0     & 0     & 1     & 1000   \\
    11    & 0     & 1     & 1     & 1     & 1000   \\
    12    & 0     & 0     & 0     & 1     & 1000   \\
    13    & 1     & 1     & 0     & 1     & 1000   \\
    14    & 1     & 0     & 1     & 1     & 1000   \\
    15    & 0     & 1     & 0     & 1     & 1000   \\
    16    & 1     & 1     & 1     & 1     & 1000   \\
    17    & 0     & 0     & 1     & 0     & 2000   \\
    18    & 1     & 0     & 0     & 0     & 2000   \\
    19    & 0     & 1     & 1     & 0     & 2000   \\
    20    & 0     & 0     & 0     & 0     & 2000   \\
    21    & 1     & 1     & 0     & 0     & 2000   \\
    22    & 1     & 0     & 1     & 0     & 2000   \\
    23    & 0     & 1     & 0     & 0     & 2000   \\
    24    & 1     & 1     & 1     & 0     & 2000   \\
    25    & 0     & 0     & 1     & 1     & 2000   \\
    26    & 1     & 0     & 0     & 1     & 2000   \\
    27    & 0     & 1     & 1     & 1     & 2000   \\
    28    & 0     & 0     & 0     & 1     & 2000   \\
    29    & 1     & 1     & 0     & 1     & 2000   \\
    30    & 1     & 0     & 1     & 1     & 2000   \\
    31    & 0     & 1     & 0     & 1     & 2000   \\
    32    & 1     & 1     & 1     & 1     & 2000   \\
\hline
\end{tabular}
\end{adjustbox}
\end{table}

\subsection{Local average treatment effects averaged over different dimensions}\label{appendix_specifications}

The following Tables~\ref{Effect homogeneity}-\ref{Larger sample size} evaluate the LATE estimators with respect to different properties like the absolute bias or the root mean squared error. The tables refer to different scenarios that vary in terms of effect heterogeneity, instrument selectivity, instrument strength, outcome distribution, and sample size. The estimators are sorted according to their performance in terms of coverage rates for 95\% confidence intervals.

\begin{sidewaystable}[ht]
\centering
\begin{threeparttable}
\caption{Effect homogeneity}
\label{Effect homogeneity}
\small
\begin{tabular}{rrrrrrrrrrrrrrrrrrr}
\toprule
& cov & r & int & d & r & bias & d & r & sd & d & r & rmse & d & r & seb & d & r \\
\midrule
  $radmatchx^{probit}$ & 96.0 & 6 & 525.3 & 26 & 8 & 2.0 & 94 & 6 & 17.8 & 41 & 9 & 18.0 & 40 & 9 & 6.9 & 99 & 9 \\
  $radmatchx^{cbps}$ & 96.1 & 6 & 563.3 & 35 & 8 & 1.9 & 80 & 6 & 15.7 & 25 & 8 & 15.8 & 23 & 8 & 8.3 & 140 & 8 \\
  $means$ & 93.8 & 9 & 464.1 & 11 & 4 & 5.7 & 447 & 11 & 14.4 & 15 & 3 & 15.8 & 23 & 7 & 11.3 & 227 & 5 \\
  $pairmatch^{x}$ & 96.3 & 7 & 525.6 & 26 & 8 & 1.3 & 27 & 4 & 18.1 & 44 & 8 & 18.2 & 41 & 8 & 11.8 & 242 & 8 \\
  $randforest$ & 96.4 & 6 & 452.6 & 8 & 3 & 1.0 & 0 & 4 & 14.7 & 17 & 6 & 14.7 & 14 & 5 & 3.5 & 0 & 6 \\
  $reg^{kernel}$ & 96.5 & 8 & 653.7 & 56 & 6 & 2.4 & 130 & 9 & 12.6 & 0 & 5 & 12.9 & -0 & 5 & 8.6 & 147 & 6 \\
  $radmatch^{probit}$ & 96.5 & 5 & 706.5 & 69 & 10 & 2.2 & 113 & 7 & 22.2 & 76 & 10 & 22.3 & 73 & 10 & 21.7 & 526 & 9 \\
  $ipw^{cbps}$ & 97.0 & 6 & 575.7 & 38 & 6 & 2.0 & 90 & 7 & 13.5 & 7 & 4 & 13.7 & 6 & 4 & 8.9 & 156 & 5 \\
  $tsls$ & 97.1 & 4 & 516.6 & 23 & 4 & 1.2 & 18 & 4 & 14.5 & 15 & 4 & 14.6 & 13 & 4 & 7.6 & 118 & 5 \\
  $reg$ & 97.1 & 4 & 418.5 & 0 & 4 & 1.2 & 14 & 4 & 13.2 & 5 & 4 & 13.3 & 4 & 3 & 8.1 & 134 & 4 \\
  $dr^{probit}$ & 97.2 & 5 & 509.8 & 22 & 5 & 1.4 & 38 & 5 & 13.9 & 11 & 4 & 14.0 & 9 & 4 & 7.9 & 127 & 3 \\
\bottomrule
\end{tabular}
\begin{tablenotes}
\item \scriptsize Notes: `cov' and `int' refer to coverage rate and length of confidence interval, `bias' denotes the absolute bias from the true treatment effect, `sd' is the standard deviation of the estimator, `rmse' stands for root mean squared error, and `seb' indicates the median bias of the estimated bootstrap standard error. `diff' indicates the difference to the left best performer in percent (\%) and `r' is the average rank.
\end{tablenotes}
\end{threeparttable}
\end{sidewaystable}

\begin{sidewaystable}[ht]
\centering
\begin{threeparttable}
\caption{Effect heterogeneity}
\label{Effect heterogeneity}
\small
\begin{tabular}{rrrrrrrrrrrrrrrrrrr}
\toprule
& cov & r & int & d & r & bias & d & r & sd & d & r & rmse & d & r & seb & d & r \\
\midrule
  $tsls$ & 94.9 & 2 & 5.1 & 4 & 3 & 0.2 & 0 & 4 & 1.3 & 3 & 2 & 1.3 & 0 & 1 & -0.0 & 5 & 3 \\
  $reg$ & 94.9 & 2 & 5.2 & 6 & 5 & 0.2 & 14 & 4 & 1.3 & 5 & 4 & 1.3 & 2 & 2 & -0.0 & 6 & 3 \\
  $dr^{probit}$ & 94.8 & 2 & 5.2 & 7 & 6 & 0.2 & 28 & 5 & 1.4 & 5 & 4 & 1.3 & 3 & 3 & -0.0 & 0 & 4 \\
  $ipw^{cbps}$ & 94.0 & 4 & 5.1 & 5 & 5 & 0.4 & 106 & 8 & 1.3 & 4 & 4 & 1.4 & 4 & 4 & -0.0 & 13 & 3 \\
  $randforest$ & 93.6 & 6 & 5.0 & 2 & 2 & 0.2 & 26 & 5 & 1.4 & 7 & 6 & 1.4 & 4 & 4 & -0.1 & 244 & 7 \\
  $radmatch^{probit}$ & 93.6 & 6 & 6.4 & 31 & 11 & 0.3 & 72 & 5 & 1.8 & 37 & 11 & 1.8 & 35 & 10 & -0.1 & 383 & 8 \\
  $reg^{kernel}$ & 93.2 & 7 & 5.1 & 5 & 5 & 0.4 & 150 & 9 & 1.5 & 15 & 6 & 1.5 & 16 & 7 & -0.2 & 503 & 6 \\
  $pairmatch^{x}$ & 92.8 & 8 & 5.7 & 16 & 9 & 0.2 & 17 & 4 & 1.6 & 24 & 9 & 1.6 & 21 & 9 & -0.1 & 405 & 10 \\
  $radmatchx^{cbps}$ & 92.7 & 8 & 5.7 & 18 & 9 & 0.3 & 103 & 6 & 1.6 & 24 & 8 & 1.6 & 24 & 8 & -0.1 & 358 & 8 \\
  $radmatchx^{probit}$ & 92.2 & 9 & 5.8 & 18 & 10 & 0.3 & 88 & 5 & 1.6 & 26 & 10 & 1.6 & 26 & 9 & -0.2 & 422 & 10 \\
  $means$ & 87.1 & 11 & 4.9 & 0 & 2 & 1.1 & 539 & 11 & 1.3 & 0 & 2 & 1.7 & 29 & 8 & -0.0 & 43 & 3 \\
\bottomrule
\end{tabular}
\begin{tablenotes}
\item \scriptsize Notes: `cov' and `int' refer to coverage rate and length of confidence interval, `bias' denotes the absolute bias from the true treatment effect, `sd' is the standard deviation of the estimator, `rmse' stands for root mean squared error, and `seb' indicates the median bias of the estimated bootstrap standard error. `diff' indicates the difference to the left best performer in percent (\%) and `r' is the average rank.
\end{tablenotes}
\end{threeparttable}
\end{sidewaystable}

\begin{sidewaystable}[ht]
\centering
\begin{threeparttable}
\caption{Empirically observed instrument selection}
\label{Empirically observed instrument selection}
\small
\begin{tabular}{rrrrrrrrrrrrrrrrrrr}
\toprule
& cov & r & int & d & r & bias & d & r & sd & d & r & rmse & d & r & seb & d & r \\
\midrule
  $randforest$ & 95.0 & 7 & 202.8 & 0 & 1 & 0.6 & 0 & 4 & 7.5 & 5 & 6 & 7.5 & 5 & 4 & 2.0 & 0 & 8 \\
  $radmatchx^{cbps}$ & 95.0 & 6 & 337.4 & 66 & 9 & 0.7 & 30 & 3 & 9.3 & 31 & 8 & 9.3 & 31 & 8 & 3.9 & 94 & 8 \\
  $pairmatch^{x}$ & 94.7 & 8 & 245.7 & 21 & 8 & 0.9 & 66 & 4 & 9.7 & 37 & 9 & 9.8 & 37 & 9 & 5.1 & 152 & 10 \\
  $radmatch^{probit}$ & 95.4 & 5 & 292.2 & 44 & 10 & 1.0 & 85 & 5 & 9.5 & 33 & 10 & 9.5 & 33 & 9 & 8.3 & 315 & 7 \\
  $reg^{kernel}$ & 95.5 & 7 & 264.5 & 30 & 6 & 1.5 & 160 & 10 & 7.6 & 7 & 6 & 7.7 & 8 & 6 & 4.7 & 135 & 6 \\
  $radmatchx^{probit}$ & 94.5 & 8 & 320.4 & 58 & 9 & 0.7 & 29 & 2 & 10.4 & 47 & 9 & 10.4 & 46 & 9 & 3.3 & 62 & 9 \\
  $ipw^{cbps}$ & 95.9 & 4 & 315.1 & 55 & 6 & 1.1 & 97 & 9 & 8.8 & 24 & 5 & 8.9 & 24 & 5 & 4.3 & 112 & 4 \\
  $dr^{probit}$ & 96.0 & 4 & 297.9 & 47 & 5 & 0.9 & 67 & 7 & 7.8 & 10 & 3 & 7.8 & 10 & 3 & 4.0 & 98 & 4 \\
  $reg$ & 96.0 & 3 & 209.8 & 3 & 4 & 0.9 & 62 & 6 & 7.1 & 0 & 3 & 7.1 & 0 & 2 & 4.5 & 125 & 3 \\
  $tsls$ & 96.0 & 4 & 252.6 & 25 & 3 & 1.0 & 70 & 6 & 8.1 & 15 & 3 & 8.2 & 15 & 3 & 4.2 & 107 & 4 \\
  $means$ & 93.0 & 10 & 207.0 & 2 & 3 & 2.8 & 406 & 11 & 7.9 & 12 & 3 & 8.5 & 20 & 8 & 5.8 & 187 & 4 \\
\bottomrule
\end{tabular}
\begin{tablenotes}
\item \scriptsize Notes: `cov' and `int' refer to coverage rate and length of confidence interval, `bias' denotes the absolute bias from the true treatment effect, `sd' is the standard deviation of the estimator, `rmse' stands for root mean squared error, and `seb' indicates the median bias of the estimated bootstrap standard error. `diff' indicates the difference to the left best performer in percent (\%) and `r' is the average rank.
\end{tablenotes}
\end{threeparttable}
\end{sidewaystable}

\begin{sidewaystable}[ht]
\centering
\begin{threeparttable}
\caption{Strong instrument selection}
\label{Strong instrument selection}
\small
\begin{tabular}{rrrrrrrrrrrrrrrrrrr}
\toprule
& cov & r & int & d & r & bias & d & r & sd & d & r & rmse & d & r & seb & d & r \\
\midrule
$randforest$ & 95.1 & 5 & 254.7 & 21 & 4 & 0.7 & 55 & 5 & 8.6 & 43 & 6 & 8.6 & 39 & 4 & 1.4 & 0 & 6 \\
  $ipw^{cbps}$ & 95.1 & 6 & 265.7 & 26 & 5 & 1.2 & 174 & 7 & 6.0 & 0 & 3 & 6.2 & -0 & 4 & 4.6 & 237 & 4 \\
  $radmatch^{probit}$ & 94.7 & 5 & 420.7 & 100 & 11 & 1.5 & 229 & 7 & 14.4 & 140 & 11 & 14.5 & 135 & 11 & 13.2 & 875 & 10 \\
  $pairmatch^{x}$ & 94.4 & 7 & 285.5 & 36 & 8 & 0.6 & 33 & 4 & 10.0 & 66 & 8 & 10.0 & 62 & 8 & 6.6 & 388 & 8 \\
  $reg^{kernel}$ & 94.2 & 8 & 394.3 & 87 & 4 & 1.4 & 205 & 9 & 6.5 & 8 & 5 & 6.7 & 8 & 6 & 3.7 & 171 & 6 \\
  $tsls$ & 96.0 & 3 & 269.1 & 28 & 4 & 0.5 & 0 & 2 & 7.7 & 28 & 3 & 7.7 & 25 & 2 & 3.4 & 150 & 4 \\
  $reg$ & 96.0 & 3 & 213.9 & 2 & 5 & 0.5 & 7 & 2 & 7.5 & 25 & 4 & 7.5 & 22 & 3 & 3.5 & 162 & 4 \\
  $dr^{probit}$ & 96.0 & 3 & 217.1 & 3 & 6 & 0.7 & 61 & 3 & 7.5 & 24 & 6 & 7.5 & 22 & 4 & 3.9 & 186 & 3 \\
  $radmatchx^{cbps}$ & 93.8 & 8 & 231.7 & 10 & 8 & 1.5 & 233 & 9 & 8.0 & 32 & 8 & 8.1 & 32 & 8 & 4.3 & 214 & 8 \\
  $radmatchx^{probit}$ & 93.7 & 8 & 210.7 & 0 & 9 & 1.6 & 261 & 8 & 9.0 & 50 & 9 & 9.2 & 49 & 9 & 3.5 & 157 & 10 \\
  $means$ & 87.9 & 11 & 262.0 & 24 & 2 & 4.0 & 782 & 11 & 7.8 & 30 & 2 & 9.0 & 45 & 7 & 5.5 & 307 & 4 \\
\bottomrule
\end{tabular}
\begin{tablenotes}
\item \scriptsize Notes: `cov' and `int' refer to coverage rate and length of confidence interval, `bias' denotes the absolute bias from the true treatment effect, `sd' is the standard deviation of the estimator, `rmse' stands for root mean squared error, and `seb' indicates the median bias of the estimated bootstrap standard error. `diff' indicates the difference to the left best performer in percent (\%) and `r' is the average rank.
\end{tablenotes}
\end{threeparttable}
\end{sidewaystable}

\begin{sidewaystable}[ht]
\centering
\begin{threeparttable}
\caption{Weaker instrument strength}
\label{Weaker instrument strength}
\small
\begin{tabular}{rrrrrrrrrrrrrrrrrrr}
\toprule
& cov & r & int & d & r & bias & d & r & sd & d & r & rmse & d & r & seb & d & r \\
\midrule
  $dr^{probit}$ & 94.8 & 2 & 5.4 & 6 & 6 & 0.2 & 28 & 5 & 1.4 & 5 & 4 & 1.4 & 2 & 4 & 0.0 & 1327 & 3 \\
  $reg$ & 94.7 & 3 & 5.4 & 5 & 5 & 0.2 & 13 & 4 & 1.4 & 5 & 4 & 1.4 & 1 & 2 & -0.0 & 0 & 3 \\
  $tsls$ & 94.7 & 3 & 5.3 & 4 & 3 & 0.2 & 0 & 4 & 1.4 & 3 & 2 & 1.4 & 0 & 1 & -0.0 & 1563 & 4 \\
  $ipw^{cbps}$ & 94.1 & 3 & 5.3 & 4 & 5 & 0.4 & 105 & 8 & 1.4 & 4 & 4 & 1.4 & 4 & 4 & -0.0 & 3852 & 3 \\
  $radmatch^{probit}$ & 93.6 & 6 & 6.7 & 31 & 11 & 0.3 & 81 & 6 & 1.8 & 38 & 11 & 1.8 & 34 & 11 & -0.1 & 141524 & 8 \\
  $randforest$ & 93.4 & 7 & 5.2 & 1 & 2 & 0.2 & 20 & 4 & 1.4 & 7 & 6 & 1.4 & 3 & 4 & -0.1 & 96925 & 7 \\
  $reg^{kernel}$ & 93.1 & 7 & 5.3 & 4 & 5 & 0.4 & 143 & 9 & 1.5 & 15 & 6 & 1.6 & 15 & 6 & -0.2 & 193929 & 6 \\
  $pairmatch^{x}$ & 92.9 & 8 & 5.9 & 15 & 9 & 0.2 & 14 & 4 & 1.6 & 24 & 9 & 1.6 & 20 & 8 & -0.1 & 156350 & 9 \\
  $radmatchx^{cbps}$ & 92.6 & 8 & 6.0 & 17 & 9 & 0.3 & 93 & 6 & 1.6 & 25 & 9 & 1.7 & 23 & 8 & -0.1 & 147726 & 9 \\
  $radmatchx^{probit}$ & 92.3 & 9 & 6.0 & 17 & 10 & 0.3 & 83 & 5 & 1.7 & 27 & 10 & 1.7 & 25 & 9 & -0.1 & 172399 & 10 \\
  $means$ & 87.4 & 11 & 5.1 & 0 & 2 & 1.1 & 541 & 11 & 1.3 & 0 & 2 & 1.8 & 29 & 8 & -0.0 & 8467 & 3 \\
\bottomrule
\end{tabular}
\begin{tablenotes}
\item \scriptsize Notes: `cov' and `int' refer to coverage rate and length of confidence interval, `bias' denotes the absolute bias from the true treatment effect, `sd' is the standard deviation of the estimator, `rmse' stands for root mean squared error, and `seb' indicates the median bias of the estimated bootstrap standard error. `diff' indicates the difference to the left best performer in percent (\%) and `r' is the average rank.
\end{tablenotes}
\end{threeparttable}
\end{sidewaystable}

\begin{sidewaystable}[ht]
\centering
\begin{threeparttable}
\caption{Empirically observed instrument strength}
\label{Empirically observed instrument strength}
\small
\begin{tabular}{rrrrrrrrrrrrrrrrrrr}
\toprule
& cov & r & int & d & r & bias & d & r & sd & d & r & rmse & d & r & seb & d & r \\
\midrule
  $radmatchx^{probit}$ & 95.9 & 6 & 525.1 & 26 & 8 & 2.0 & 93 & 6 & 17.8 & 41 & 9 & 17.9 & 40 & 9 & 6.9 & 100 & 8 \\
  $pairmatch^{x}$ & 96.2 & 7 & 525.4 & 26 & 8 & 1.3 & 27 & 4 & 18.1 & 44 & 9 & 18.1 & 42 & 8 & 11.8 & 243 & 9 \\
  $radmatchx^{cbps}$ & 96.3 & 6 & 563.1 & 35 & 8 & 1.9 & 80 & 6 & 15.7 & 25 & 8 & 15.8 & 23 & 7 & 8.3 & 141 & 7 \\
  $means$ & 93.5 & 10 & 463.9 & 11 & 3 & 5.7 & 441 & 11 & 14.4 & 15 & 4 & 15.8 & 23 & 7 & 11.3 & 228 & 5 \\
  $reg^{kernel}$ & 96.5 & 8 & 653.5 & 56 & 6 & 2.4 & 129 & 9 & 12.6 & 0 & 5 & 12.8 & 0 & 6 & 8.5 & 148 & 5 \\
  $radmatch^{probit}$ & 96.6 & 5 & 706.2 & 69 & 10 & 2.2 & 110 & 6 & 22.1 & 76 & 10 & 22.2 & 73 & 10 & 21.7 & 530 & 9 \\
  $randforest$ & 96.6 & 5 & 452.4 & 8 & 3 & 1.0 & 0 & 4 & 14.7 & 17 & 6 & 14.7 & 15 & 5 & 3.4 & 0 & 6 \\
  $ipw^{cbps}$ & 96.9 & 6 & 575.5 & 38 & 6 & 2.0 & 88 & 7 & 13.5 & 7 & 4 & 13.6 & 6 & 4 & 8.8 & 157 & 5 \\
  $reg$ & 97.2 & 4 & 418.3 & 0 & 4 & 1.2 & 13 & 4 & 13.2 & 5 & 4 & 13.3 & 4 & 3 & 8.1 & 135 & 3 \\
  $dr^{probit}$ & 97.3 & 5 & 509.6 & 22 & 5 & 1.4 & 37 & 5 & 13.9 & 11 & 4 & 14.0 & 9 & 4 & 7.9 & 128 & 4 \\
  $tsls$ & 97.3 & 4 & 516.4 & 23 & 4 & 1.2 & 17 & 4 & 14.5 & 15 & 4 & 14.5 & 14 & 4 & 7.5 & 119 & 5 \\
\bottomrule
\end{tabular}
\begin{tablenotes}
\item \scriptsize Notes: `cov' and `int' refer to coverage rate and length of confidence interval, `bias' denotes the absolute bias from the true treatment effect, `sd' is the standard deviation of the estimator, `rmse' stands for root mean squared error, and `seb' indicates the median bias of the estimated bootstrap standard error. `diff' indicates the difference to the left best performer in percent (\%) and `r' is the average rank.
\end{tablenotes}
\end{threeparttable}
\end{sidewaystable}

\begin{sidewaystable}[ht]
\centering
\begin{threeparttable}
\caption{Discrete outcome}
\label{Discrete outcome}
\small
\begin{tabular}{rrrrrrrrrrrrrrrrrrr}
\toprule
& cov & r & int & d & r & bias & d & r & sd & d & r & rmse & d & r & seb & d & r \\
\midrule
  $radmatch^{probit}$ & 95.0 & 6 & 695.6 & 68 & 11 & 2.4 & 100 & 6 & 21.0 & 62 & 10 & 21.2 & 59 & 10 & 23.5 & 460 & 8 \\
  $randforest$ & 94.9 & 5 & 449.1 & 9 & 2 & 1.2 & 0 & 4 & 14.8 & 14 & 6 & 14.8 & 12 & 4 & 4.2 & 0 & 7 \\
  $ipw^{cbps}$ & 95.2 & 5 & 569.1 & 37 & 6 & 2.3 & 86 & 8 & 13.1 & 1 & 4 & 13.3 & 1 & 4 & 10.0 & 139 & 4 \\
  $reg^{kernel}$ & 94.5 & 7 & 646.1 & 56 & 5 & 2.8 & 127 & 9 & 13.0 & 0 & 6 & 13.3 & 0 & 6 & 9.0 & 115 & 7 \\
  $pairmatch^{x}$ & 94.4 & 8 & 519.0 & 25 & 8 & 1.5 & 23 & 4 & 18.5 & 43 & 9 & 18.6 & 40 & 8 & 12.2 & 190 & 9 \\
  $dr^{probit}$ & 95.9 & 4 & 504.0 & 22 & 5 & 1.6 & 33 & 5 & 14.2 & 9 & 4 & 14.3 & 7 & 3 & 8.4 & 101 & 4 \\
  $reg$ & 95.9 & 3 & 413.9 & 0 & 4 & 1.4 & 12 & 4 & 13.9 & 7 & 3 & 13.9 & 5 & 3 & 8.3 & 99 & 3 \\
  $radmatchx^{cbps}$ & 94.0 & 7 & 556.0 & 34 & 9 & 2.2 & 80 & 6 & 15.7 & 21 & 8 & 15.8 & 19 & 7 & 9.2 & 120 & 8 \\
  $tsls$ & 96.0 & 3 & 509.9 & 23 & 4 & 1.4 & 13 & 3 & 14.0 & 8 & 3 & 14.1 & 6 & 2 & 8.8 & 111 & 4 \\
  $radmatchx^{probit}$ & 93.8 & 8 & 518.3 & 25 & 9 & 2.3 & 89 & 6 & 17.5 & 35 & 9 & 17.7 & 33 & 9 & 8.1 & 93 & 9 \\
  $means$ & 87.6 & 10 & 460.2 & 11 & 3 & 6.7 & 452 & 11 & 15.2 & 17 & 3 & 16.9 & 28 & 9 & 11.3 & 170 & 5 \\
\bottomrule
\end{tabular}
\begin{tablenotes}
\item \scriptsize Notes: `cov' and `int' refer to coverage rate and length of confidence interval, `bias' denotes the absolute bias from the true treatment effect, `sd' is the standard deviation of the estimator, `rmse' stands for root mean squared error, and `seb' indicates the median bias of the estimated bootstrap standard error. `diff' indicates the difference to the left best performer in percent (\%) and `r' is the average rank.
\end{tablenotes}
\end{threeparttable}
\end{sidewaystable}

\begin{sidewaystable}[ht]
\centering
\begin{threeparttable}
\caption{Binary outcome}
\label{Binary outcome}
\small
\begin{tabular}{rrrrrrrrrrrrrrrrrrr}
\toprule
& cov & r & int & d & r & bias & d & r & sd & d & r & rmse & d & r & seb & d & r \\
\midrule
  $radmatch^{probit}$ & 95.1 & 5 & 17.3 & 106 & 11 & 0.1 & 188 & 6 & 2.9 & 420 & 10 & 2.9 & 406 & 10 & -1.9 & 23346 & 8 \\
  $randforest$ & 95.1 & 7 & 8.4 & 0 & 2 & 0.0 & 52 & 4 & 1.3 & 130 & 6 & 1.3 & 124 & 4 & -0.8 & 10024 & 7 \\
  $radmatchx^{cbps}$ & 94.8 & 7 & 13.1 & 55 & 8 & 0.0 & 24 & 6 & 1.6 & 192 & 8 & 1.6 & 184 & 8 & -1.1 & 12892 & 8 \\
  $reg^{kernel}$ & 95.2 & 7 & 12.7 & 50 & 5 & 0.1 & 152 & 10 & 1.1 & 97 & 5 & 1.1 & 92 & 6 & -0.6 & 7446 & 5 \\
  $pairmatch^{x}$ & 94.7 & 7 & 12.3 & 45 & 8 & 0.0 & 0 & 4 & 1.2 & 115 & 8 & 1.2 & 109 & 8 & -0.5 & 5728 & 9 \\
  $radmatchx^{probit}$ & 94.4 & 8 & 12.8 & 52 & 9 & 0.1 & 73 & 5 & 1.9 & 246 & 9 & 1.9 & 236 & 9 & -1.3 & 16419 & 10 \\
  $ipw^{cbps}$ & 95.8 & 4 & 11.7 & 38 & 5 & 0.1 & 155 & 8 & 1.7 & 207 & 5 & 1.7 & 200 & 4 & -1.2 & 14503 & 4 \\
  $tsls$ & 96.0 & 4 & 11.8 & 40 & 4 & 0.0 & 4 & 5 & 1.8 & 227 & 3 & 1.8 & 218 & 3 & -1.3 & 15757 & 4 \\
  $reg$ & 96.0 & 3 & 9.7 & 15 & 5 & 0.0 & 7 & 4 & 0.7 & 34 & 4 & 0.7 & 30 & 3 & -0.2 & 2870 & 3 \\
  $dr^{probit}$ & 96.1 & 3 & 11.0 & 30 & 6 & 0.0 & 40 & 5 & 1.1 & 97 & 5 & 1.1 & 92 & 4 & -0.6 & 7033 & 4 \\
  $means$ & 93.3 & 10 & 8.8 & 4 & 2 & 0.1 & 235 & 11 & 0.6 & 0 & 2 & 0.6 & 0 & 6 & -0.0 & 0 & 3 \\
\bottomrule
\end{tabular}
\begin{tablenotes}
\item \scriptsize Notes: `cov' and `int' refer to coverage rate and length of confidence interval, `bias' denotes the absolute bias from the true treatment effect, `sd' is the standard deviation of the estimator, `rmse' stands for root mean squared error, and `seb' indicates the median bias of the estimated bootstrap standard error. `diff' indicates the difference to the left best performer in percent (\%) and `r' is the average rank.
\end{tablenotes}
\end{threeparttable}
\end{sidewaystable}

\begin{sidewaystable}[ht]
\centering
\begin{threeparttable}
\caption{Small sample size}
\label{Small sample size}
\small
\begin{tabular}{rrrrrrrrrrrrrrrrrrr}
\toprule
& cov & r & int & d & r & bias & d & r & sd & d & r & rmse & d & r & seb & d & r \\
\midrule
  $reg^{kernel}$ & 94.8 & 7 & 556.6 & 68 & 6 & 1.5 & 141 & 10 & 9.8 & 3 & 6 & 9.9 & 4 & 6 & 8.4 & 68 & 7 \\
  $radmatch^{probit}$ & 95.2 & 5 & 457.0 & 38 & 10 & 1.3 & 102 & 6 & 14.3 & 49 & 10 & 14.3 & 49 & 10 & 17.2 & 244 & 8 \\
  $randforest$ & 94.8 & 7 & 351.8 & 6 & 2 & 0.6 & 0 & 5 & 10.3 & 7 & 5 & 10.3 & 7 & 4 & 5.0 & 0 & 7 \\
  $radmatchx^{cbps}$ & 94.5 & 7 & 435.4 & 32 & 8 & 0.9 & 51 & 6 & 11.0 & 15 & 8 & 11.0 & 15 & 8 & 8.6 & 73 & 8 \\
  $pairmatch^{x}$ & 94.5 & 8 & 362.5 & 10 & 8 & 0.8 & 22 & 3 & 12.4 & 29 & 9 & 12.4 & 29 & 9 & 12.5 & 150 & 9 \\
  $ipw^{cbps}$ & 95.5 & 4 & 469.7 & 42 & 6 & 1.2 & 90 & 8 & 10.8 & 13 & 4 & 10.9 & 14 & 4 & 8.2 & 65 & 4 \\
  $radmatchx^{probit}$ & 94.3 & 7 & 390.2 & 18 & 8 & 1.1 & 70 & 5 & 12.4 & 29 & 9 & 12.4 & 30 & 9 & 7.9 & 59 & 9 \\
  $reg$ & 95.8 & 4 & 330.5 & -0 & 4 & 0.7 & 6 & 4 & 9.6 & 0 & 3 & 9.6 & 0 & 2 & 8.7 & 74 & 3 \\
  $dr^{probit}$ & 95.8 & 4 & 420.3 & 27 & 6 & 0.8 & 36 & 5 & 10.4 & 9 & 5 & 10.4 & 9 & 4 & 8.2 & 63 & 3 \\
  $tsls$ & 95.9 & 3 & 429.1 & 30 & 5 & 0.7 & 9 & 4 & 11.1 & 16 & 3 & 11.1 & 16 & 3 & 7.9 & 58 & 4 \\
  $means$ & 92.5 & 10 & 356.1 & 8 & 3 & 3.4 & 446 & 11 & 11.5 & 20 & 3 & 12.0 & 26 & 7 & 10.2 & 105 & 4 \\
\bottomrule
\end{tabular}
\begin{tablenotes}
\item \scriptsize Notes: `cov' and `int' refer to coverage rate and length of confidence interval, `bias' denotes the absolute bias from the true treatment effect, `sd' is the standard deviation of the estimator, `rmse' stands for root mean squared error, and `seb' indicates the median bias of the estimated bootstrap standard error. `diff' indicates the difference to the left best performer in percent (\%) and `r' is the average rank.
\end{tablenotes}
\end{threeparttable}
\end{sidewaystable}

\begin{sidewaystable}[ht]
\centering
\begin{threeparttable}
\caption{Larger sample size}
\label{Larger sample size}
\small
\begin{tabular}{rrrrrrrrrrrrrrrrrrr}
\toprule
& cov & r & int & d & r & bias & d & r & sd & d & r & rmse & d & r & seb & d & r \\
\midrule
  $radmatch^{probit}$ & 94.9 & 6 & 255.9 & 176 & 11 & 1.3 & 98 & 7 & 9.7 & 141 & 10 & 9.8 & 134 & 10 & 4.3 & 27420 & 9 \\
  $reg^{kernel}$ & 94.9 & 7 & 102.3 & 10 & 5 & 1.3 & 109 & 9 & 4.2 & 6 & 6 & 4.5 & 7 & 6 & -0.0 & 0 & 5 \\
  $randforest$ & 95.3 & 6 & 105.8 & 14 & 3 & 0.6 & 0 & 3 & 5.8 & 44 & 6 & 5.8 & 39 & 5 & -1.6 & 10318 & 7 \\
  $pairmatch^{x}$ & 94.6 & 7 & 168.7 & 82 & 8 & 0.8 & 21 & 4 & 7.4 & 84 & 9 & 7.4 & 77 & 8 & -0.8 & 5196 & 9 \\
  $ipw^{cbps}$ & 95.5 & 5 & 111.1 & 20 & 5 & 1.2 & 81 & 7 & 4.0 & 0 & 4 & 4.2 & 0 & 4 & 0.6 & 3665 & 4 \\
  $radmatchx^{cbps}$ & 94.3 & 7 & 133.7 & 44 & 9 & 1.3 & 102 & 6 & 6.3 & 57 & 9 & 6.4 & 54 & 8 & -0.5 & 2893 & 8 \\
  $tsls$ & 96.1 & 4 & 92.6 & 0 & 3 & 0.7 & 13 & 4 & 4.7 & 18 & 3 & 4.8 & 15 & 2 & -0.4 & 2354 & 4 \\
  $radmatchx^{probit}$ & 93.9 & 8 & 140.9 & 52 & 9 & 1.3 & 102 & 6 & 7.1 & 77 & 10 & 7.2 & 72 & 9 & -1.2 & 7523 & 10 \\
  $reg$ & 96.2 & 3 & 93.1 & 1 & 4 & 0.7 & 14 & 4 & 5.0 & 25 & 4 & 5.1 & 21 & 3 & -0.6 & 3883 & 3 \\
  $dr^{probit}$ & 96.2 & 4 & 94.7 & 2 & 5 & 0.8 & 28 & 5 & 4.9 & 21 & 4 & 4.9 & 18 & 3 & -0.3 & 1993 & 4 \\
  $means$ & 88.4 & 10 & 112.9 & 22 & 3 & 3.4 & 435 & 11 & 4.3 & 6 & 2 & 5.5 & 31 & 8 & 1.0 & 6554 & 4 \\
\bottomrule
\end{tabular}
\begin{tablenotes}
\item \scriptsize Notes: `cov' and `int' refer to coverage rate and length of confidence interval, `bias' denotes the absolute bias from the true treatment effect, `sd' is the standard deviation of the estimator, `rmse' stands for root mean squared error, and `seb' indicates the median bias of the estimated bootstrap standard error. `diff' indicates the difference to the left best performer in percent (\%) and `r' is the average rank.
\end{tablenotes}
\end{threeparttable}
\end{sidewaystable}

\clearpage \newpage

\subsection{Local average treatment effects for each DGP}
\label{appendix_dgps}

The following Tables~\ref{DGP 1}-\ref{DGP 32} provide the performance of the LATE estimators separately by each DGP. The estimators are sorted according to their performance in terms of coverage rates of 95\% confidence intervals.

\begin{sidewaystable}[ht]
\centering
\begin{threeparttable}
\caption{DGP 1}
\label{DGP 1}
\small
\begin{tabular}{rrrrrrrrrrrrrrrrrrr}
\toprule
& cov & int & d & r & bias & d & r & sd & d & r & rmse & d & r & seb & d & r & nsimp & nsimse \\
\midrule
  $pairmatch^{probit}$ & 94.6 & 10.8 & 24 & 18 & 0.3 & 257 & 10 & 2.8 & 16 & 16 & 2.8 & 16 & 16 & -0.0 & 0 & 1 & 2000 & 2000 \\
  $pairmatch^{cbps}$ & 94.3 & 10.8 & 24 & 19 & 0.3 & 319 & 14 & 2.8 & 18 & 18 & 2.8 & 18 & 18 & -0.1 & 306 & 2 & 2000 & 2000 \\
  $radmatch^{cbps}$ & 93.8 & 10.3 & 18 & 15 & 0.3 & 323 & 15 & 2.8 & 16 & 17 & 2.8 & 16 & 17 & -0.1 & 973 & 12 & 2000 & 2000 \\
  $ipw^{cbps}$ & 93.8 & 9.2 & 6 & 7 & 0.4 & 384 & 16 & 2.4 & 1 & 8 & 2.4 & 2 & 6 & -0.1 & 426 & 7 & 2000 & 2000 \\
  $reg$ & 93.7 & 9.1 & 5 & 3 & 0.3 & 254 & 9 & 2.4 & 0 & 2 & 2.4 & 0 & 2 & -0.1 & 397 & 5 & 2000 & 2000 \\
  $ipw^{probit}$ & 93.7 & 9.1 & 5 & 6 & 0.3 & 242 & 7 & 2.4 & 0 & 5 & 2.4 & 0 & 4 & -0.1 & 426 & 8 & 2000 & 2000 \\
  $dr^{probit}$ & 93.7 & 9.1 & 5 & 5 & 0.3 & 263 & 11 & 2.4 & 0 & 4 & 2.4 & 0 & 5 & -0.1 & 428 & 9 & 2000 & 2000 \\
  $tsls$ & 93.6 & 9.1 & 5 & 2 & 0.3 & 237 & 5 & 2.4 & 0 & 1 & 2.4 & 0 & 1 & -0.1 & 385 & 4 & 2000 & 2000 \\
  $radmatch^{probit}$ & 93.6 & 10.3 & 18 & 14 & 0.3 & 244 & 8 & 2.7 & 15 & 15 & 2.7 & 15 & 14 & -0.1 & 820 & 10 & 2000 & 2000 \\
  $dr^{cbps}$ & 93.6 & 9.1 & 5 & 4 & 0.3 & 264 & 12 & 2.4 & 0 & 3 & 2.4 & 0 & 3 & -0.1 & 420 & 6 & 2000 & 2000 \\
  $dr^{lc}$ & 96.8 & 18.5 & 113 & 21 & 0.5 & 546 & 18 & 2.4 & 1 & 9 & 2.5 & 3 & 8 & 0.3 & 1964 & 17 & 2000 & 1955 \\
  $ipw^{lc}$ & 96.8 & 14.8 & 70 & 20 & 0.8 & 931 & 20 & 2.4 & 1 & 6 & 2.5 & 6 & 10 & 0.3 & 2089 & 18 & 2000 & 1993 \\
  $radmatchx^{cbps}$ & 92.7 & 9.7 & 12 & 11 & 0.2 & 120 & 3 & 2.7 & 12 & 12 & 2.7 & 11 & 11 & -0.2 & 1329 & 13 & 2000 & 2000 \\
  $reg^{kernel}$ & 92.7 & 9.2 & 6 & 9 & 0.6 & 607 & 19 & 2.5 & 4 & 11 & 2.5 & 6 & 9 & -0.1 & 828 & 11 & 2000 & 1968 \\
  $randforest$ & 92.3 & 8.7 & 0 & 1 & 0.3 & 240 & 6 & 2.4 & 2 & 10 & 2.5 & 2 & 7 & -0.2 & 1587 & 16 & 2000 & 2000 \\
  $radmatchx^{probit}$ & 92.3 & 9.7 & 12 & 10 & 0.1 & 69 & 2 & 2.7 & 13 & 13 & 2.7 & 12 & 12 & -0.2 & 1535 & 15 & 2000 & 2000 \\
  $pairmatch^{x}$ & 92.3 & 9.8 & 12 & 12 & 0.2 & 177 & 4 & 2.7 & 14 & 14 & 2.7 & 13 & 13 & -0.2 & 1513 & 14 & 2000 & 2000 \\
  $means$ & 90.2 & 9.2 & 6 & 8 & 1.4 & 1633 & 21 & 2.4 & 1 & 7 & 2.8 & 16 & 15 & -0.1 & 307 & 3 & 2000 & 2000 \\
  $pairmatch^{lc}$ & 80.2 & 10.3 & 19 & 16 & 0.3 & 312 & 13 & 4.5 & 91 & 19 & 4.5 & 90 & 19 & -1.9 & 14427 & 19 & 2000 & 2000 \\
  $radmatchx^{lc}$ & 75.4 & 9.9 & 13 & 13 & 0.1 & 0 & 1 & 4.7 & 99 & 20 & 4.7 & 98 & 20 & -2.2 & 16864 & 20 & 2000 & 2000 \\
  $radmatch^{lc}$ & 73.6 & 10.4 & 19 & 17 & 0.4 & 406 & 17 & 5.2 & 119 & 21 & 5.2 & 118 & 21 & -2.6 & 19571 & 21 & 2000 & 2000 \\
\bottomrule
\end{tabular}
\begin{tablenotes}
\item \scriptsize Notes: `cov' and `int' refer to coverage rate and length of confidence interval, `bias' denotes the absolute bias from the true treatment effect, `sd' is the standard
deviation of the estimator, `rmse' stands for root mean squared error, and `seb' indicates the median bias of the estimated bootstrap standard error. `diff' indicates the difference to the left best performer in percent (\%) and `r' is the rank. `nsimp' shows the number of simulations with point estimations smaller than $|10^{10}|$ and `nsimse' denotes the number of simulations used to compute `cov', `int', and `seb'. Conditional on `nsimp', simulations with standard errors greater than 150 times the (true) standard error are not considered (i.e.simulations with non-converging bootstrap standard errors are dropped).
\end{tablenotes}
\end{threeparttable}
\end{sidewaystable}

\begin{sidewaystable}[ht]
\centering
\begin{threeparttable}
\caption{DGP 2}
\label{DGP 2}
\small
\begin{tabular}{rrrrrrrrrrrrrrrrrrr}
\toprule
& cov & int & d & r & bias & d & r & sd & d & r & rmse & d & r & seb & d & r & nsimp & nsimse \\
\midrule
  $dr^{probit}$ & 94.8 & 15.8 & 4 & 7 & 0.3 & 171 & 11 & 4.0 & 1 & 6 & 4.0 & 1 & 5 & 0.1 & 96 & 9 & 2000 & 2000 \\
  $ipw^{probit}$ & 94.8 & 15.7 & 4 & 5 & 0.3 & 144 & 9 & 4.0 & 1 & 3 & 4.0 & 0 & 2 & 0.0 & 81 & 4 & 2000 & 2000 \\
  $dr^{cbps}$ & 94.8 & 15.7 & 4 & 6 & 0.3 & 175 & 13 & 4.0 & 1 & 5 & 4.0 & 1 & 4 & 0.0 & 86 & 5 & 2000 & 2000 \\
  $reg$ & 94.8 & 15.7 & 4 & 4 & 0.3 & 165 & 10 & 4.0 & 1 & 4 & 4.0 & 0 & 3 & 0.0 & 87 & 6 & 2000 & 2000 \\
  $tsls$ & 94.8 & 15.7 & 4 & 3 & 0.3 & 143 & 8 & 3.9 & 0 & 2 & 3.9 & 0 & 1 & 0.0 & 61 & 2 & 2000 & 2000 \\
  $ipw^{cbps}$ & 94.7 & 15.8 & 4 & 8 & 0.5 & 310 & 17 & 4.0 & 1 & 7 & 4.0 & 1 & 6 & 0.1 & 94 & 8 & 2000 & 2000 \\
  $radmatch^{cbps}$ & 94.6 & 17.7 & 17 & 17 & 0.4 & 249 & 15 & 4.5 & 16 & 14 & 4.6 & 16 & 15 & -0.0 & 80 & 3 & 2000 & 2000 \\
  $pairmatch^{probit}$ & 95.6 & 18.6 & 23 & 19 & 0.2 & 35 & 2 & 4.6 & 18 & 17 & 4.6 & 17 & 16 & 0.1 & 249 & 11 & 2000 & 2000 \\
  $reg^{kernel}$ & 94.4 & 15.9 & 5 & 9 & 0.9 & 585 & 19 & 4.1 & 4 & 11 & 4.2 & 6 & 10 & -0.0 & 88 & 7 & 2000 & 1970 \\
  $pairmatch^{cbps}$ & 95.7 & 18.6 & 23 & 18 & 0.5 & 266 & 16 & 4.6 & 18 & 16 & 4.6 & 17 & 18 & 0.1 & 290 & 12 & 2000 & 2000 \\
  $radmatchx^{cbps}$ & 94.2 & 16.8 & 11 & 11 & 0.3 & 130 & 7 & 4.5 & 14 & 12 & 4.5 & 13 & 12 & -0.2 & 596 & 13 & 2000 & 2000 \\
  $radmatch^{probit}$ & 94.2 & 17.7 & 17 & 16 & 0.2 & 61 & 3 & 4.6 & 16 & 15 & 4.6 & 15 & 14 & -0.1 & 181 & 10 & 2000 & 2000 \\
  $pairmatch^{x}$ & 93.8 & 17.1 & 13 & 13 & 0.3 & 128 & 6 & 4.6 & 18 & 18 & 4.6 & 17 & 17 & -0.3 & 1045 & 16 & 2000 & 2000 \\
  $randforest$ & 93.7 & 15.1 & 0 & 1 & 0.4 & 227 & 14 & 4.1 & 4 & 10 & 4.1 & 4 & 8 & -0.2 & 791 & 15 & 2000 & 2000 \\
  $radmatchx^{probit}$ & 93.6 & 16.8 & 11 & 12 & 0.1 & 0 & 1 & 4.5 & 14 & 13 & 4.5 & 14 & 13 & -0.2 & 736 & 14 & 2000 & 2000 \\
  $ipw^{lc}$ & 96.6 & 21.8 & 44 & 20 & 1.2 & 879 & 20 & 4.0 & 1 & 8 & 4.1 & 5 & 9 & 0.6 & 2101 & 17 & 2000 & 1989 \\
  $dr^{lc}$ & 96.9 & 31.7 & 110 & 21 & 0.8 & 502 & 18 & 4.0 & 3 & 9 & 4.1 & 4 & 7 & 0.6 & 2193 & 18 & 2000 & 1954 \\
  $means$ & 91.6 & 15.5 & 3 & 2 & 2.1 & 1604 & 21 & 3.9 & 0 & 1 & 4.5 & 13 & 11 & 0.0 & 0 & 1 & 2000 & 2000 \\
  $pairmatch^{lc}$ & 79.2 & 17.4 & 15 & 14 & 0.3 & 110 & 4 & 8.4 & 114 & 20 & 8.4 & 112 & 20 & -4.0 & 14979 & 19 & 2000 & 2000 \\
  $radmatchx^{lc}$ & 75.3 & 16.8 & 11 & 10 & 0.3 & 126 & 5 & 8.2 & 110 & 19 & 8.2 & 109 & 19 & -4.0 & 15086 & 20 & 2000 & 2000 \\
  $radmatch^{lc}$ & 74.2 & 17.5 & 16 & 15 & 0.3 & 174 & 12 & 9.2 & 136 & 21 & 9.3 & 134 & 21 & -4.8 & 18271 & 21 & 2000 & 2000 \\
\bottomrule
\end{tabular}
\begin{tablenotes}
\item \scriptsize Notes: `cov' and `int' refer to coverage rate and length of confidence interval, `bias' denotes the absolute bias from the true treatment effect, `sd' is the standard
deviation of the estimator, `rmse' stands for root mean squared error, and `seb' indicates the median bias of the estimated bootstrap standard error. `diff' indicates the difference to the left best performer in percent (\%) and `r' is the rank. `nsimp' shows the number of simulations with point estimations smaller than $|10^{10}|$ and `nsimse' denotes the number of simulations used to compute `cov', `int', and `seb'. Conditional on `nsimp', simulations with standard errors greater than 150 times the (true) standard error are not considered (i.e.simulations with non-converging bootstrap standard errors are dropped).
\end{tablenotes}
\end{threeparttable}
\end{sidewaystable}

\begin{sidewaystable}[ht]
\centering
\begin{threeparttable}
\caption{DGP 3}
\label{DGP 3}
\small
\begin{tabular}{rrrrrrrrrrrrrrrrrrr}
\toprule
& cov & int & d & r & bias & d & r & sd & d & r & rmse & d & r & seb & d & r & nsimp & nsimse \\
\midrule
  $tsls$ & 95.0 & 9.5 & 4 & 5 & 0.1 & 0 & 1 & 2.5 & 4 & 4 & 2.5 & 0 & 1 & -0.1 & 17 & 7 & 2000 & 2000 \\
  $dr^{cbps}$ & 94.4 & 9.8 & 7 & 8 & 0.2 & 189 & 3 & 2.5 & 6 & 8 & 2.6 & 2 & 5 & -0.1 & 2 & 2 & 2000 & 2000 \\
  $ipw^{probit}$ & 94.3 & 9.6 & 5 & 6 & 0.4 & 437 & 9 & 2.5 & 4 & 5 & 2.5 & 1 & 3 & -0.1 & 6 & 4 & 2000 & 2000 \\
  $dr^{probit}$ & 94.2 & 9.8 & 7 & 9 & 0.2 & 189 & 4 & 2.6 & 6 & 9 & 2.6 & 2 & 6 & -0.1 & 0 & 1 & 2000 & 2000 \\
  $reg$ & 94.1 & 9.7 & 6 & 7 & 0.1 & 73 & 2 & 2.5 & 5 & 6 & 2.5 & 1 & 2 & -0.1 & 10 & 5 & 2000 & 2000 \\
  $pairmatch^{probit}$ & 94.0 & 12.6 & 38 & 19 & 0.6 & 605 & 12 & 3.3 & 38 & 15 & 3.4 & 34 & 15 & -0.1 & 78 & 8 & 2000 & 2000 \\
  $ipw^{cbps}$ & 93.8 & 9.5 & 4 & 4 & 0.6 & 662 & 14 & 2.5 & 4 & 3 & 2.6 & 3 & 7 & -0.1 & 13 & 6 & 2000 & 2000 \\
  $randforest$ & 92.7 & 9.1 & 0 & 1 & 0.3 & 237 & 7 & 2.5 & 5 & 7 & 2.5 & 2 & 4 & -0.2 & 244 & 10 & 2000 & 2000 \\
  $pairmatch^{cbps}$ & 92.7 & 12.5 & 37 & 18 & 0.6 & 605 & 11 & 3.3 & 38 & 16 & 3.4 & 35 & 16 & -0.2 & 153 & 9 & 2000 & 2000 \\
  $radmatch^{probit}$ & 92.6 & 12.4 & 36 & 17 & 0.6 & 686 & 15 & 3.4 & 42 & 18 & 3.5 & 39 & 18 & -0.3 & 353 & 12 & 2000 & 2000 \\
  $radmatch^{cbps}$ & 92.0 & 12.3 & 35 & 16 & 0.6 & 632 & 13 & 3.4 & 40 & 17 & 3.4 & 37 & 17 & -0.3 & 331 & 11 & 2000 & 2000 \\
  $radmatchx^{probit}$ & 91.8 & 10.8 & 18 & 15 & 0.6 & 697 & 17 & 3.0 & 26 & 14 & 3.1 & 24 & 13 & -0.3 & 387 & 14 & 2000 & 2000 \\
  $reg^{kernel}$ & 91.8 & 9.5 & 4 & 3 & 0.7 & 767 & 19 & 2.9 & 22 & 12 & 3.0 & 21 & 11 & -0.5 & 765 & 16 & 2000 & 1866 \\
  $pairmatch^{x}$ & 91.6 & 10.5 & 15 & 13 & 0.3 & 221 & 6 & 2.9 & 22 & 11 & 3.0 & 18 & 10 & -0.3 & 368 & 13 & 2000 & 2000 \\
  $radmatchx^{cbps}$ & 91.5 & 10.7 & 18 & 14 & 0.7 & 740 & 18 & 3.0 & 25 & 13 & 3.1 & 23 & 12 & -0.3 & 393 & 15 & 2000 & 2000 \\
  $ipw^{lc}$ & 99.9 & 69.9 & 665 & 20 & 1.0 & 1202 & 20 & 2.5 & 3 & 2 & 2.7 & 7 & 8 & 6.4 & 10272 & 20 & 2000 & 1644 \\
  $dr^{lc}$ & 100.0 & 405.2 & 4337 & 21 & 0.6 & 690 & 16 & 2.6 & 9 & 10 & 2.7 & 8 & 9 & 36.0 & 58554 & 21 & 2000 & 1264 \\
  $means$ & 85.6 & 9.2 & 1 & 2 & 2.1 & 2531 & 21 & 2.4 & 0 & 1 & 3.2 & 27 & 14 & -0.1 & 4 & 3 & 2000 & 2000 \\
  $pairmatch^{lc}$ & 66.8 & 10.4 & 14 & 11 & 0.2 & 207 & 5 & 6.1 & 155 & 20 & 6.1 & 146 & 20 & -3.5 & 5582 & 18 & 2000 & 2000 \\
  $radmatchx^{lc}$ & 65.5 & 10.0 & 9 & 10 & 0.3 & 263 & 8 & 5.9 & 146 & 19 & 5.9 & 137 & 19 & -3.4 & 5406 & 17 & 2000 & 2000 \\
  $radmatch^{lc}$ & 60.0 & 10.4 & 14 & 12 & 0.4 & 451 & 10 & 6.6 & 173 & 21 & 6.6 & 164 & 21 & -3.9 & 6296 & 19 & 2000 & 2000 \\
\bottomrule
\end{tabular}
\begin{tablenotes}
\item \scriptsize Notes: `cov' and `int' refer to coverage rate and length of confidence interval, `bias' denotes the absolute bias from the true treatment effect, `sd' is the standard
deviation of the estimator, `rmse' stands for root mean squared error, and `seb' indicates the median bias of the estimated bootstrap standard error. `diff' indicates the difference to the left best performer in percent (\%) and `r' is the rank. `nsimp' shows the number of simulations with point estimations smaller than $|10^{10}|$ and `nsimse' denotes the number of simulations used to compute `cov', `int', and `seb'. Conditional on `nsimp', simulations with standard errors greater than 150 times the (true) standard error are not considered (i.e.simulations with non-converging bootstrap standard errors are dropped).
\end{tablenotes}
\end{threeparttable}
\end{sidewaystable}

\begin{sidewaystable}[ht]
\centering
\begin{threeparttable}
\caption{DGP 4}
\label{DGP 4}
\small
\begin{tabular}{rrrrrrrrrrrrrrrrrrr}
\toprule
& cov & int & d & r & bias & d & r & sd & d & r & rmse & d & r & seb & d & r & nsimp & nsimse \\
\midrule
  $radmatch^{lc}$ & 95.4 & 4500.7 & 94 & 20 & 3.2 & 155 & 5 & 119.8 & 108 & 20 & 119.8 & 107 & 20 & 104.1 & 265 & 14 & 1970 & 1968 \\
  $pairmatch^{lc}$ & 96.3 & 3517.3 & 52 & 11 & 1.3 & 0 & 1 & 107.7 & 87 & 19 & 107.7 & 86 & 19 & 120.2 & 321 & 16 & 1973 & 1962 \\
  $pairmatch^{probit}$ & 99.7 & 5040.8 & 118 & 21 & 5.2 & 312 & 8 & 88.4 & 53 & 17 & 88.5 & 53 & 17 & 128.9 & 352 & 18 & 1980 & 1978 \\
  $randforest$ & 99.8 & 2478.2 & 7 & 3 & 3.0 & 138 & 3 & 67.1 & 16 & 6 & 67.2 & 16 & 6 & 41.5 & 46 & 2 & 1992 & 1974 \\
  $radmatchx^{probit}$ & 99.8 & 3787.2 & 64 & 13 & 3.0 & 136 & 2 & 78.8 & 37 & 12 & 78.9 & 37 & 12 & 76.5 & 168 & 5 & 1987 & 1984 \\
  $ipw^{cbps}$ & 99.8 & 3934.4 & 70 & 15 & 7.6 & 501 & 16 & 74.4 & 29 & 10 & 74.7 & 29 & 10 & 82.1 & 188 & 9 & 1991 & 1990 \\
  $means$ & 99.8 & 2314.1 & 0 & 1 & 16.7 & 1221 & 21 & 80.5 & 40 & 13 & 82.2 & 42 & 14 & 83.0 & 191 & 11 & 1985 & 1975 \\
  $radmatch^{cbps}$ & 99.8 & 3696.9 & 60 & 12 & 6.2 & 390 & 12 & 85.1 & 48 & 16 & 85.3 & 48 & 16 & 124.8 & 338 & 17 & 1987 & 1977 \\
  $reg$ & 99.8 & 2441.8 & 6 & 2 & 5.3 & 319 & 9 & 60.6 & 5 & 2 & 60.8 & 5 & 2 & 77.2 & 171 & 6 & 1995 & 1978 \\
  $radmatch^{probit}$ & 99.8 & 3225.3 & 39 & 8 & 8.4 & 563 & 17 & 88.8 & 54 & 18 & 89.1 & 54 & 18 & 109.9 & 286 & 15 & 1990 & 1979 \\
  $radmatchx^{cbps}$ & 99.8 & 3971.1 & 72 & 18 & 3.1 & 143 & 4 & 66.3 & 15 & 5 & 66.3 & 15 & 4 & 82.3 & 189 & 10 & 1985 & 1981 \\
  $dr^{cbps}$ & 99.8 & 3002.7 & 30 & 5 & 4.7 & 271 & 7 & 57.6 & 0 & 1 & 57.8 & 0 & 1 & 88.5 & 210 & 12 & 1991 & 1983 \\
  $ipw^{lc}$ & 99.8 & 3947.6 & 71 & 16 & 11.4 & 804 & 20 & 70.8 & 23 & 8 & 71.7 & 24 & 8 & 195.7 & 586 & 21 & 1995 & 1985 \\
  $tsls$ & 99.8 & 3150.0 & 36 & 6 & 6.1 & 382 & 11 & 67.5 & 17 & 7 & 67.8 & 17 & 7 & 80.4 & 182 & 7 & 1993 & 1986 \\
  $ipw^{probit}$ & 99.8 & 3952.7 & 71 & 17 & 4.2 & 234 & 6 & 72.7 & 26 & 9 & 72.9 & 26 & 9 & 73.4 & 157 & 3 & 1993 & 1992 \\
  $dr^{probit}$ & 99.8 & 3873.1 & 67 & 14 & 5.6 & 342 & 10 & 65.8 & 14 & 3 & 66.0 & 14 & 3 & 74.1 & 160 & 4 & 1993 & 1993 \\
  $reg^{kernel}$ & 99.9 & 3313.5 & 43 & 9 & 9.9 & 688 & 19 & 66.1 & 15 & 4 & 66.8 & 16 & 5 & 81.8 & 187 & 8 & 1994 & 1959 \\
  $pairmatch^{x}$ & 99.9 & 2775.0 & 20 & 4 & 6.8 & 438 & 15 & 80.5 & 40 & 14 & 80.8 & 40 & 13 & 91.3 & 220 & 13 & 1990 & 1978 \\
  $pairmatch^{cbps}$ & 99.9 & 4201.6 & 82 & 19 & 6.6 & 423 & 14 & 84.0 & 46 & 15 & 84.2 & 46 & 15 & 153.8 & 439 & 20 & 1980 & 1975 \\
  $dr^{lc}$ & 100.0 & 3203.3 & 38 & 7 & 8.5 & 571 & 18 & 75.5 & 31 & 11 & 76.0 & 31 & 11 & 149.5 & 424 & 19 & 1995 & 1955 \\
  $radmatchx^{lc}$ & 89.7 & 3488.8 & 51 & 10 & 6.2 & 395 & 13 & 127.0 & 121 & 21 & 127.2 & 120 & 21 & -28.5 & 0 & 1 & 1975 & 1973 \\
\bottomrule
\end{tabular}
\begin{tablenotes}
\item \scriptsize Notes: `cov' and `int' refer to coverage rate and length of confidence interval, `bias' denotes the absolute bias from the true treatment effect, `sd' is the standard
deviation of the estimator, `rmse' stands for root mean squared error, and `seb' indicates the median bias of the estimated bootstrap standard error. `diff' indicates the difference to the left best performer in percent (\%) and `r' is the rank. `nsimp' shows the number of simulations with point estimations smaller than $|10^{10}|$ and `nsimse' denotes the number of simulations used to compute `cov', `int', and `seb'. Conditional on `nsimp', simulations with standard errors greater than 150 times the (true) standard error are not considered (i.e.simulations with non-converging bootstrap standard errors are dropped).
\end{tablenotes}
\end{threeparttable}
\end{sidewaystable}

\begin{sidewaystable}[ht]
\centering
\begin{threeparttable}
\caption{DGP 5}
\label{DGP 5}
\small
\begin{tabular}{rrrrrrrrrrrrrrrrrrr}
\toprule
& cov & int & d & r & bias & d & r & sd & d & r & rmse & d & r & seb & d & r & nsimp & nsimse \\
\midrule
  $tsls$ & 94.7 & 14.4 & 11 & 4 & 0.7 & 37 & 2 & 3.6 & 11 & 3 & 3.7 & 0 & 1 & 0.0 & 928 & 7 & 2000 & 2000 \\
  $reg$ & 94.3 & 14.8 & 15 & 10 & 0.8 & 59 & 3 & 3.7 & 15 & 7 & 3.8 & 4 & 3 & 0.0 & 345 & 5 & 2000 & 2000 \\
  $dr^{probit}$ & 93.8 & 15.1 & 17 & 12 & 0.8 & 75 & 4 & 3.8 & 17 & 9 & 3.9 & 6 & 6 & 0.0 & 199 & 3 & 2000 & 2000 \\
  $pairmatch^{probit}$ & 93.8 & 19.7 & 53 & 19 & 1.2 & 158 & 12 & 5.1 & 57 & 16 & 5.2 & 43 & 16 & -0.2 & 5624 & 10 & 2000 & 2000 \\
  $dr^{cbps}$ & 93.8 & 15.0 & 17 & 11 & 0.8 & 78 & 6 & 3.8 & 17 & 8 & 3.9 & 6 & 5 & 0.0 & 221 & 4 & 2000 & 2000 \\
  $ipw^{probit}$ & 93.7 & 14.6 & 13 & 8 & 1.1 & 130 & 8 & 3.7 & 13 & 6 & 3.9 & 5 & 4 & 0.0 & 0 & 1 & 2000 & 2000 \\
  $radmatch^{probit}$ & 93.6 & 19.3 & 50 & 17 & 1.3 & 170 & 13 & 5.2 & 60 & 18 & 5.4 & 45 & 18 & -0.4 & 13333 & 12 & 2000 & 2000 \\
  $pairmatch^{cbps}$ & 93.5 & 19.4 & 51 & 18 & 1.2 & 148 & 10 & 5.0 & 54 & 15 & 5.1 & 40 & 15 & -0.1 & 4095 & 8 & 2000 & 2000 \\
  $randforest$ & 93.3 & 13.9 & 8 & 2 & 0.9 & 94 & 7 & 3.7 & 13 & 5 & 3.8 & 3 & 2 & -0.1 & 5373 & 9 & 2000 & 2000 \\
  $radmatch^{cbps}$ & 92.6 & 19.0 & 48 & 16 & 1.2 & 152 & 11 & 5.1 & 57 & 17 & 5.3 & 43 & 17 & -0.4 & 13262 & 11 & 2000 & 2000 \\
  $ipw^{cbps}$ & 92.4 & 14.4 & 12 & 5 & 1.4 & 185 & 14 & 3.7 & 12 & 4 & 3.9 & 6 & 7 & -0.0 & 22 & 2 & 2000 & 2000 \\
  $pairmatch^{x}$ & 92.3 & 16.5 & 28 & 13 & 0.8 & 77 & 5 & 4.6 & 40 & 12 & 4.6 & 26 & 11 & -0.4 & 14139 & 13 & 2000 & 2000 \\
  $reg^{kernel}$ & 91.8 & 14.2 & 10 & 3 & 1.4 & 196 & 16 & 4.1 & 26 & 11 & 4.3 & 18 & 10 & -0.5 & 18965 & 16 & 2000 & 1871 \\
  $radmatchx^{probit}$ & 90.3 & 16.7 & 30 & 15 & 1.5 & 209 & 17 & 4.7 & 44 & 14 & 4.9 & 34 & 14 & -0.5 & 17905 & 15 & 2000 & 2000 \\
  $ipw^{lc}$ & 99.8 & 126.3 & 880 & 20 & 2.0 & 312 & 20 & 3.5 & 7 & 2 & 4.0 & 9 & 8 & 9.6 & 359462 & 20 & 2000 & 1589 \\
  $radmatchx^{cbps}$ & 90.1 & 16.6 & 29 & 14 & 1.5 & 223 & 19 & 4.6 & 42 & 13 & 4.9 & 32 & 13 & -0.4 & 15508 & 14 & 2000 & 2000 \\
  $dr^{lc}$ & 100.0 & 696.7 & 5306 & 21 & 1.4 & 193 & 15 & 3.8 & 18 & 10 & 4.1 & 11 & 9 & 148.5 & 5558002 & 21 & 2000 & 1362 \\
  $means$ & 83.6 & 12.9 & 0 & 1 & 3.3 & 594 & 21 & 3.3 & 0 & 1 & 4.7 & 27 & 12 & 0.0 & 824 & 6 & 2000 & 2000 \\
  $pairmatch^{lc}$ & 65.6 & 14.5 & 13 & 7 & 1.1 & 130 & 9 & 9.9 & 205 & 20 & 10.0 & 172 & 20 & -6.2 & 233179 & 18 & 2000 & 2000 \\
  $radmatchx^{lc}$ & 64.8 & 14.4 & 12 & 6 & 0.5 & 0 & 1 & 8.9 & 174 & 19 & 8.9 & 143 & 19 & -5.3 & 196777 & 17 & 2000 & 2000 \\
  $radmatch^{lc}$ & 59.2 & 14.6 & 13 & 9 & 1.5 & 218 & 18 & 10.0 & 206 & 21 & 10.1 & 174 & 21 & -6.3 & 234489 & 19 & 2000 & 2000 \\
\bottomrule
\end{tabular}
\begin{tablenotes}
\item \scriptsize Notes: `cov' and `int' refer to coverage rate and length of confidence interval, `bias' denotes the absolute bias from the true treatment effect, `sd' is the standard
deviation of the estimator, `rmse' stands for root mean squared error, and `seb' indicates the median bias of the estimated bootstrap standard error. `diff' indicates the difference to the left best performer in percent (\%) and `r' is the rank. `nsimp' shows the number of simulations with point estimations smaller than $|10^{10}|$ and `nsimse' denotes the number of simulations used to compute `cov', `int', and `seb'. Conditional on `nsimp', simulations with standard errors greater than 150 times the (true) standard error are not considered (i.e.simulations with non-converging bootstrap standard errors are dropped).
\end{tablenotes}
\end{threeparttable}
\end{sidewaystable}

\begin{sidewaystable}[ht]
\centering
\begin{threeparttable}
\caption{DGP 6}
\label{DGP 6}
\small
\begin{tabular}{rrrrrrrrrrrrrrrrrrr}
\toprule
& cov & int & d & r & bias & d & r & sd & d & r & rmse & d & r & seb & d & r & nsimp & nsimse \\
\midrule
  $ipw^{cbps}$ & 94.8 & 8.1 & 5 & 7 & 0.3 & 259 & 17 & 2.7 & 1 & 6 & 2.1 & 1 & 6 & -0.6 & 77 & 6 & 2000 & 2000 \\
  $pairmatch^{cbps}$ & 95.3 & 9.5 & 23 & 19 & 0.2 & 224 & 16 & 3.0 & 10 & 15 & 2.4 & 18 & 18 & -0.5 & 47 & 3 & 2000 & 2000 \\
  $tsls$ & 94.7 & 8.0 & 3 & 2 & 0.2 & 123 & 9 & 2.7 & 0 & 1 & 2.0 & 0 & 1 & -0.6 & 77 & 7 & 2000 & 2000 \\
  $dr^{probit}$ & 94.7 & 8.0 & 4 & 5 & 0.2 & 132 & 11 & 2.7 & 0 & 4 & 2.0 & 0 & 3 & -0.7 & 78 & 10 & 2000 & 2000 \\
  $reg$ & 94.7 & 8.0 & 4 & 3 & 0.2 & 133 & 12 & 2.7 & 0 & 2 & 2.0 & 0 & 2 & -0.7 & 77 & 8 & 2000 & 2000 \\
  $ipw^{probit}$ & 94.6 & 8.0 & 4 & 6 & 0.2 & 112 & 7 & 2.7 & 0 & 5 & 2.0 & 0 & 5 & -0.7 & 77 & 9 & 2000 & 2000 \\
  $dr^{cbps}$ & 94.6 & 8.0 & 4 & 4 & 0.2 & 137 & 13 & 2.7 & 0 & 3 & 2.0 & 0 & 4 & -0.7 & 79 & 11 & 2000 & 2000 \\
  $radmatch^{cbps}$ & 94.4 & 9.1 & 17 & 15 & 0.2 & 215 & 15 & 2.9 & 9 & 14 & 2.4 & 16 & 14 & -0.6 & 70 & 5 & 2000 & 2000 \\
  $pairmatch^{probit}$ & 95.6 & 9.5 & 23 & 18 & 0.1 & 24 & 3 & 3.0 & 12 & 18 & 2.4 & 17 & 17 & -0.6 & 58 & 4 & 2000 & 2000 \\
  $radmatch^{probit}$ & 94.0 & 9.0 & 17 & 14 & 0.1 & 51 & 4 & 3.0 & 10 & 17 & 2.3 & 16 & 13 & -0.7 & 81 & 12 & 2000 & 2000 \\
  $radmatchx^{cbps}$ & 93.9 & 8.6 & 11 & 11 & 0.2 & 117 & 8 & 2.9 & 8 & 12 & 2.3 & 13 & 11 & -0.7 & 95 & 14 & 2000 & 2000 \\
  $reg^{kernel}$ & 93.8 & 8.4 & 8 & 9 & 0.4 & 458 & 19 & 2.8 & 4 & 11 & 2.2 & 6 & 9 & -0.7 & 96 & 15 & 2000 & 1973 \\
  $pairmatch^{x}$ & 93.8 & 8.7 & 12 & 12 & 0.2 & 107 & 6 & 3.0 & 10 & 16 & 2.4 & 17 & 16 & -0.8 & 107 & 17 & 2000 & 2000 \\
  $ipw^{lc}$ & 96.3 & 12.1 & 56 & 20 & 0.7 & 767 & 20 & 2.7 & 1 & 7 & 2.2 & 6 & 10 & -0.4 & -0 & 1 & 2000 & 1991 \\
  $randforest$ & 93.6 & 7.7 & 0 & 1 & 0.2 & 201 & 14 & 2.7 & 2 & 10 & 2.1 & 4 & 8 & -0.8 & 110 & 18 & 2000 & 2000 \\
  $dr^{lc}$ & 96.7 & 15.8 & 104 & 21 & 0.4 & 407 & 18 & 2.7 & 2 & 9 & 2.1 & 4 & 7 & -0.4 & 5 & 2 & 2000 & 1959 \\
  $radmatchx^{probit}$ & 93.0 & 8.6 & 11 & 10 & 0.1 & 0 & 1 & 2.9 & 8 & 13 & 2.3 & 13 & 12 & -0.7 & 98 & 16 & 2000 & 2000 \\
  $means$ & 91.0 & 8.1 & 5 & 8 & 1.2 & 1415 & 21 & 2.7 & 1 & 8 & 2.4 & 16 & 15 & -0.7 & 82 & 13 & 2000 & 2000 \\
  $pairmatch^{lc}$ & 80.7 & 9.1 & 17 & 16 & 0.1 & 16 & 2 & 4.5 & 68 & 19 & 4.1 & 104 & 19 & -2.2 & 500 & 19 & 2000 & 2000 \\
  $radmatchx^{lc}$ & 76.0 & 8.7 & 12 & 13 & 0.2 & 124 & 10 & 4.6 & 71 & 20 & 4.2 & 108 & 20 & -2.4 & 555 & 20 & 2000 & 2000 \\
  $radmatch^{lc}$ & 74.8 & 9.1 & 18 & 17 & 0.2 & 100 & 5 & 5.0 & 87 & 21 & 4.7 & 131 & 21 & -2.7 & 637 & 21 & 2000 & 2000 \\
\bottomrule
\end{tabular}
\begin{tablenotes}
\item \scriptsize Notes: `cov' and `int' refer to coverage rate and length of confidence interval, `bias' denotes the absolute bias from the true treatment effect, `sd' is the standard
deviation of the estimator, `rmse' stands for root mean squared error, and `seb' indicates the median bias of the estimated bootstrap standard error. `diff' indicates the difference to the left best performer in percent (\%) and `r' is the rank. `nsimp' shows the number of simulations with point estimations smaller than $|10^{10}|$ and `nsimse' denotes the number of simulations used to compute `cov', `int', and `seb'. Conditional on `nsimp', simulations with standard errors greater than 150 times the (true) standard error are not considered (i.e.simulations with non-converging bootstrap standard errors are dropped).
\end{tablenotes}
\end{threeparttable}
\end{sidewaystable}

\begin{sidewaystable}[ht]
\centering
\begin{threeparttable}
\caption{DGP 7}
\label{DGP 7}
\small
\begin{tabular}{rrrrrrrrrrrrrrrrrrr}
\toprule
& cov & int & d & r & bias & d & r & sd & d & r & rmse & d & r & seb & d & r & nsimp & nsimse \\
\midrule
  $pairmatch^{lc}$ & 94.1 & 4598.4 & 146 & 15 & 2.4 & 2656 & 3 & 131.0 & 156 & 19 & 131.0 & 155 & 19 & 109.2 & 165 & 14 & 1974 & 1971 \\
  $radmatch^{lc}$ & 93.6 & 5664.4 & 203 & 20 & 1.5 & 1602 & 2 & 132.0 & 158 & 20 & 132.0 & 157 & 20 & 106.1 & 157 & 13 & 1962 & 1962 \\
  $pairmatch^{probit}$ & 99.7 & 4874.0 & 161 & 16 & 10.6 & 12005 & 19 & 95.7 & 87 & 17 & 96.3 & 87 & 17 & 172.5 & 318 & 18 & 1982 & 1978 \\
  $radmatchx^{probit}$ & 99.7 & 2225.9 & 19 & 2 & 10.2 & 11453 & 18 & 71.2 & 39 & 11 & 71.9 & 40 & 11 & 72.8 & 76 & 8 & 1991 & 1984 \\
  $radmatch^{cbps}$ & 99.8 & 5302.2 & 184 & 17 & 7.3 & 8222 & 13 & 99.3 & 94 & 18 & 99.6 & 94 & 18 & 170.7 & 314 & 17 & 1978 & 1976 \\
  $radmatch^{probit}$ & 99.8 & 3838.6 & 106 & 13 & 8.0 & 9041 & 14 & 95.6 & 87 & 16 & 95.9 & 87 & 16 & 178.5 & 333 & 19 & 1986 & 1981 \\
  $tsls$ & 99.8 & 3493.4 & 87 & 12 & 2.8 & 3060 & 4 & 66.2 & 29 & 7 & 66.2 & 29 & 6 & 67.4 & 63 & 5 & 1994 & 1993 \\
  $randforest$ & 99.8 & 2989.8 & 60 & 8 & 4.2 & 4656 & 7 & 72.8 & 42 & 12 & 72.9 & 42 & 12 & 41.2 & 0 & 1 & 1995 & 1994 \\
  $dr^{probit}$ & 99.8 & 2641.5 & 42 & 4 & 5.3 & 5910 & 10 & 67.6 & 32 & 9 & 67.8 & 32 & 9 & 66.8 & 62 & 4 & 1990 & 1972 \\
  $dr^{cbps}$ & 99.8 & 2421.6 & 30 & 3 & 4.4 & 4874 & 8 & 66.1 & 29 & 5 & 66.2 & 29 & 5 & 72.0 & 75 & 7 & 1989 & 1980 \\
  $pairmatch^{x}$ & 99.8 & 2815.7 & 51 & 7 & 3.0 & 3358 & 6 & 83.9 & 64 & 14 & 83.9 & 63 & 14 & 115.2 & 179 & 15 & 1989 & 1980 \\
  $radmatchx^{cbps}$ & 99.8 & 2754.3 & 48 & 6 & 8.1 & 9155 & 15 & 66.7 & 30 & 8 & 67.2 & 31 & 7 & 73.7 & 79 & 9 & 1991 & 1983 \\
  $reg^{kernel}$ & 99.9 & 5359.0 & 187 & 18 & 8.7 & 9747 & 17 & 58.7 & 15 & 3 & 59.3 & 16 & 3 & 62.5 & 52 & 2 & 1991 & 1892 \\
  $pairmatch^{cbps}$ & 99.9 & 4426.6 & 137 & 14 & 5.6 & 6262 & 11 & 93.3 & 82 & 15 & 93.5 & 82 & 15 & 159.7 & 287 & 16 & 1981 & 1979 \\
  $reg$ & 99.9 & 2657.9 & 42 & 5 & 2.9 & 3228 & 5 & 65.0 & 27 & 4 & 65.1 & 27 & 4 & 66.5 & 61 & 3 & 1992 & 1987 \\
  $ipw^{lc}$ & 99.9 & 5380.6 & 188 & 19 & 13.7 & 15458 & 20 & 66.1 & 29 & 6 & 67.4 & 31 & 8 & 463.4 & 1024 & 20 & 1994 & 1980 \\
  $ipw^{probit}$ & 100.0 & 3144.4 & 68 & 9 & 4.7 & 5269 & 9 & 51.2 & 0 & 1 & 51.4 & -0 & 1 & 76.2 & 85 & 10 & 1991 & 1990 \\
  $ipw^{cbps}$ & 100.0 & 3364.2 & 80 & 11 & 6.6 & 7404 & 12 & 56.7 & 11 & 2 & 57.1 & 11 & 2 & 68.6 & 66 & 6 & 1991 & 1990 \\
  $dr^{lc}$ & 100.0 & 8846.7 & 374 & 21 & 8.3 & 9347 & 16 & 70.0 & 37 & 10 & 70.5 & 37 & 10 & 513.6 & 1146 & 21 & 1994 & 1889 \\
  $means$ & 100.0 & 3212.4 & 72 & 10 & 24.8 & 28138 & 21 & 78.6 & 54 & 13 & 82.4 & 60 & 13 & 82.0 & 99 & 11 & 1991 & 1984 \\
  $radmatchx^{lc}$ & 79.2 & 1866.5 & 0 & 1 & 0.1 & 0 & 1 & 138.9 & 172 & 21 & 138.9 & 170 & 21 & -95.0 & 130 & 12 & 1968 & 1967 \\
\bottomrule
\end{tabular}
\begin{tablenotes}
\item \scriptsize Notes: `cov' and `int' refer to coverage rate and length of confidence interval, `bias' denotes the absolute bias from the true treatment effect, `sd' is the standard
deviation of the estimator, `rmse' stands for root mean squared error, and `seb' indicates the median bias of the estimated bootstrap standard error. `diff' indicates the difference to the left best performer in percent (\%) and `r' is the rank. `nsimp' shows the number of simulations with point estimations smaller than $|10^{10}|$ and `nsimse' denotes the number of simulations used to compute `cov', `int', and `seb'. Conditional on `nsimp', simulations with standard errors greater than 150 times the (true) standard error are not considered (i.e.simulations with non-converging bootstrap standard errors are dropped).
\end{tablenotes}
\end{threeparttable}
\end{sidewaystable}

\begin{sidewaystable}[ht]
\centering
\begin{threeparttable}
\caption{DGP 8}
\label{DGP 8}
\small
\begin{tabular}{rrrrrrrrrrrrrrrrrrr}
\toprule
& cov & int & d & r & bias & d & r & sd & d & r & rmse & d & r & seb & d & r & nsimp & nsimse \\
\midrule
  $tsls$ & 95.0 & 8.5 & 5 & 5 & 0.2 & 1 & 2 & 2.2 & 5 & 3 & 2.2 & 0 & 1 & 0.0 & 969 & 6 & 2000 & 2000 \\
  $reg$ & 94.4 & 8.7 & 6 & 7 & 0.3 & 26 & 3 & 2.2 & 7 & 6 & 2.2 & 2 & 2 & -0.0 & 0 & 1 & 2000 & 2000 \\
  $dr^{probit}$ & 94.4 & 8.8 & 7 & 9 & 0.4 & 55 & 4 & 2.2 & 8 & 9 & 2.2 & 3 & 6 & 0.0 & 624 & 4 & 2000 & 2000 \\
  $dr^{cbps}$ & 94.3 & 8.7 & 7 & 8 & 0.4 & 56 & 5 & 2.2 & 8 & 8 & 2.2 & 3 & 4 & 0.0 & 282 & 2 & 2000 & 2000 \\
  $ipw^{probit}$ & 94.0 & 8.6 & 5 & 6 & 0.5 & 123 & 9 & 2.2 & 6 & 5 & 2.2 & 3 & 3 & 0.0 & 691 & 5 & 2000 & 2000 \\
  $pairmatch^{cbps}$ & 94.0 & 11.2 & 38 & 18 & 0.7 & 193 & 10 & 2.9 & 41 & 15 & 3.0 & 37 & 15 & -0.1 & 6688 & 9 & 2000 & 2000 \\
  $pairmatch^{probit}$ & 93.6 & 11.3 & 39 & 19 & 0.7 & 194 & 11 & 2.9 & 42 & 17 & 3.0 & 38 & 16 & -0.1 & 6507 & 8 & 2000 & 2000 \\
  $randforest$ & 93.3 & 8.3 & 2 & 2 & 0.4 & 86 & 7 & 2.2 & 7 & 7 & 2.2 & 3 & 5 & -0.1 & 8662 & 10 & 2000 & 2000 \\
  $ipw^{cbps}$ & 93.2 & 8.5 & 5 & 4 & 0.7 & 196 & 12 & 2.2 & 5 & 4 & 2.3 & 5 & 7 & 0.0 & 327 & 3 & 2000 & 2000 \\
  $radmatch^{cbps}$ & 93.0 & 11.0 & 35 & 16 & 0.7 & 204 & 14 & 2.9 & 42 & 16 & 3.0 & 38 & 17 & -0.2 & 14189 & 12 & 2000 & 2000 \\
  $radmatch^{probit}$ & 92.8 & 11.1 & 37 & 17 & 0.7 & 212 & 16 & 3.0 & 45 & 18 & 3.1 & 41 & 18 & -0.2 & 17701 & 14 & 2000 & 2000 \\
  $reg^{kernel}$ & 92.7 & 8.5 & 5 & 3 & 0.7 & 208 & 15 & 2.3 & 10 & 11 & 2.4 & 9 & 10 & -0.1 & 10309 & 11 & 1999 & 1880 \\
  $pairmatch^{x}$ & 92.2 & 9.5 & 16 & 13 & 0.4 & 58 & 6 & 2.6 & 27 & 13 & 2.6 & 22 & 11 & -0.2 & 20258 & 15 & 2000 & 2000 \\
  $radmatchx^{cbps}$ & 91.6 & 9.6 & 18 & 14 & 0.8 & 243 & 19 & 2.6 & 27 & 12 & 2.7 & 25 & 12 & -0.2 & 16330 & 13 & 2000 & 2000 \\
  $radmatchx^{probit}$ & 91.3 & 9.6 & 18 & 15 & 0.8 & 229 & 18 & 2.6 & 29 & 14 & 2.8 & 27 & 13 & -0.2 & 20345 & 16 & 2000 & 2000 \\
  $ipw^{lc}$ & 99.9 & 57.0 & 599 & 20 & 1.1 & 351 & 20 & 2.1 & 3 & 2 & 2.4 & 9 & 9 & 5.4 & 497997 & 20 & 2000 & 1648 \\
  $dr^{lc}$ & 100.0 & 350.1 & 4193 & 21 & 0.7 & 196 & 13 & 2.2 & 9 & 10 & 2.3 & 8 & 8 & 32.6 & 3026625 & 21 & 1999 & 1276 \\
  $means$ & 84.4 & 8.2 & 0 & 1 & 2.0 & 729 & 21 & 2.1 & 0 & 1 & 2.9 & 31 & 14 & 0.0 & 1802 & 7 & 2000 & 2000 \\
  $pairmatch^{lc}$ & 68.2 & 9.2 & 13 & 11 & 0.5 & 120 & 8 & 5.6 & 171 & 20 & 5.6 & 157 & 20 & -3.2 & 297620 & 18 & 2000 & 2000 \\
  $radmatchx^{lc}$ & 64.8 & 8.8 & 8 & 10 & 0.2 & 0 & 1 & 5.2 & 155 & 19 & 5.2 & 141 & 19 & -3.0 & 276921 & 17 & 2000 & 2000 \\
  $radmatch^{lc}$ & 62.9 & 9.2 & 13 & 12 & 0.8 & 217 & 17 & 5.9 & 189 & 21 & 6.0 & 175 & 21 & -3.6 & 332536 & 19 & 2000 & 2000 \\
\bottomrule
\end{tabular}
\begin{tablenotes}
\item \scriptsize Notes: `cov' and `int' refer to coverage rate and length of confidence interval, `bias' denotes the absolute bias from the true treatment effect, `sd' is the standard
deviation of the estimator, `rmse' stands for root mean squared error, and `seb' indicates the median bias of the estimated bootstrap standard error. `diff' indicates the difference to the left best performer in percent (\%) and `r' is the rank. `nsimp' shows the number of simulations with point estimations smaller than $|10^{10}|$ and `nsimse' denotes the number of simulations used to compute `cov', `int', and `seb'. Conditional on `nsimp', simulations with standard errors greater than 150 times the (true) standard error are not considered (i.e.simulations with non-converging bootstrap standard errors are dropped).
\end{tablenotes}
\end{threeparttable}
\end{sidewaystable}

\begin{sidewaystable}[ht]
\centering
\begin{threeparttable}
\caption{DGP 9}
\label{DGP 9}
\small
\begin{tabular}{rrrrrrrrrrrrrrrrrrr}
\toprule
& cov & int & d & r & bias & d & r & sd & d & r & rmse & d & r & seb & d & r & nsimp & nsimse \\
\midrule
  $dr^{probit}$ & 95.0 & 0.2 & 4 & 6 & 0.0 & 199 & 9 & 0.1 & 0 & 4 & 0.1 & 0 & 4 & -0.0 & 57 & 7 & 2000 & 2000 \\
  $ipw^{cbps}$ & 95.0 & 0.2 & 4 & 8 & 0.0 & 325 & 14 & 0.1 & 0 & 8 & 0.1 & 1 & 6 & -0.0 & 9 & 2 & 2000 & 2000 \\
  $dr^{cbps}$ & 95.0 & 0.2 & 4 & 3 & 0.0 & 199 & 10 & 0.1 & 0 & 3 & 0.1 & 0 & 3 & -0.0 & 61 & 9 & 2000 & 2000 \\
  $pairmatch^{cbps}$ & 95.0 & 0.2 & 22 & 18 & 0.0 & 327 & 15 & 0.1 & 18 & 18 & 0.1 & 18 & 18 & -0.0 & 51 & 5 & 2000 & 2000 \\
  $radmatch^{cbps}$ & 95.0 & 0.2 & 17 & 16 & 0.0 & 316 & 13 & 0.1 & 16 & 16 & 0.1 & 16 & 16 & -0.0 & 534 & 12 & 2000 & 2000 \\
  $reg$ & 95.0 & 0.2 & 4 & 2 & 0.0 & 189 & 6 & 0.1 & 0 & 1 & 0.1 & 0 & 1 & -0.0 & 42 & 4 & 2000 & 2000 \\
  $pairmatch^{probit}$ & 95.1 & 0.2 & 23 & 19 & 0.0 & 194 & 7 & 0.1 & 16 & 17 & 0.1 & 16 & 17 & 0.0 & 0 & 1 & 2000 & 2000 \\
  $ipw^{probit}$ & 94.9 & 0.2 & 4 & 7 & 0.0 & 215 & 12 & 0.1 & 0 & 6 & 0.1 & 0 & 5 & -0.0 & 56 & 6 & 2000 & 2000 \\
  $tsls$ & 94.8 & 0.2 & 4 & 4 & 0.0 & 206 & 11 & 0.1 & 0 & 2 & 0.1 & 0 & 2 & -0.0 & 33 & 3 & 2000 & 2000 \\
  $reg^{kernel}$ & 94.5 & 0.2 & 5 & 9 & 0.0 & 654 & 19 & 0.1 & 2 & 11 & 0.1 & 3 & 10 & -0.0 & 91 & 10 & 2000 & 2000 \\
  $radmatch^{probit}$ & 94.0 & 0.2 & 17 & 15 & 0.0 & 198 & 8 & 0.1 & 15 & 15 & 0.1 & 14 & 15 & -0.0 & 367 & 11 & 2000 & 2000 \\
  $pairmatch^{x}$ & 93.8 & 0.2 & 12 & 13 & 0.0 & 103 & 4 & 0.1 & 13 & 14 & 0.1 & 13 & 14 & -0.0 & 909 & 16 & 2000 & 2000 \\
  $radmatchx^{cbps}$ & 93.7 & 0.2 & 11 & 10 & 0.0 & 44 & 3 & 0.1 & 11 & 12 & 0.1 & 11 & 12 & -0.0 & 822 & 13 & 2000 & 2000 \\
  $means$ & 93.6 & 0.2 & 4 & 5 & 0.0 & 1321 & 21 & 0.1 & 0 & 5 & 0.1 & 6 & 11 & -0.0 & 60 & 8 & 2000 & 2000 \\
  $radmatchx^{probit}$ & 93.5 & 0.2 & 11 & 11 & 0.0 & 0 & 1 & 0.1 & 12 & 13 & 0.1 & 12 & 13 & -0.0 & 874 & 14 & 2000 & 2000 \\
  $randforest$ & 93.4 & 0.2 & 0 & 1 & 0.0 & 178 & 5 & 0.1 & 2 & 10 & 0.1 & 2 & 7 & -0.0 & 884 & 15 & 2000 & 2000 \\
  $ipw^{lc}$ & 97.4 & 0.3 & 60 & 20 & 0.0 & 804 & 20 & 0.1 & 0 & 7 & 0.1 & 3 & 9 & 0.0 & 1915 & 18 & 2000 & 1992 \\
  $dr^{lc}$ & 97.6 & 0.4 & 91 & 21 & 0.0 & 531 & 17 & 0.1 & 1 & 9 & 0.1 & 2 & 8 & 0.0 & 1815 & 17 & 2000 & 1984 \\
  $pairmatch^{lc}$ & 78.4 & 0.2 & 17 & 14 & 0.0 & 481 & 16 & 0.1 & 101 & 19 & 0.1 & 101 & 19 & -0.0 & 13476 & 19 & 2000 & 2000 \\
  $radmatchx^{lc}$ & 74.1 & 0.2 & 12 & 12 & 0.0 & 10 & 2 & 0.1 & 106 & 20 & 0.1 & 106 & 20 & -0.1 & 15054 & 20 & 2000 & 2000 \\
  $radmatch^{lc}$ & 71.7 & 0.2 & 17 & 17 & 0.0 & 582 & 18 & 0.1 & 133 & 21 & 0.1 & 133 & 21 & -0.1 & 18336 & 21 & 2000 & 2000 \\
\bottomrule
\end{tabular}
\begin{tablenotes}
\item \scriptsize Notes: `cov' and `int' refer to coverage rate and length of confidence interval, `bias' denotes the absolute bias from the true treatment effect, `sd' is the standard
deviation of the estimator, `rmse' stands for root mean squared error, and `seb' indicates the median bias of the estimated bootstrap standard error. `diff' indicates the difference to the left best performer in percent (\%) and `r' is the rank. `nsimp' shows the number of simulations with point estimations smaller than $|10^{10}|$ and `nsimse' denotes the number of simulations used to compute `cov', `int', and `seb'. Conditional on `nsimp', simulations with standard errors greater than 150 times the (true) standard error are not considered (i.e.simulations with non-converging bootstrap standard errors are dropped).
\end{tablenotes}
\end{threeparttable}
\end{sidewaystable}

\begin{sidewaystable}[ht]
\centering
\begin{threeparttable}
\caption{DGP 10}
\label{DGP 10}
\small
\begin{tabular}{rrrrrrrrrrrrrrrrrrr}
\toprule
& cov & int & d & r & bias & d & r & sd & d & r & rmse & d & r & seb & d & r & nsimp & nsimse \\
\midrule
  $pairmatch^{probit}$ & 95.5 & 0.5 & 22 & 19 & 0.0 & 72 & 3 & 0.1 & 18 & 16 & 0.1 & 16 & 17 & -0.0 & 0 & 1 & 2000 & 2000 \\
  $pairmatch^{cbps}$ & 94.5 & 0.5 & 22 & 18 & 0.0 & 315 & 13 & 0.1 & 18 & 18 & 0.1 & 17 & 18 & -0.0 & 322 & 2 & 2000 & 2000 \\
  $dr^{probit}$ & 94.4 & 0.4 & 4 & 8 & 0.0 & 262 & 8 & 0.1 & 2 & 7 & 0.1 & 0 & 5 & -0.0 & 1222 & 3 & 2000 & 2000 \\
  $dr^{cbps}$ & 94.4 & 0.4 & 4 & 7 & 0.0 & 265 & 9 & 0.1 & 2 & 6 & 0.1 & 0 & 4 & -0.0 & 1329 & 6 & 2000 & 2000 \\
  $tsls$ & 94.4 & 0.4 & 3 & 3 & 0.0 & 289 & 11 & 0.1 & 1 & 2 & 0.1 & 0 & 1 & -0.0 & 1656 & 8 & 2000 & 2000 \\
  $radmatch^{probit}$ & 94.3 & 0.5 & 16 & 16 & 0.0 & 128 & 4 & 0.1 & 16 & 14 & 0.1 & 14 & 14 & -0.0 & 4754 & 11 & 2000 & 2000 \\
  $reg$ & 94.3 & 0.4 & 4 & 6 & 0.0 & 244 & 7 & 0.1 & 2 & 5 & 0.1 & 0 & 3 & -0.0 & 1224 & 4 & 2000 & 2000 \\
  $ipw^{probit}$ & 94.2 & 0.4 & 3 & 4 & 0.0 & 285 & 10 & 0.1 & 2 & 4 & 0.1 & 0 & 2 & -0.0 & 1306 & 5 & 2000 & 2000 \\
  $ipw^{cbps}$ & 94.1 & 0.4 & 4 & 5 & 0.0 & 477 & 15 & 0.1 & 2 & 8 & 0.1 & 1 & 6 & -0.0 & 1621 & 7 & 2000 & 2000 \\
  $radmatch^{cbps}$ & 94.1 & 0.5 & 16 & 17 & 0.0 & 310 & 12 & 0.1 & 16 & 15 & 0.1 & 15 & 15 & -0.0 & 5478 & 12 & 2000 & 2000 \\
  $reg^{kernel}$ & 94.0 & 0.4 & 5 & 9 & 0.0 & 1014 & 19 & 0.1 & 4 & 10 & 0.1 & 4 & 10 & -0.0 & 2700 & 10 & 2000 & 2000 \\
  $ipw^{lc}$ & 96.2 & 0.5 & 36 & 20 & 0.0 & 1250 & 20 & 0.1 & 1 & 3 & 0.1 & 2 & 8 & 0.0 & 16720 & 17 & 2000 & 1989 \\
  $radmatchx^{cbps}$ & 93.6 & 0.4 & 10 & 11 & 0.0 & 177 & 5 & 0.1 & 14 & 12 & 0.1 & 12 & 12 & -0.0 & 9528 & 13 & 2000 & 2000 \\
  $dr^{lc}$ & 96.5 & 0.7 & 80 & 21 & 0.0 & 835 & 18 & 0.1 & 2 & 9 & 0.1 & 2 & 7 & 0.0 & 19120 & 18 & 2000 & 1979 \\
  $radmatchx^{probit}$ & 93.3 & 0.4 & 11 & 12 & 0.0 & 0 & 1 & 0.1 & 14 & 13 & 0.1 & 12 & 13 & -0.0 & 9795 & 14 & 2000 & 2000 \\
  $means$ & 93.2 & 0.4 & 1 & 2 & 0.0 & 1983 & 21 & 0.1 & 0 & 1 & 0.1 & 4 & 11 & -0.0 & 2423 & 9 & 2000 & 2000 \\
  $pairmatch^{x}$ & 92.8 & 0.4 & 12 & 13 & 0.0 & 190 & 6 & 0.1 & 18 & 17 & 0.1 & 16 & 16 & -0.0 & 13635 & 16 & 2000 & 2000 \\
  $randforest$ & 92.6 & 0.4 & 0 & 1 & 0.0 & 353 & 14 & 0.1 & 5 & 11 & 0.1 & 3 & 9 & -0.0 & 11295 & 15 & 2000 & 2000 \\
  $pairmatch^{lc}$ & 78.6 & 0.5 & 14 & 14 & 0.0 & 564 & 16 & 0.2 & 107 & 20 & 0.2 & 104 & 20 & -0.1 & 144307 & 19 & 2000 & 2000 \\
  $radmatchx^{lc}$ & 75.0 & 0.4 & 9 & 10 & 0.0 & 3 & 2 & 0.2 & 105 & 19 & 0.2 & 102 & 19 & -0.1 & 147467 & 20 & 2000 & 2000 \\
  $radmatch^{lc}$ & 71.9 & 0.5 & 14 & 15 & 0.0 & 715 & 17 & 0.2 & 131 & 21 & 0.2 & 128 & 21 & -0.1 & 179991 & 21 & 2000 & 2000 \\
\bottomrule
\end{tabular}
\begin{tablenotes}
\item \scriptsize Notes: `cov' and `int' refer to coverage rate and length of confidence interval, `bias' denotes the absolute bias from the true treatment effect, `sd' is the standard
deviation of the estimator, `rmse' stands for root mean squared error, and `seb' indicates the median bias of the estimated bootstrap standard error. `diff' indicates the difference to the left best performer in percent (\%) and `r' is the rank. `nsimp' shows the number of simulations with point estimations smaller than $|10^{10}|$ and `nsimse' denotes the number of simulations used to compute `cov', `int', and `seb'. Conditional on `nsimp', simulations with standard errors greater than 150 times the (true) standard error are not considered (i.e.simulations with non-converging bootstrap standard errors are dropped).
\end{tablenotes}
\end{threeparttable}
\end{sidewaystable}

\begin{sidewaystable}[ht]
\centering
\begin{threeparttable}
\caption{DGP 11}
\label{DGP 11}
\small
\begin{tabular}{rrrrrrrrrrrrrrrrrrr}
\toprule
& cov & int & d & r & bias & d & r & sd & d & r & rmse & d & r & seb & d & r & nsimp & nsimse \\
\midrule
  $dr^{probit}$ & 95.0 & 0.2 & 7 & 9 & 0.0 & 133 & 3 & 0.1 & 6 & 10 & 0.1 & 1 & 8 & -0.0 & 8 & 2 & 2000 & 2000 \\
  $ipw^{cbps}$ & 94.9 & 0.2 & 5 & 4 & 0.0 & 1556 & 15 & 0.1 & 4 & 3 & 0.1 & 0 & 1 & -0.0 & 12 & 4 & 2000 & 2000 \\
  $ipw^{probit}$ & 95.2 & 0.2 & 6 & 7 & 0.0 & 667 & 6 & 0.1 & 5 & 5 & 0.1 & 0 & 3 & -0.0 & 0 & 1 & 2000 & 2000 \\
  $pairmatch^{cbps}$ & 94.8 & 0.3 & 36 & 18 & 0.0 & 816 & 9 & 0.1 & 37 & 15 & 0.1 & 31 & 15 & -0.0 & 163 & 9 & 2000 & 2000 \\
  $dr^{cbps}$ & 95.2 & 0.2 & 7 & 8 & 0.0 & 97 & 2 & 0.1 & 6 & 8 & 0.1 & 1 & 6 & -0.0 & 12 & 3 & 2000 & 2000 \\
  $reg$ & 94.8 & 0.2 & 5 & 6 & 0.0 & 596 & 5 & 0.1 & 5 & 6 & 0.1 & 1 & 4 & -0.0 & 43 & 7 & 2000 & 2000 \\
  $pairmatch^{probit}$ & 94.7 & 0.3 & 37 & 19 & 0.0 & 896 & 11 & 0.1 & 37 & 16 & 0.1 & 31 & 16 & -0.0 & 65 & 8 & 2000 & 2000 \\
  $tsls$ & 94.6 & 0.2 & 5 & 5 & 0.0 & 866 & 10 & 0.1 & 5 & 4 & 0.1 & 0 & 2 & -0.0 & 42 & 6 & 2000 & 2000 \\
  $radmatch^{cbps}$ & 93.7 & 0.3 & 33 & 16 & 0.0 & 689 & 7 & 0.1 & 39 & 17 & 0.1 & 33 & 17 & -0.0 & 413 & 12 & 2000 & 2000 \\
  $radmatch^{probit}$ & 93.4 & 0.3 & 35 & 17 & 0.0 & 1082 & 12 & 0.1 & 41 & 18 & 0.1 & 35 & 18 & -0.0 & 490 & 14 & 2000 & 2000 \\
  $reg^{kernel}$ & 93.3 & 0.2 & 2 & 3 & 0.0 & 2581 & 18 & 0.1 & 7 & 11 & 0.1 & 3 & 11 & -0.0 & 318 & 10 & 2000 & 2000 \\
  $radmatchx^{cbps}$ & 93.2 & 0.2 & 17 & 14 & 0.0 & 1623 & 16 & 0.1 & 23 & 13 & 0.1 & 18 & 13 & -0.0 & 448 & 13 & 2000 & 2000 \\
  $randforest$ & 93.2 & 0.2 & 1 & 2 & 0.0 & 236 & 4 & 0.1 & 6 & 9 & 0.1 & 1 & 7 & -0.0 & 341 & 11 & 2000 & 2000 \\
  $pairmatch^{x}$ & 92.8 & 0.2 & 15 & 13 & 0.0 & 0 & 1 & 0.1 & 23 & 12 & 0.1 & 17 & 12 & -0.0 & 547 & 15 & 2000 & 2000 \\
  $radmatchx^{probit}$ & 92.6 & 0.2 & 18 & 15 & 0.0 & 1293 & 13 & 0.1 & 26 & 14 & 0.1 & 20 & 14 & -0.0 & 572 & 16 & 2000 & 2000 \\
  $means$ & 92.1 & 0.2 & 0 & 1 & 0.0 & 7255 & 21 & 0.1 & 0 & 1 & 0.1 & 3 & 10 & -0.0 & 41 & 5 & 2000 & 2000 \\
  $ipw^{lc}$ & 100.0 & 1.6 & 662 & 20 & 0.0 & 3506 & 19 & 0.1 & 4 & 2 & 0.1 & 1 & 5 & 0.1 & 16478 & 20 & 2000 & 1638 \\
  $dr^{lc}$ & 100.0 & 8.7 & 4154 & 21 & 0.0 & 2158 & 17 & 0.1 & 6 & 7 & 0.1 & 2 & 9 & 0.8 & 88121 & 21 & 2000 & 1320 \\
  $pairmatch^{lc}$ & 66.7 & 0.2 & 13 & 11 & 0.0 & 794 & 8 & 0.1 & 165 & 20 & 0.1 & 153 & 20 & -0.1 & 9425 & 18 & 2000 & 2000 \\
  $radmatchx^{lc}$ & 66.0 & 0.2 & 8 & 10 & 0.0 & 3824 & 20 & 0.1 & 150 & 19 & 0.1 & 140 & 19 & -0.1 & 8782 & 17 & 2000 & 2000 \\
  $radmatch^{lc}$ & 61.7 & 0.2 & 13 & 12 & 0.0 & 1365 & 14 & 0.1 & 180 & 21 & 0.1 & 168 & 21 & -0.1 & 10363 & 19 & 2000 & 2000 \\
\bottomrule
\end{tabular}
\begin{tablenotes}
\item \scriptsize Notes: `cov' and `int' refer to coverage rate and length of confidence interval, `bias' denotes the absolute bias from the true treatment effect, `sd' is the standard
deviation of the estimator, `rmse' stands for root mean squared error, and `seb' indicates the median bias of the estimated bootstrap standard error. `diff' indicates the difference to the left best performer in percent (\%) and `r' is the rank. `nsimp' shows the number of simulations with point estimations smaller than $|10^{10}|$ and `nsimse' denotes the number of simulations used to compute `cov', `int', and `seb'. Conditional on `nsimp', simulations with standard errors greater than 150 times the (true) standard error are not considered (i.e.simulations with non-converging bootstrap standard errors are dropped).
\end{tablenotes}
\end{threeparttable}
\end{sidewaystable}

\begin{sidewaystable}[ht]
\centering
\begin{threeparttable}
\caption{DGP 12}
\label{DGP 12}
\small
\begin{tabular}{rrrrrrrrrrrrrrrrrrr}
\toprule
& cov & int & d & r & bias & d & r & sd & d & r & rmse & d & r & seb & d & r & nsimp & nsimse \\
\midrule
  $pairmatch^{lc}$ & 94.9 & 86.1 & 89 & 15 & 0.3 & 523 & 13 & 12.1 & 503 & 13 & 12.1 & 498 & 13 & -7.1 & 1544 & 14 & 2000 & 1993 \\
  $radmatch^{lc}$ & 93.9 & 121.2 & 166 & 20 & 0.3 & 720 & 15 & 15.4 & 669 & 16 & 15.4 & 663 & 16 & -10.5 & 2336 & 16 & 1999 & 1999 \\
  $radmatchx^{cbps}$ & 99.2 & 111.8 & 146 & 19 & 0.0 & 0 & 1 & 18.0 & 796 & 17 & 18.0 & 788 & 17 & -14.5 & 3265 & 18 & 1999 & 1998 \\
  $pairmatch^{probit}$ & 99.3 & 142.2 & 212 & 21 & 0.7 & 1569 & 20 & 19.6 & 878 & 18 & 19.6 & 871 & 18 & -14.4 & 3241 & 17 & 1999 & 1999 \\
  $pairmatch^{cbps}$ & 99.4 & 87.6 & 92 & 17 & 0.2 & 285 & 6 & 12.2 & 510 & 15 & 12.2 & 505 & 15 & -7.1 & 1541 & 13 & 2000 & 1995 \\
  $radmatchx^{probit}$ & 99.6 & 109.0 & 139 & 18 & 0.4 & 906 & 18 & 22.0 & 995 & 20 & 22.0 & 987 & 20 & -18.4 & 4168 & 20 & 1999 & 1998 \\
  $radmatch^{probit}$ & 99.6 & 75.5 & 66 & 10 & 0.1 & 233 & 5 & 6.3 & 213 & 7 & 6.3 & 210 & 7 & -1.8 & 321 & 5 & 2000 & 1992 \\
  $radmatch^{cbps}$ & 99.6 & 85.1 & 87 & 14 & 0.2 & 317 & 9 & 8.1 & 304 & 11 & 8.1 & 301 & 11 & -3.4 & 698 & 8 & 2000 & 1993 \\
  $ipw^{cbps}$ & 99.6 & 83.2 & 83 & 13 & 0.8 & 1847 & 21 & 22.3 & 1014 & 21 & 22.3 & 1006 & 21 & -18.9 & 4263 & 21 & 2000 & 1998 \\
  $pairmatch^{x}$ & 99.7 & 72.9 & 60 & 9 & 0.1 & 226 & 4 & 8.6 & 327 & 12 & 8.6 & 324 & 12 & -4.5 & 940 & 11 & 2000 & 1994 \\
  $randforest$ & 99.7 & 50.3 & 11 & 2 & 0.0 & 16 & 2 & 3.6 & 79 & 4 & 3.6 & 77 & 4 & -1.0 & 125 & 3 & 2000 & 1978 \\
  $reg^{kernel}$ & 99.8 & 69.6 & 53 & 7 & 0.3 & 543 & 14 & 8.0 & 298 & 10 & 8.0 & 295 & 10 & -4.9 & 1035 & 12 & 2000 & 1995 \\
  $ipw^{probit}$ & 99.8 & 78.0 & 71 & 11 & 0.2 & 316 & 8 & 5.7 & 182 & 5 & 5.7 & 180 & 5 & -2.4 & 449 & 6 & 1999 & 1998 \\
  $reg$ & 99.8 & 58.6 & 29 & 3 & 0.2 & 430 & 11 & 3.5 & 73 & 3 & 3.5 & 72 & 3 & -0.4 & 0 & 1 & 2000 & 1986 \\
  $ipw^{lc}$ & 99.8 & 82.3 & 81 & 12 & 0.3 & 736 & 16 & 6.1 & 205 & 6 & 6.1 & 203 & 6 & -0.5 & 8 & 2 & 2000 & 1990 \\
  $tsls$ & 99.8 & 71.5 & 57 & 8 & 0.0 & 18 & 3 & 6.9 & 244 & 8 & 6.9 & 241 & 8 & -3.5 & 711 & 9 & 2000 & 1994 \\
  $dr^{probit}$ & 99.8 & 86.6 & 90 & 16 & 0.4 & 956 & 19 & 12.2 & 506 & 14 & 12.2 & 502 & 14 & -9.0 & 1984 & 15 & 1998 & 1998 \\
  $means$ & 99.9 & 45.5 & -0 & 1 & 0.3 & 514 & 12 & 2.0 & 0 & 1 & 2.0 & 0 & 1 & 1.4 & 231 & 4 & 1991 & 1978 \\
  $dr^{cbps}$ & 99.9 & 66.3 & 46 & 5 & 0.2 & 305 & 7 & 7.1 & 253 & 9 & 7.1 & 250 & 9 & -3.9 & 805 & 10 & 2000 & 1990 \\
  $dr^{lc}$ & 99.9 & 59.1 & 30 & 4 & 0.2 & 339 & 10 & 2.1 & 7 & 2 & 2.1 & 6 & 2 & 3.0 & 584 & 7 & 2000 & 1981 \\
  $radmatchx^{lc}$ & 88.5 & 67.4 & 48 & 6 & 0.4 & 750 & 17 & 20.5 & 921 & 19 & 20.5 & 913 & 19 & -18.4 & 4154 & 19 & 2000 & 1997 \\
\bottomrule
\end{tabular}
\begin{tablenotes}
\item \scriptsize Notes: `cov' and `int' refer to coverage rate and length of confidence interval, `bias' denotes the absolute bias from the true treatment effect, `sd' is the standard
deviation of the estimator, `rmse' stands for root mean squared error, and `seb' indicates the median bias of the estimated bootstrap standard error. `diff' indicates the difference to the left best performer in percent (\%) and `r' is the rank. `nsimp' shows the number of simulations with point estimations smaller than $|10^{10}|$ and `nsimse' denotes the number of simulations used to compute `cov', `int', and `seb'. Conditional on `nsimp', simulations with standard errors greater than 150 times the (true) standard error are not considered (i.e.simulations with non-converging bootstrap standard errors are dropped).
\end{tablenotes}
\end{threeparttable}
\end{sidewaystable}

\begin{sidewaystable}[ht]
\centering
\begin{threeparttable}
\caption{DGP 13}
\label{DGP 13}
\small
\begin{tabular}{rrrrrrrrrrrrrrrrrrr}
\toprule
& cov & int & d & r & bias & d & r & sd & d & r & rmse & d & r & seb & d & r & nsimp & nsimse \\
\midrule
  $dr^{cbps}$ & 95.0 & 0.4 & 17 & 11 & 0.0 & 291 & 6 & 0.1 & 17 & 10 & 0.1 & 4 & 9 & -0.0 & 152 & 4 & 2000 & 2000 \\
  $dr^{probit}$ & 95.2 & 0.4 & 18 & 12 & 0.0 & 284 & 5 & 0.1 & 18 & 11 & 0.1 & 5 & 11 & -0.0 & 161 & 5 & 2000 & 2000 \\
  $ipw^{probit}$ & 94.8 & 0.4 & 14 & 9 & 0.0 & 418 & 11 & 0.1 & 15 & 7 & 0.1 & 3 & 7 & -0.0 & 185 & 6 & 2000 & 2000 \\
  $pairmatch^{probit}$ & 95.2 & 0.5 & 51 & 19 & 0.0 & 397 & 10 & 0.1 & 53 & 16 & 0.1 & 36 & 16 & -0.0 & 364 & 8 & 2000 & 2000 \\
  $pairmatch^{cbps}$ & 94.6 & 0.5 & 49 & 18 & 0.0 & 367 & 8 & 0.1 & 51 & 15 & 0.1 & 35 & 15 & -0.0 & 592 & 9 & 2000 & 2000 \\
  $reg$ & 94.6 & 0.4 & 15 & 10 & 0.0 & 254 & 4 & 0.1 & 15 & 8 & 0.1 & 2 & 4 & -0.0 & 0 & 1 & 2000 & 2000 \\
  $tsls$ & 94.4 & 0.4 & 13 & 5 & 0.0 & 220 & 2 & 0.1 & 12 & 3 & 0.1 & 0 & 1 & -0.0 & 22 & 3 & 2000 & 2000 \\
  $radmatch^{cbps}$ & 94.2 & 0.5 & 46 & 16 & 0.0 & 381 & 9 & 0.1 & 53 & 17 & 0.1 & 37 & 17 & -0.0 & 1398 & 13 & 2000 & 2000 \\
  $ipw^{cbps}$ & 94.2 & 0.4 & 13 & 6 & 0.0 & 545 & 15 & 0.1 & 14 & 5 & 0.1 & 3 & 6 & -0.0 & 245 & 7 & 2000 & 2000 \\
  $radmatch^{probit}$ & 94.0 & 0.5 & 48 & 17 & 0.0 & 427 & 12 & 0.1 & 55 & 18 & 0.1 & 39 & 18 & -0.0 & 1386 & 12 & 2000 & 2000 \\
  $randforest$ & 93.4 & 0.4 & 9 & 3 & 0.0 & 355 & 7 & 0.1 & 15 & 9 & 0.1 & 3 & 5 & -0.0 & 1038 & 11 & 2000 & 2000 \\
  $radmatchx^{probit}$ & 93.1 & 0.4 & 30 & 15 & 0.0 & 529 & 14 & 0.1 & 42 & 14 & 0.1 & 27 & 14 & -0.0 & 2107 & 15 & 2000 & 2000 \\
  $radmatchx^{cbps}$ & 92.8 & 0.4 & 29 & 13 & 0.0 & 575 & 16 & 0.1 & 39 & 12 & 0.1 & 25 & 13 & -0.0 & 1864 & 14 & 2000 & 2000 \\
  $pairmatch^{x}$ & 92.8 & 0.4 & 29 & 14 & 0.0 & 222 & 3 & 0.1 & 41 & 13 & 0.1 & 25 & 12 & -0.0 & 2150 & 16 & 2000 & 2000 \\
  $reg^{kernel}$ & 92.0 & 0.4 & 9 & 2 & 0.0 & 760 & 18 & 0.1 & 13 & 4 & 0.1 & 4 & 8 & -0.0 & 749 & 10 & 2000 & 2000 \\
  $means$ & 91.2 & 0.3 & -0 & 1 & 0.0 & 1283 & 21 & 0.1 & 0 & 1 & 0.1 & 1 & 3 & -0.0 & 19 & 2 & 2000 & 2000 \\
  $dr^{lc}$ & 99.9 & 18.4 & 5334 & 21 & 0.0 & 650 & 17 & 0.1 & 15 & 6 & 0.1 & 5 & 10 & 3.8 & 739653 & 21 & 2000 & 1396 \\
  $ipw^{lc}$ & 99.9 & 3.3 & 884 & 20 & 0.0 & 812 & 20 & 0.1 & 8 & 2 & 0.1 & 1 & 2 & 0.3 & 50533 & 20 & 2000 & 1599 \\
  $radmatchx^{lc}$ & 63.0 & 0.4 & 12 & 4 & 0.0 & 0 & 1 & 0.2 & 172 & 19 & 0.2 & 140 & 19 & -0.1 & 27026 & 17 & 2000 & 2000 \\
  $pairmatch^{lc}$ & 61.8 & 0.4 & 13 & 7 & 0.0 & 495 & 13 & 0.3 & 208 & 20 & 0.3 & 173 & 20 & -0.2 & 32883 & 18 & 2000 & 2000 \\
  $radmatch^{lc}$ & 57.3 & 0.4 & 13 & 8 & 0.0 & 766 & 19 & 0.3 & 208 & 21 & 0.3 & 174 & 21 & -0.2 & 32935 & 19 & 2000 & 2000 \\
\bottomrule
\end{tabular}
\begin{tablenotes}
\item \scriptsize Notes: `cov' and `int' refer to coverage rate and length of confidence interval, `bias' denotes the absolute bias from the true treatment effect, `sd' is the standard
deviation of the estimator, `rmse' stands for root mean squared error, and `seb' indicates the median bias of the estimated bootstrap standard error. `diff' indicates the difference to the left best performer in percent (\%) and `r' is the rank. `nsimp' shows the number of simulations with point estimations smaller than $|10^{10}|$ and `nsimse' denotes the number of simulations used to compute `cov', `int', and `seb'. Conditional on `nsimp', simulations with standard errors greater than 150 times the (true) standard error are not considered (i.e.simulations with non-converging bootstrap standard errors are dropped).
\end{tablenotes}
\end{threeparttable}
\end{sidewaystable}

\begin{sidewaystable}[ht]
\centering
\begin{threeparttable}
\caption{DGP 14}
\label{DGP 14}
\small
\begin{tabular}{rrrrrrrrrrrrrrrrrrr}
\toprule
& cov & int & d & r & bias & d & r & sd & d & r & rmse & d & r & seb & d & r & nsimp & nsimse \\
\midrule
  $pairmatch^{probit}$ & 95.0 & 0.3 & 21 & 18 & 0.0 & 1648 & 3 & 0.1 & 16 & 17 & 0.1 & 16 & 17 & -0.0 & 0 & 1 & 2000 & 2000 \\
  $reg$ & 94.2 & 0.2 & 3 & 3 & 0.0 & 3381 & 7 & 0.1 & 0 & 1 & 0.1 & 0 & 1 & -0.0 & 1345 & 7 & 2000 & 2000 \\
  $tsls$ & 94.2 & 0.2 & 3 & 4 & 0.0 & 3875 & 11 & 0.1 & 0 & 3 & 0.1 & 0 & 4 & -0.0 & 1239 & 4 & 2000 & 2000 \\
  $ipw^{probit}$ & 94.1 & 0.2 & 3 & 7 & 0.0 & 3813 & 10 & 0.1 & 0 & 5 & 0.1 & 0 & 5 & -0.0 & 1308 & 6 & 2000 & 2000 \\
  $dr^{probit}$ & 94.1 & 0.2 & 3 & 6 & 0.0 & 3552 & 8 & 0.1 & 0 & 4 & 0.1 & 0 & 3 & -0.0 & 1230 & 3 & 2000 & 2000 \\
  $dr^{cbps}$ & 94.1 & 0.2 & 3 & 5 & 0.0 & 3588 & 9 & 0.1 & 0 & 2 & 0.1 & 0 & 2 & -0.0 & 1290 & 5 & 2000 & 2000 \\
  $radmatch^{probit}$ & 94.0 & 0.2 & 16 & 15 & 0.0 & 2266 & 4 & 0.1 & 14 & 14 & 0.1 & 14 & 14 & -0.0 & 4239 & 11 & 2000 & 2000 \\
  $ipw^{cbps}$ & 94.0 & 0.2 & 4 & 8 & 0.0 & 5798 & 15 & 0.1 & 1 & 7 & 0.1 & 1 & 6 & -0.0 & 1487 & 8 & 2000 & 2000 \\
  $pairmatch^{cbps}$ & 94.0 & 0.3 & 21 & 19 & 0.0 & 4207 & 13 & 0.1 & 17 & 18 & 0.1 & 17 & 18 & -0.0 & 776 & 2 & 2000 & 2000 \\
  $ipw^{lc}$ & 96.3 & 0.3 & 55 & 20 & 0.0 & 13712 & 20 & 0.1 & 1 & 9 & 0.1 & 3 & 9 & 0.0 & 13095 & 18 & 2000 & 1991 \\
  $reg^{kernel}$ & 93.6 & 0.2 & 4 & 9 & 0.0 & 11231 & 19 & 0.1 & 2 & 10 & 0.1 & 4 & 10 & -0.0 & 2707 & 10 & 2000 & 2000 \\
  $dr^{lc}$ & 96.4 & 0.4 & 108 & 21 & 0.0 & 9377 & 17 & 0.1 & 1 & 8 & 0.1 & 2 & 7 & 0.0 & 13022 & 17 & 2000 & 1985 \\
  $radmatchx^{cbps}$ & 93.6 & 0.2 & 10 & 10 & 0.0 & 2873 & 5 & 0.1 & 12 & 12 & 0.1 & 12 & 12 & -0.0 & 7717 & 13 & 2000 & 2000 \\
  $radmatch^{cbps}$ & 93.4 & 0.2 & 16 & 17 & 0.0 & 4109 & 12 & 0.1 & 15 & 15 & 0.1 & 15 & 15 & -0.0 & 4502 & 12 & 2000 & 2000 \\
  $radmatchx^{probit}$ & 93.0 & 0.2 & 10 & 11 & 0.0 & 1022 & 2 & 0.1 & 12 & 13 & 0.1 & 12 & 13 & -0.0 & 7760 & 14 & 2000 & 2000 \\
  $pairmatch^{x}$ & 92.6 & 0.2 & 11 & 13 & 0.0 & 3014 & 6 & 0.1 & 16 & 16 & 0.1 & 16 & 16 & -0.0 & 11099 & 16 & 2000 & 2000 \\
  $means$ & 92.5 & 0.2 & 3 & 2 & 0.0 & 21512 & 21 & 0.1 & 0 & 6 & 0.1 & 6 & 11 & -0.0 & 2007 & 9 & 2000 & 2000 \\
  $randforest$ & 92.4 & 0.2 & 0 & 1 & 0.0 & 4516 & 14 & 0.1 & 3 & 11 & 0.1 & 3 & 8 & -0.0 & 8640 & 15 & 2000 & 2000 \\
  $pairmatch^{lc}$ & 79.2 & 0.2 & 15 & 14 & 0.0 & 7806 & 16 & 0.1 & 99 & 19 & 0.1 & 99 & 19 & -0.0 & 105015 & 19 & 2000 & 2000 \\
  $radmatchx^{lc}$ & 74.6 & 0.2 & 10 & 12 & 0.0 & 0 & 1 & 0.1 & 106 & 20 & 0.1 & 105 & 20 & -0.1 & 118774 & 20 & 2000 & 2000 \\
  $radmatch^{lc}$ & 72.3 & 0.2 & 16 & 16 & 0.0 & 10211 & 18 & 0.1 & 129 & 21 & 0.1 & 129 & 21 & -0.1 & 140473 & 21 & 2000 & 2000 \\
\bottomrule
\end{tabular}
\begin{tablenotes}
\item \scriptsize Notes: `cov' and `int' refer to coverage rate and length of confidence interval, `bias' denotes the absolute bias from the true treatment effect, `sd' is the standard
deviation of the estimator, `rmse' stands for root mean squared error, and `seb' indicates the median bias of the estimated bootstrap standard error. `diff' indicates the difference to the left best performer in percent (\%) and `r' is the rank. `nsimp' shows the number of simulations with point estimations smaller than $|10^{10}|$ and `nsimse' denotes the number of simulations used to compute `cov', `int', and `seb'. Conditional on `nsimp', simulations with standard errors greater than 150 times the (true) standard error are not considered (i.e.simulations with non-converging bootstrap standard errors are dropped).
\end{tablenotes}
\end{threeparttable}
\end{sidewaystable}

\begin{sidewaystable}[ht]
\centering
\begin{threeparttable}
\caption{DGP 15}
\label{DGP 15}
\small
\begin{tabular}{rrrrrrrrrrrrrrrrrrr}
\toprule
& cov & int & d & r & bias & d & r & sd & d & r & rmse & d & r & seb & d & r & nsimp & nsimse \\
\midrule
  $pairmatch^{lc}$ & 92.9 & 100.1 & 191 & 16 & 0.3 & 20802 & 16 & 14.6 & 629 & 16 & 14.6 & 629 & 16 & -10.0 & 325332 & 17 & 1999 & 1997 \\
  $radmatch^{lc}$ & 91.4 & 113.6 & 230 & 18 & 0.2 & 10325 & 12 & 17.0 & 753 & 19 & 17.0 & 752 & 19 & -12.4 & 405403 & 19 & 1999 & 1999 \\
  $pairmatch^{probit}$ & 99.0 & 97.9 & 184 & 15 & 0.3 & 19871 & 15 & 13.8 & 589 & 15 & 13.8 & 590 & 15 & -7.6 & 249521 & 14 & 2000 & 1993 \\
  $pairmatch^{cbps}$ & 99.3 & 139.9 & 307 & 20 & 0.4 & 27850 & 20 & 14.7 & 636 & 17 & 14.7 & 636 & 17 & -8.9 & 290960 & 15 & 1999 & 1998 \\
  $radmatch^{cbps}$ & 99.5 & 125.2 & 264 & 19 & 0.6 & 39671 & 21 & 23.5 & 1078 & 21 & 23.5 & 1079 & 21 & -17.4 & 569058 & 21 & 1997 & 1995 \\
  $radmatch^{probit}$ & 99.5 & 91.0 & 165 & 13 & 0.4 & 22711 & 17 & 15.0 & 650 & 18 & 15.0 & 650 & 18 & -9.0 & 293805 & 16 & 2000 & 1997 \\
  $radmatchx^{probit}$ & 99.5 & 47.5 & 38 & 3 & 0.1 & 7527 & 9 & 4.7 & 136 & 9 & 4.7 & 136 & 9 & -1.6 & 53564 & 8 & 2000 & 1991 \\
  $randforest$ & 99.5 & 45.8 & 33 & 2 & 0.1 & 9290 & 11 & 2.7 & 37 & 3 & 2.7 & 37 & 3 & -0.0 & 0 & 1 & 2000 & 1991 \\
  $pairmatch^{x}$ & 99.6 & 63.3 & 84 & 10 & 0.0 & 1553 & 3 & 4.0 & 100 & 6 & 4.0 & 100 & 6 & 0.6 & 19956 & 4 & 2000 & 1994 \\
  $ipw^{lc}$ & 99.6 & 110.6 & 221 & 17 & 0.0 & 1310 & 2 & 4.6 & 129 & 8 & 4.6 & 129 & 8 & 6.9 & 226034 & 13 & 2000 & 1983 \\
  $ipw^{probit}$ & 99.7 & 47.8 & 39 & 4 & 0.0 & 2383 & 5 & 2.4 & 22 & 2 & 2.4 & 22 & 2 & 0.3 & 8624 & 3 & 2000 & 1981 \\
  $dr^{cbps}$ & 99.7 & 56.4 & 64 & 7 & 0.1 & 4617 & 6 & 4.0 & 98 & 5 & 4.0 & 98 & 5 & -0.9 & 29663 & 6 & 2000 & 1993 \\
  $reg$ & 99.7 & 62.3 & 81 & 9 & 0.1 & 4996 & 7 & 6.3 & 213 & 12 & 6.3 & 213 & 12 & -3.3 & 106430 & 12 & 2000 & 1998 \\
  $radmatchx^{cbps}$ & 99.7 & 54.7 & 59 & 5 & 0.1 & 8887 & 10 & 4.1 & 106 & 7 & 4.1 & 106 & 7 & -1.1 & 35861 & 7 & 2000 & 1987 \\
  $tsls$ & 99.7 & 84.2 & 145 & 12 & 0.2 & 13764 & 13 & 19.7 & 886 & 20 & 19.7 & 886 & 20 & -16.7 & 545671 & 20 & 2000 & 1999 \\
  $dr^{probit}$ & 99.8 & 55.8 & 62 & 6 & 0.0 & 0 & 1 & 3.2 & 59 & 4 & 3.2 & 59 & 4 & -0.3 & 8076 & 2 & 2000 & 1977 \\
  $dr^{lc}$ & 99.8 & 166.7 & 385 & 21 & 0.3 & 18617 & 14 & 11.3 & 467 & 13 & 11.3 & 467 & 13 & 2.2 & 72836 & 10 & 2000 & 1987 \\
  $means$ & 99.8 & 60.7 & 77 & 8 & 0.4 & 27114 & 19 & 5.0 & 152 & 10 & 5.0 & 152 & 10 & -1.8 & 60077 & 9 & 1998 & 1989 \\
  $reg^{kernel}$ & 99.8 & 95.4 & 177 & 14 & 0.4 & 23661 & 18 & 5.4 & 168 & 11 & 5.4 & 169 & 11 & -2.7 & 88938 & 11 & 1999 & 1997 \\
  $ipw^{cbps}$ & 99.8 & 66.0 & 92 & 11 & 0.0 & 2232 & 4 & 2.0 & 0 & 1 & 2.0 & 0 & 1 & 0.7 & 23230 & 5 & 1999 & 1998 \\
  $radmatchx^{lc}$ & 76.4 & 34.4 & 0 & 1 & 0.1 & 5385 & 8 & 12.8 & 540 & 14 & 12.8 & 540 & 14 & -11.9 & 387831 & 18 & 2000 & 1999 \\
\bottomrule
\end{tabular}
\begin{tablenotes}
\item \scriptsize Notes: `cov' and `int' refer to coverage rate and length of confidence interval, `bias' denotes the absolute bias from the true treatment effect, `sd' is the standard
deviation of the estimator, `rmse' stands for root mean squared error, and `seb' indicates the median bias of the estimated bootstrap standard error. `diff' indicates the difference to the left best performer in percent (\%) and `r' is the rank. `nsimp' shows the number of simulations with point estimations smaller than $|10^{10}|$ and `nsimse' denotes the number of simulations used to compute `cov', `int', and `seb'. Conditional on `nsimp', simulations with standard errors greater than 150 times the (true) standard error are not considered (i.e.simulations with non-converging bootstrap standard errors are dropped).
\end{tablenotes}
\end{threeparttable}
\end{sidewaystable}

\begin{sidewaystable}[ht]
\centering
\begin{threeparttable}
\caption{DGP 16}
\label{DGP 16}
\small
\begin{tabular}{rrrrrrrrrrrrrrrrrrr}
\toprule
& cov & int & d & r & bias & d & r & sd & d & r & rmse & d & r & seb & d & r & nsimp & nsimse \\
\midrule
  $tsls$ & 94.9 & 0.2 & 6 & 5 & 0.0 & 85 & 2 & 0.1 & 5 & 3 & 0.1 & 0 & 1 & 0.0 & 116 & 2 & 2000 & 2000 \\
  $reg$ & 94.8 & 0.2 & 6 & 6 & 0.0 & 87 & 3 & 0.1 & 7 & 6 & 0.1 & 1 & 2 & -0.0 & 240 & 3 & 2000 & 2000 \\
  $pairmatch^{probit}$ & 94.8 & 0.3 & 38 & 19 & 0.0 & 314 & 11 & 0.1 & 41 & 17 & 0.1 & 35 & 16 & -0.0 & 2461 & 8 & 2000 & 2000 \\
  $ipw^{probit}$ & 94.6 & 0.2 & 7 & 7 & 0.0 & 239 & 9 & 0.1 & 7 & 8 & 0.1 & 2 & 3 & -0.0 & 435 & 4 & 2000 & 2000 \\
  $dr^{probit}$ & 94.5 & 0.2 & 8 & 9 & 0.0 & 127 & 5 & 0.1 & 9 & 11 & 0.1 & 3 & 8 & -0.0 & 731 & 6 & 2000 & 2000 \\
  $pairmatch^{cbps}$ & 94.4 & 0.3 & 37 & 18 & 0.0 & 313 & 10 & 0.1 & 40 & 15 & 0.1 & 34 & 15 & -0.0 & 2971 & 9 & 2000 & 2000 \\
  $dr^{cbps}$ & 94.4 & 0.2 & 8 & 8 & 0.0 & 129 & 6 & 0.1 & 8 & 10 & 0.1 & 3 & 5 & -0.0 & 856 & 7 & 2000 & 2000 \\
  $ipw^{cbps}$ & 94.2 & 0.2 & 6 & 4 & 0.0 & 346 & 14 & 0.1 & 6 & 4 & 0.1 & 3 & 4 & -0.0 & 597 & 5 & 2000 & 2000 \\
  $radmatch^{probit}$ & 93.8 & 0.3 & 36 & 17 & 0.0 & 340 & 13 & 0.1 & 42 & 18 & 0.1 & 36 & 18 & -0.0 & 5914 & 13 & 2000 & 2000 \\
  $radmatch^{cbps}$ & 93.8 & 0.3 & 34 & 16 & 0.0 & 331 & 12 & 0.1 & 40 & 16 & 0.1 & 35 & 17 & -0.0 & 5903 & 12 & 2000 & 2000 \\
  $randforest$ & 93.7 & 0.2 & 3 & 3 & 0.0 & 214 & 7 & 0.1 & 7 & 9 & 0.1 & 3 & 6 & -0.0 & 4083 & 11 & 2000 & 2000 \\
  $reg^{kernel}$ & 93.6 & 0.2 & 3 & 2 & 0.0 & 499 & 19 & 0.1 & 7 & 7 & 0.1 & 5 & 10 & -0.0 & 3661 & 10 & 2000 & 2000 \\
  $radmatchx^{cbps}$ & 93.2 & 0.3 & 18 & 14 & 0.0 & 393 & 16 & 0.1 & 25 & 12 & 0.1 & 20 & 13 & -0.0 & 6891 & 14 & 2000 & 2000 \\
  $pairmatch^{x}$ & 92.8 & 0.2 & 17 & 13 & 0.0 & 112 & 4 & 0.1 & 26 & 13 & 0.1 & 20 & 12 & -0.0 & 8901 & 15 & 2000 & 2000 \\
  $radmatchx^{probit}$ & 92.1 & 0.3 & 18 & 15 & 0.0 & 365 & 15 & 0.1 & 28 & 14 & 0.1 & 24 & 14 & -0.0 & 9590 & 16 & 2000 & 2000 \\
  $means$ & 91.4 & 0.2 & 0 & 1 & 0.0 & 954 & 21 & 0.1 & 0 & 1 & 0.1 & 7 & 11 & -0.0 & 0 & 1 & 2000 & 2000 \\
  $ipw^{lc}$ & 99.9 & 1.6 & 629 & 20 & 0.0 & 558 & 20 & 0.1 & 3 & 2 & 0.1 & 3 & 7 & 0.1 & 237479 & 20 & 2000 & 1642 \\
  $dr^{lc}$ & 100.0 & 8.9 & 4050 & 21 & 0.0 & 422 & 18 & 0.1 & 6 & 5 & 0.1 & 4 & 9 & 0.8 & 1258701 & 21 & 2000 & 1338 \\
  $pairmatch^{lc}$ & 68.6 & 0.2 & 13 & 11 & 0.0 & 221 & 8 & 0.1 & 162 & 20 & 0.1 & 149 & 20 & -0.1 & 136363 & 18 & 2000 & 2000 \\
  $radmatchx^{lc}$ & 67.0 & 0.2 & 8 & 10 & 0.0 & 0 & 1 & 0.1 & 149 & 19 & 0.1 & 136 & 19 & -0.1 & 129025 & 17 & 2000 & 2000 \\
  $radmatch^{lc}$ & 62.8 & 0.2 & 13 & 12 & 0.0 & 396 & 17 & 0.2 & 183 & 21 & 0.2 & 169 & 21 & -0.1 & 155289 & 19 & 2000 & 2000 \\
\bottomrule
\end{tabular}
\begin{tablenotes}
\item \scriptsize Notes: `cov' and `int' refer to coverage rate and length of confidence interval, `bias' denotes the absolute bias from the true treatment effect, `sd' is the standard
deviation of the estimator, `rmse' stands for root mean squared error, and `seb' indicates the median bias of the estimated bootstrap standard error. `diff' indicates the difference to the left best performer in percent (\%) and `r' is the rank. `nsimp' shows the number of simulations with point estimations smaller than $|10^{10}|$ and `nsimse' denotes the number of simulations used to compute `cov', `int', and `seb'. Conditional on `nsimp', simulations with standard errors greater than 150 times the (true) standard error are not considered (i.e.simulations with non-converging bootstrap standard errors are dropped).
\end{tablenotes}
\end{threeparttable}
\end{sidewaystable}

\begin{sidewaystable}[ht]
\centering
\begin{threeparttable}
\caption{DGP 17}
\label{DGP 17}
\small
\begin{tabular}{rrrrrrrrrrrrrrrrrrr}
\toprule
& cov & int & d & r & bias & d & r & sd & d & r & rmse & d & r & seb & d & r & nsimp & nsimse \\
\midrule
  $radmatch^{cbps}$ & 94.7 & 7.4 & 17 & 17 & 0.5 & 50 & 12 & 1.9 & 14 & 13 & 1.9 & 14 & 10 & 0.0 & 871 & 3 & 1000 & 1000 \\
  $reg$ & 94.4 & 6.4 & 2 & 3 & 0.4 & 33 & 7 & 1.6 & 0 & 1 & 1.7 & 0 & 3 & -0.0 & 994 & 4 & 1000 & 1000 \\
  $ipw^{probit}$ & 94.3 & 6.4 & 2 & 6 & 0.4 & 31 & 6 & 1.6 & 0 & 4 & 1.7 & 0 & 2 & -0.0 & 1156 & 6 & 1000 & 1000 \\
  $dr^{cbps}$ & 94.3 & 6.4 & 2 & 4 & 0.4 & 36 & 8 & 1.6 & 0 & 2 & 1.7 & 0 & 4 & -0.0 & 1198 & 8 & 1000 & 1000 \\
  $pairmatch^{cbps}$ & 95.7 & 7.6 & 20 & 19 & 0.5 & 59 & 15 & 1.9 & 14 & 12 & 1.9 & 14 & 11 & 0.1 & 12317 & 11 & 1000 & 1000 \\
  $tsls$ & 94.3 & 6.4 & 2 & 2 & 0.4 & 30 & 5 & 1.6 & 0 & 5 & 1.7 & 0 & 1 & -0.0 & 1679 & 9 & 1000 & 1000 \\
  $dr^{probit}$ & 94.2 & 6.4 & 2 & 5 & 0.4 & 36 & 9 & 1.6 & 0 & 3 & 1.7 & 0 & 5 & -0.0 & 1107 & 5 & 1000 & 1000 \\
  $ipw^{cbps}$ & 94.1 & 6.5 & 3 & 8 & 0.5 & 50 & 13 & 1.7 & 1 & 7 & 1.7 & 2 & 7 & -0.0 & 1183 & 7 & 1000 & 1000 \\
  $pairmatch^{probit}$ & 93.8 & 7.5 & 19 & 18 & 0.5 & 53 & 14 & 1.9 & 17 & 15 & 2.0 & 17 & 15 & -0.0 & 0 & 1 & 1000 & 1000 \\
  $radmatch^{probit}$ & 93.7 & 7.3 & 16 & 15 & 0.5 & 41 & 10 & 1.9 & 18 & 16 & 2.0 & 17 & 16 & -0.1 & 12426 & 12 & 1000 & 1000 \\
  $reg^{kernel}$ & 93.5 & 6.5 & 3 & 7 & 0.7 & 108 & 17 & 1.7 & 5 & 11 & 1.8 & 9 & 9 & -0.1 & 11435 & 10 & 1000 & 991 \\
  $randforest$ & 92.9 & 6.3 & 0 & 1 & 0.3 & 0 & 1 & 1.7 & 3 & 10 & 1.7 & 1 & 6 & -0.1 & 14343 & 13 & 1000 & 1000 \\
  $pairmatch^{x}$ & 92.6 & 6.9 & 10 & 10 & 0.4 & 18 & 4 & 1.9 & 17 & 14 & 1.9 & 15 & 12 & -0.1 & 26318 & 15 & 1000 & 1000 \\
  $radmatchx^{cbps}$ & 92.4 & 7.1 & 12 & 12 & 0.3 & 7 & 3 & 1.9 & 18 & 17 & 2.0 & 16 & 14 & -0.1 & 25397 & 14 & 1000 & 1000 \\
  $ipw^{lc}$ & 97.8 & 12.0 & 91 & 20 & 1.0 & 215 & 19 & 1.7 & 2 & 8 & 2.0 & 16 & 13 & 0.8 & 134704 & 17 & 1000 & 998 \\
  $radmatchx^{probit}$ & 92.2 & 7.0 & 12 & 11 & 0.3 & 1 & 2 & 2.0 & 20 & 18 & 2.0 & 18 & 17 & -0.2 & 31170 & 16 & 1000 & 1000 \\
  $dr^{lc}$ & 98.9 & 20.3 & 223 & 21 & 0.7 & 104 & 16 & 1.7 & 2 & 9 & 1.8 & 6 & 8 & 1.0 & 186122 & 18 & 1000 & 987 \\
  $means$ & 84.7 & 6.5 & 3 & 9 & 1.5 & 375 & 21 & 1.7 & 1 & 6 & 2.3 & 34 & 18 & 0.0 & 187 & 2 & 1000 & 1000 \\
  $radmatchx^{lc}$ & 56.8 & 7.1 & 13 & 13 & 0.5 & 49 & 11 & 4.6 & 179 & 19 & 4.6 & 172 & 19 & -2.8 & 491729 & 19 & 1000 & 1000 \\
  $pairmatch^{lc}$ & 54.4 & 7.2 & 14 & 14 & 1.0 & 193 & 18 & 4.8 & 191 & 20 & 4.9 & 187 & 20 & -3.0 & 525920 & 20 & 1000 & 1000 \\
  $radmatch^{lc}$ & 49.2 & 7.3 & 16 & 16 & 1.2 & 260 & 20 & 5.5 & 235 & 21 & 5.6 & 231 & 21 & -3.6 & 645246 & 21 & 1000 & 1000 \\
\bottomrule
\end{tabular}
\begin{tablenotes}
\item \scriptsize Notes: `cov' and `int' refer to coverage rate and length of confidence interval, `bias' denotes the absolute bias from the true treatment effect, `sd' is the standard
deviation of the estimator, `rmse' stands for root mean squared error, and `seb' indicates the median bias of the estimated bootstrap standard error. `diff' indicates the difference to the left best performer in percent (\%) and `r' is the rank. `nsimp' shows the number of simulations with point estimations smaller than $|10^{10}|$ and `nsimse' denotes the number of simulations used to compute `cov', `int', and `seb'. Conditional on `nsimp', simulations with standard errors greater than 150 times the (true) standard error are not considered (i.e.simulations with non-converging bootstrap standard errors are dropped).
\end{tablenotes}
\end{threeparttable}
\end{sidewaystable}

\begin{sidewaystable}[ht]
\centering
\begin{threeparttable}
\caption{DGP 18}
\label{DGP 18}
\small
\begin{tabular}{rrrrrrrrrrrrrrrrrrr}
\toprule
& cov & int & d & r & bias & d & r & sd & d & r & rmse & d & r & seb & d & r & nsimp & nsimse \\
\midrule
  $ipw^{cbps}$ & 95.1 & 11.1 & 2 & 8 & 0.4 & 232 & 15 & 2.8 & 1 & 7 & 2.8 & 1 & 6 & 0.0 & 191 & 4 & 1000 & 1000 \\
  $reg$ & 95.2 & 11.0 & 1 & 5 & 0.3 & 162 & 10 & 2.8 & 1 & 6 & 2.8 & 1 & 5 & -0.0 & 0 & 1 & 1000 & 1000 \\
  $dr^{probit}$ & 95.3 & 11.0 & 1 & 7 & 0.3 & 173 & 11 & 2.8 & 0 & 4 & 2.8 & 0 & 3 & 0.0 & 285 & 5 & 1000 & 1000 \\
  $dr^{cbps}$ & 95.3 & 11.0 & 1 & 6 & 0.3 & 177 & 13 & 2.8 & 0 & 5 & 2.8 & 0 & 4 & 0.0 & 158 & 3 & 1000 & 1000 \\
  $tsls$ & 95.4 & 11.0 & 1 & 3 & 0.3 & 136 & 7 & 2.8 & 0 & 2 & 2.8 & 0 & 1 & -0.0 & 13 & 2 & 1000 & 1000 \\
  $radmatch^{probit}$ & 94.6 & 12.5 & 15 & 16 & 0.2 & 99 & 5 & 3.3 & 20 & 17 & 3.3 & 19 & 16 & -0.2 & 8491 & 11 & 1000 & 1000 \\
  $ipw^{probit}$ & 95.5 & 11.0 & 1 & 4 & 0.3 & 142 & 8 & 2.8 & 0 & 3 & 2.8 & 0 & 2 & 0.0 & 378 & 6 & 1000 & 1000 \\
  $pairmatch^{probit}$ & 95.5 & 12.9 & 19 & 18 & 0.2 & 122 & 6 & 3.2 & 16 & 13 & 3.3 & 16 & 12 & 0.0 & 2492 & 8 & 1000 & 1000 \\
  $pairmatch^{cbps}$ & 95.5 & 13.0 & 19 & 19 & 0.3 & 151 & 9 & 3.2 & 16 & 11 & 3.2 & 16 & 10 & 0.1 & 3937 & 10 & 1000 & 1000 \\
  $radmatch^{cbps}$ & 94.2 & 12.6 & 15 & 17 & 0.3 & 175 & 12 & 3.2 & 16 & 12 & 3.3 & 16 & 13 & -0.1 & 2772 & 9 & 1000 & 1000 \\
  $radmatchx^{cbps}$ & 93.9 & 12.1 & 11 & 14 & 0.1 & 15 & 2 & 3.3 & 17 & 14 & 3.3 & 16 & 11 & -0.2 & 9307 & 12 & 1000 & 1000 \\
  $randforest$ & 93.8 & 10.9 & 0 & 2 & 0.2 & 78 & 4 & 2.9 & 5 & 10 & 2.9 & 5 & 7 & -0.2 & 9496 & 13 & 1000 & 1000 \\
  $reg^{kernel}$ & 93.5 & 11.1 & 2 & 9 & 0.8 & 573 & 19 & 3.6 & 29 & 18 & 3.7 & 31 & 18 & -0.8 & 42791 & 16 & 1000 & 987 \\
  $radmatchx^{probit}$ & 93.0 & 12.1 & 11 & 13 & 0.1 & 0 & 1 & 3.3 & 18 & 15 & 3.3 & 18 & 14 & -0.2 & 12265 & 14 & 1000 & 1000 \\
  $pairmatch^{x}$ & 92.5 & 12.0 & 11 & 12 & 0.2 & 59 & 3 & 3.3 & 19 & 16 & 3.3 & 19 & 15 & -0.3 & 14465 & 15 & 1000 & 1000 \\
  $ipw^{lc}$ & 98.9 & 20.4 & 88 & 20 & 1.3 & 1018 & 20 & 2.9 & 2 & 8 & 3.1 & 11 & 9 & 1.2 & 68686 & 17 & 1000 & 998 \\
  $dr^{lc}$ & 99.1 & 34.6 & 218 & 21 & 0.7 & 531 & 18 & 2.9 & 5 & 9 & 3.0 & 7 & 8 & 1.7 & 96375 & 18 & 1000 & 980 \\
  $means$ & 88.5 & 10.9 & 0 & 1 & 2.1 & 1773 & 21 & 2.8 & 0 & 1 & 3.5 & 24 & 17 & -0.0 & 1045 & 7 & 1000 & 1000 \\
  $radmatchx^{lc}$ & 60.1 & 12.0 & 11 & 11 & 0.4 & 219 & 14 & 7.9 & 184 & 19 & 7.9 & 183 & 19 & -4.9 & 270863 & 19 & 1000 & 1000 \\
  $pairmatch^{lc}$ & 57.7 & 12.0 & 10 & 10 & 0.4 & 248 & 16 & 8.4 & 203 & 20 & 8.4 & 202 & 20 & -5.4 & 300684 & 20 & 1000 & 1000 \\
  $radmatch^{lc}$ & 53.7 & 12.3 & 13 & 15 & 0.5 & 376 & 17 & 9.5 & 239 & 21 & 9.5 & 238 & 21 & -6.3 & 353212 & 21 & 1000 & 1000 \\
\bottomrule
\end{tabular}
\begin{tablenotes}
\item \scriptsize Notes: `cov' and `int' refer to coverage rate and length of confidence interval, `bias' denotes the absolute bias from the true treatment effect, `sd' is the standard
deviation of the estimator, `rmse' stands for root mean squared error, and `seb' indicates the median bias of the estimated bootstrap standard error. `diff' indicates the difference to the left best performer in percent (\%) and `r' is the rank. `nsimp' shows the number of simulations with point estimations smaller than $|10^{10}|$ and `nsimse' denotes the number of simulations used to compute `cov', `int', and `seb'. Conditional on `nsimp', simulations with standard errors greater than 150 times the (true) standard error are not considered (i.e.simulations with non-converging bootstrap standard errors are dropped).
\end{tablenotes}
\end{threeparttable}
\end{sidewaystable}

\begin{sidewaystable}[ht]
\centering
\begin{threeparttable}
\caption{DGP 19}
\label{DGP 19}
\small
\begin{tabular}{rrrrrrrrrrrrrrrrrrr}
\toprule
& cov & int & d & r & bias & d & r & sd & d & r & rmse & d & r & seb & d & r & nsimp & nsimse \\
\midrule
  $reg$ & 95.0 & 6.8 & 5 & 7 & 0.2 & 32 & 2 & 1.7 & 6 & 6 & 1.8 & 2 & 2 & -0.0 & 1449 & 7 & 1000 & 1000 \\
  $dr^{probit}$ & 95.2 & 6.9 & 6 & 9 & 0.3 & 84 & 6 & 1.8 & 7 & 8 & 1.8 & 3 & 4 & -0.0 & 205 & 2 & 1000 & 1000 \\
  $dr^{cbps}$ & 95.2 & 6.9 & 6 & 8 & 0.3 & 83 & 5 & 1.8 & 7 & 7 & 1.8 & 3 & 3 & -0.0 & 349 & 3 & 1000 & 1000 \\
  $tsls$ & 94.4 & 6.7 & 3 & 3 & 0.2 & 0 & 1 & 1.7 & 5 & 2 & 1.7 & -0 & 1 & -0.0 & 869 & 4 & 1000 & 1000 \\
  $ipw^{probit}$ & 94.2 & 6.7 & 4 & 5 & 0.5 & 202 & 8 & 1.7 & 5 & 3 & 1.8 & 4 & 6 & -0.0 & 0 & 1 & 1000 & 1000 \\
  $pairmatch^{cbps}$ & 93.9 & 9.4 & 44 & 16 & 0.6 & 276 & 13 & 2.5 & 49 & 15 & 2.5 & 47 & 14 & -0.1 & 12857 & 8 & 1000 & 1000 \\
  $pairmatch^{x}$ & 93.5 & 7.5 & 15 & 13 & 0.3 & 54 & 4 & 2.1 & 26 & 12 & 2.1 & 20 & 10 & -0.2 & 24962 & 11 & 1000 & 1000 \\
  $randforest$ & 93.4 & 6.6 & 2 & 2 & 0.2 & 40 & 3 & 1.8 & 8 & 9 & 1.8 & 4 & 5 & -0.1 & 13160 & 9 & 1000 & 1000 \\
  $pairmatch^{probit}$ & 93.3 & 9.5 & 46 & 17 & 0.7 & 324 & 17 & 2.5 & 52 & 16 & 2.6 & 50 & 15 & -0.1 & 14227 & 10 & 1000 & 1000 \\
  $ipw^{cbps}$ & 93.0 & 6.7 & 3 & 4 & 0.7 & 297 & 15 & 1.7 & 5 & 4 & 1.8 & 7 & 7 & -0.0 & 1040 & 5 & 1000 & 1000 \\
  $radmatch^{probit}$ & 92.6 & 9.7 & 50 & 19 & 0.8 & 354 & 19 & 2.6 & 61 & 18 & 2.8 & 59 & 18 & -0.2 & 29448 & 15 & 1000 & 1000 \\
  $radmatch^{cbps}$ & 92.5 & 9.5 & 46 & 18 & 0.6 & 277 & 14 & 2.6 & 58 & 17 & 2.7 & 55 & 16 & -0.2 & 28640 & 14 & 1000 & 1000 \\
  $reg^{kernel}$ & 91.9 & 6.7 & 4 & 6 & 0.5 & 218 & 9 & 1.9 & 14 & 11 & 2.0 & 13 & 9 & -0.2 & 25629 & 12 & 1000 & 939 \\
  $radmatchx^{probit}$ & 91.8 & 8.0 & 23 & 15 & 0.7 & 344 & 18 & 2.2 & 36 & 14 & 2.4 & 37 & 13 & -0.2 & 32308 & 16 & 1000 & 1000 \\
  $radmatchx^{cbps}$ & 90.7 & 7.9 & 22 & 14 & 0.7 & 320 & 16 & 2.2 & 34 & 13 & 2.3 & 34 & 12 & -0.2 & 27791 & 13 & 1000 & 1000 \\
  $ipw^{lc}$ & 100.0 & 95.8 & 1374 & 20 & 1.2 & 593 & 20 & 1.7 & 5 & 5 & 2.1 & 21 & 11 & 9.7 & 1503142 & 20 & 1000 & 977 \\
  $dr^{lc}$ & 100.0 & 526.2 & 7998 & 21 & 0.6 & 254 & 10 & 1.8 & 12 & 10 & 1.9 & 12 & 8 & 131.2 & 20306303 & 21 & 1000 & 366 \\
  $means$ & 74.2 & 6.5 & 0 & 1 & 2.2 & 1192 & 21 & 1.6 & 0 & 1 & 2.7 & 57 & 17 & 0.0 & 1222 & 6 & 1000 & 1000 \\
  $radmatchx^{lc}$ & 47.6 & 7.2 & 11 & 11 & 0.6 & 265 & 12 & 5.6 & 238 & 19 & 5.6 & 224 & 19 & -3.7 & 575406 & 17 & 1000 & 1000 \\
  $radmatch^{lc}$ & 41.3 & 7.3 & 12 & 12 & 0.6 & 256 & 11 & 6.4 & 287 & 20 & 6.4 & 270 & 20 & -4.5 & 698490 & 18 & 1000 & 1000 \\
  $pairmatch^{lc}$ & 32.6 & 7.1 & 9 & 10 & 0.3 & 85 & 7 & 7.1 & 329 & 21 & 7.1 & 309 & 21 & -5.2 & 812111 & 19 & 1000 & 1000 \\
\bottomrule
\end{tabular}
\begin{tablenotes}
\item \scriptsize Notes: `cov' and `int' refer to coverage rate and length of confidence interval, `bias' denotes the absolute bias from the true treatment effect, `sd' is the standard
deviation of the estimator, `rmse' stands for root mean squared error, and `seb' indicates the median bias of the estimated bootstrap standard error. `diff' indicates the difference to the left best performer in percent (\%) and `r' is the rank. `nsimp' shows the number of simulations with point estimations smaller than $|10^{10}|$ and `nsimse' denotes the number of simulations used to compute `cov', `int', and `seb'. Conditional on `nsimp', simulations with standard errors greater than 150 times the (true) standard error are not considered (i.e.simulations with non-converging bootstrap standard errors are dropped).
\end{tablenotes}
\end{threeparttable}
\end{sidewaystable}

\begin{sidewaystable}[ht]
\centering
\begin{threeparttable}
\caption{DGP 20}
\label{DGP 20}
\small
\begin{tabular}{rrrrrrrrrrrrrrrrrrr}
\toprule
& cov & int & d & r & bias & d & r & sd & d & r & rmse & d & r & seb & d & r & nsimp & nsimse \\
\midrule
  $radmatchx^{probit}$ & 99.2 & 1143.3 & 77 & 11 & 7.3 & 160 & 14 & 46.4 & 77 & 15 & 47.0 & 65 & 15 & -3.4 & 334 & 6 & 999 & 996 \\
  $radmatchx^{cbps}$ & 99.5 & 1229.3 & 90 & 12 & 7.2 & 157 & 12 & 45.6 & 74 & 14 & 46.2 & 62 & 14 & -2.6 & 228 & 4 & 997 & 994 \\
  $pairmatch^{probit}$ & 99.6 & 1255.2 & 94 & 13 & 6.7 & 139 & 7 & 42.4 & 62 & 13 & 42.9 & 51 & 13 & 27.9 & 3415 & 16 & 1000 & 984 \\
  $radmatch^{probit}$ & 99.6 & 1280.9 & 98 & 15 & 6.6 & 136 & 5 & 37.3 & 42 & 11 & 37.8 & 33 & 11 & 26.6 & 3254 & 15 & 999 & 986 \\
  $randforest$ & 99.6 & 647.0 & 0 & 1 & 4.3 & 53 & 2 & 32.6 & 25 & 8 & 32.9 & 15 & 7 & -6.8 & 754 & 10 & 1000 & 989 \\
  $pairmatch^{x}$ & 99.7 & 997.3 & 54 & 10 & 6.5 & 133 & 4 & 48.0 & 83 & 17 & 48.4 & 70 & 17 & -3.7 & 366 & 8 & 1000 & 991 \\
  $means$ & 99.9 & 877.8 & 36 & 8 & 18.7 & 564 & 21 & 28.4 & 8 & 3 & 34.0 & 19 & 8 & 8.5 & 968 & 11 & 1000 & 983 \\
  $reg$ & 99.9 & 780.2 & 21 & 7 & 7.2 & 155 & 11 & 33.5 & 28 & 9 & 34.2 & 20 & 9 & -3.5 & 335 & 7 & 1000 & 986 \\
  $pairmatch^{cbps}$ & 99.9 & 1280.1 & 98 & 14 & 9.1 & 222 & 16 & 47.0 & 80 & 16 & 47.9 & 68 & 16 & 19.1 & 2300 & 14 & 999 & 988 \\
  $tsls$ & 99.9 & 745.7 & 15 & 4 & 7.4 & 163 & 15 & 39.7 & 52 & 12 & 40.3 & 41 & 12 & -9.6 & 1105 & 13 & 1000 & 988 \\
  $dr^{probit}$ & 99.9 & 733.8 & 13 & 3 & 7.1 & 152 & 9 & 30.9 & 18 & 5 & 31.6 & 11 & 4 & -0.9 & 7 & 2 & 1000 & 989 \\
  $dr^{cbps}$ & 99.9 & 747.0 & 15 & 5 & 7.2 & 154 & 10 & 31.8 & 22 & 7 & 32.6 & 14 & 6 & -1.7 & 117 & 3 & 1000 & 989 \\
  $ipw^{probit}$ & 99.9 & 722.2 & 12 & 2 & 7.2 & 157 & 13 & 34.3 & 31 & 10 & 35.0 & 23 & 10 & -3.1 & 295 & 5 & 1000 & 992 \\
  $radmatch^{cbps}$ & 99.9 & 1429.5 & 121 & 16 & 6.7 & 138 & 6 & 53.0 & 103 & 18 & 53.4 & 87 & 18 & 8.8 & 1004 & 12 & 997 & 994 \\
  $ipw^{cbps}$ & 100.0 & 946.4 & 46 & 9 & 6.9 & 144 & 8 & 27.7 & 6 & 2 & 28.5 & 0 & 1 & 6.1 & 672 & 9 & 999 & 995 \\
  $ipw^{lc}$ & 100.0 & 2681.6 & 314 & 21 & 13.2 & 370 & 20 & 26.2 & 0 & 1 & 29.3 & 3 & 2 & 166.9 & 20929 & 21 & 999 & 984 \\
  $dr^{lc}$ & 100.0 & 1737.5 & 169 & 18 & 9.5 & 236 & 18 & 31.0 & 19 & 6 & 32.4 & 14 & 5 & 166.2 & 20833 & 20 & 1000 & 972 \\
  $reg^{kernel}$ & 100.0 & 774.7 & 20 & 6 & 9.2 & 228 & 17 & 29.5 & 13 & 4 & 30.9 & 8 & 3 & 0.8 & 0 & 1 & 1000 & 987 \\
  $pairmatch^{lc}$ & 82.9 & 1968.3 & 204 & 20 & 5.8 & 106 & 3 & 113.5 & 334 & 19 & 113.6 & 298 & 19 & -57.8 & 7178 & 17 & 989 & 985 \\
  $radmatch^{lc}$ & 82.5 & 1953.2 & 202 & 19 & 11.3 & 302 & 19 & 136.5 & 422 & 20 & 136.9 & 380 & 20 & -72.9 & 9083 & 18 & 981 & 981 \\
  $radmatchx^{lc}$ & 71.7 & 1673.8 & 159 & 17 & 2.8 & 0 & 1 & 140.3 & 436 & 21 & 140.3 & 392 & 21 & -115.8 & 14492 & 19 & 981 & 981 \\
\bottomrule
\end{tabular}
\begin{tablenotes}
\item \scriptsize Notes: `cov' and `int' refer to coverage rate and length of confidence interval, `bias' denotes the absolute bias from the true treatment effect, `sd' is the standard
deviation of the estimator, `rmse' stands for root mean squared error, and `seb' indicates the median bias of the estimated bootstrap standard error. `diff' indicates the difference to the left best performer in percent (\%) and `r' is the rank. `nsimp' shows the number of simulations with point estimations smaller than $|10^{10}|$ and `nsimse' denotes the number of simulations used to compute `cov', `int', and `seb'. Conditional on `nsimp', simulations with standard errors greater than 150 times the (true) standard error are not considered (i.e.simulations with non-converging bootstrap standard errors are dropped).
\end{tablenotes}
\end{threeparttable}
\end{sidewaystable}

\begin{sidewaystable}[ht]
\centering
\begin{threeparttable}
\caption{DGP 21}
\label{DGP 21}
\small
\begin{tabular}{rrrrrrrrrrrrrrrrrrr}
\toprule
& cov & int & d & r & bias & d & r & sd & d & r & rmse & d & r & seb & d & r & nsimp & nsimse \\
\midrule
  $dr^{probit}$ & 95.1 & 10.5 & 16 & 12 & 0.8 & 81 & 7 & 2.6 & 11 & 8 & 2.7 & 6 & 5 & 0.1 & 239 & 7 & 1000 & 1000 \\
  $tsls$ & 94.9 & 10.1 & 11 & 5 & 0.6 & 39 & 2 & 2.5 & 7 & 3 & 2.6 & 0 & 1 & 0.1 & 134 & 4 & 1000 & 1000 \\
  $dr^{cbps}$ & 94.8 & 10.5 & 16 & 11 & 0.8 & 83 & 8 & 2.6 & 11 & 7 & 2.7 & 5 & 4 & 0.1 & 246 & 8 & 1000 & 1000 \\
  $reg$ & 95.3 & 10.4 & 15 & 9 & 0.7 & 56 & 3 & 2.5 & 10 & 6 & 2.6 & 3 & 2 & 0.1 & 228 & 6 & 1000 & 1000 \\
  $pairmatch^{cbps}$ & 93.6 & 14.5 & 61 & 16 & 1.3 & 180 & 13 & 3.7 & 61 & 15 & 4.0 & 54 & 14 & -0.1 & 292 & 9 & 1000 & 1000 \\
  $randforest$ & 93.6 & 10.1 & 11 & 4 & 0.7 & 60 & 4 & 2.6 & 12 & 9 & 2.7 & 5 & 3 & -0.0 & 32 & 2 & 1000 & 1000 \\
  $ipw^{probit}$ & 93.4 & 10.2 & 13 & 7 & 1.1 & 145 & 9 & 2.5 & 8 & 5 & 2.8 & 7 & 6 & 0.1 & 137 & 5 & 1000 & 1000 \\
  $pairmatch^{probit}$ & 92.9 & 14.8 & 63 & 18 & 1.3 & 176 & 11 & 3.9 & 69 & 16 & 4.1 & 60 & 16 & -0.3 & 827 & 13 & 1000 & 1000 \\
  $ipw^{cbps}$ & 92.3 & 10.1 & 12 & 6 & 1.4 & 197 & 16 & 2.5 & 8 & 4 & 2.9 & 11 & 7 & 0.1 & 95 & 3 & 1000 & 1000 \\
  $radmatch^{cbps}$ & 92.3 & 14.8 & 63 & 17 & 1.3 & 177 & 12 & 4.0 & 71 & 17 & 4.2 & 62 & 17 & -0.3 & 1026 & 15 & 1000 & 1000 \\
  $pairmatch^{x}$ & 92.1 & 11.7 & 29 & 13 & 0.8 & 72 & 6 & 3.1 & 35 & 12 & 3.2 & 26 & 11 & -0.2 & 494 & 10 & 1000 & 1000 \\
  $radmatch^{probit}$ & 91.6 & 15.1 & 67 & 19 & 1.3 & 176 & 10 & 4.3 & 83 & 18 & 4.4 & 72 & 18 & -0.5 & 1715 & 16 & 1000 & 1000 \\
  $reg^{kernel}$ & 90.9 & 10.0 & 11 & 3 & 1.3 & 187 & 14 & 2.7 & 18 & 11 & 3.0 & 18 & 9 & -0.2 & 656 & 11 & 1000 & 941 \\
  $ipw^{lc}$ & 100.0 & 160.5 & 1676 & 20 & 2.0 & 338 & 20 & 2.4 & 5 & 2 & 3.2 & 23 & 10 & 16.7 & 61775 & 20 & 1000 & 962 \\
  $dr^{lc}$ & 100.0 & 882.1 & 9659 & 21 & 1.4 & 196 & 15 & 2.7 & 16 & 10 & 3.0 & 18 & 8 & 222.2 & 820715 & 21 & 1000 & 470 \\
  $radmatchx^{probit}$ & 89.9 & 12.4 & 37 & 15 & 1.6 & 248 & 19 & 3.4 & 47 & 14 & 3.8 & 47 & 13 & -0.3 & 979 & 14 & 1000 & 1000 \\
  $radmatchx^{cbps}$ & 89.9 & 12.3 & 36 & 14 & 1.6 & 247 & 18 & 3.3 & 44 & 13 & 3.7 & 44 & 12 & -0.2 & 740 & 12 & 1000 & 1000 \\
  $means$ & 68.8 & 9.0 & 0 & 1 & 3.3 & 622 & 21 & 2.3 & 0 & 1 & 4.0 & 57 & 15 & -0.0 & 0 & 1 & 1000 & 1000 \\
  $radmatchx^{lc}$ & 49.1 & 10.4 & 15 & 10 & 0.5 & 0 & 1 & 8.4 & 261 & 19 & 8.4 & 226 & 19 & -5.7 & 21142 & 17 & 1000 & 1000 \\
  $radmatch^{lc}$ & 40.9 & 10.2 & 13 & 8 & 1.4 & 200 & 17 & 9.7 & 318 & 20 & 9.8 & 281 & 20 & -7.1 & 26245 & 18 & 1000 & 1000 \\
  $pairmatch^{lc}$ & 37.0 & 9.9 & 10 & 2 & 0.7 & 63 & 5 & 11.1 & 377 & 21 & 11.1 & 332 & 21 & -8.6 & 31641 & 19 & 1000 & 1000 \\
\bottomrule
\end{tabular}
\begin{tablenotes}
\item \scriptsize Notes: `cov' and `int' refer to coverage rate and length of confidence interval, `bias' denotes the absolute bias from the true treatment effect, `sd' is the standard
deviation of the estimator, `rmse' stands for root mean squared error, and `seb' indicates the median bias of the estimated bootstrap standard error. `diff' indicates the difference to the left best performer in percent (\%) and `r' is the rank. `nsimp' shows the number of simulations with point estimations smaller than $|10^{10}|$ and `nsimse' denotes the number of simulations used to compute `cov', `int', and `seb'. Conditional on `nsimp', simulations with standard errors greater than 150 times the (true) standard error are not considered (i.e.simulations with non-converging bootstrap standard errors are dropped).
\end{tablenotes}
\end{threeparttable}
\end{sidewaystable}

\begin{sidewaystable}[ht]
\centering
\begin{threeparttable}
\caption{DGP 22}
\label{DGP 22}
\small
\begin{tabular}{rrrrrrrrrrrrrrrrrrr}
\toprule
& cov & int & d & r & bias & d & r & sd & d & r & rmse & d & r & seb & d & r & nsimp & nsimse \\
\midrule
  $ipw^{cbps}$ & 95.1 & 5.8 & 2 & 9 & 0.2 & 251 & 15 & 1.5 & 1 & 6 & 1.5 & 1 & 6 & 0.0 & 146 & 5 & 1000 & 1000 \\
  $dr^{cbps}$ & 95.1 & 5.8 & 1 & 4 & 0.2 & 188 & 13 & 1.5 & 0 & 2 & 1.5 & 0 & 4 & 0.0 & 84 & 3 & 1000 & 1000 \\
  $reg$ & 95.1 & 5.8 & 1 & 3 & 0.2 & 173 & 10 & 1.5 & 0 & 5 & 1.5 & 0 & 5 & -0.0 & 0 & 1 & 1000 & 1000 \\
  $dr^{probit}$ & 95.2 & 5.8 & 1 & 5 & 0.2 & 184 & 11 & 1.5 & 0 & 1 & 1.5 & 0 & 1 & 0.0 & 173 & 6 & 1000 & 1000 \\
  $pairmatch^{cbps}$ & 95.2 & 6.8 & 19 & 19 & 0.1 & 165 & 9 & 1.7 & 16 & 11 & 1.7 & 16 & 10 & 0.0 & 2570 & 10 & 1000 & 1000 \\
  $ipw^{probit}$ & 95.3 & 5.8 & 1 & 6 & 0.1 & 153 & 8 & 1.5 & 0 & 3 & 1.5 & 0 & 2 & 0.0 & 125 & 4 & 1000 & 1000 \\
  $pairmatch^{probit}$ & 95.3 & 6.8 & 18 & 18 & 0.1 & 135 & 6 & 1.7 & 16 & 14 & 1.7 & 16 & 12 & 0.0 & 1472 & 8 & 1000 & 1000 \\
  $tsls$ & 95.3 & 5.8 & 1 & 2 & 0.1 & 148 & 7 & 1.5 & 0 & 4 & 1.5 & 0 & 3 & -0.0 & 16 & 2 & 1000 & 1000 \\
  $radmatch^{probit}$ & 94.5 & 6.6 & 14 & 16 & 0.1 & 110 & 5 & 1.8 & 19 & 17 & 1.8 & 19 & 16 & -0.1 & 5846 & 11 & 1000 & 1000 \\
  $radmatch^{cbps}$ & 94.0 & 6.6 & 15 & 17 & 0.2 & 187 & 12 & 1.7 & 16 & 13 & 1.7 & 16 & 13 & -0.0 & 2105 & 9 & 1000 & 1000 \\
  $radmatchx^{cbps}$ & 94.0 & 6.4 & 11 & 12 & 0.1 & 19 & 2 & 1.7 & 16 & 12 & 1.7 & 16 & 11 & -0.1 & 6635 & 13 & 1000 & 1000 \\
  $randforest$ & 93.6 & 5.7 & -0 & 1 & 0.1 & 85 & 4 & 1.6 & 5 & 10 & 1.6 & 5 & 7 & -0.1 & 6583 & 12 & 1000 & 1000 \\
  $reg^{kernel}$ & 93.5 & 5.8 & 2 & 8 & 0.4 & 606 & 19 & 1.9 & 28 & 18 & 1.9 & 30 & 18 & -0.4 & 29647 & 16 & 1000 & 987 \\
  $radmatchx^{probit}$ & 93.2 & 6.3 & 11 & 11 & 0.1 & 0 & 1 & 1.7 & 18 & 15 & 1.7 & 17 & 14 & -0.1 & 8687 & 14 & 1000 & 1000 \\
  $pairmatch^{x}$ & 91.9 & 6.3 & 10 & 10 & 0.1 & 60 & 3 & 1.7 & 19 & 16 & 1.7 & 18 & 15 & -0.1 & 10594 & 15 & 1000 & 1000 \\
  $ipw^{lc}$ & 98.9 & 10.8 & 89 & 20 & 0.7 & 1098 & 20 & 1.5 & 3 & 8 & 1.7 & 12 & 9 & 0.7 & 50770 & 17 & 1000 & 998 \\
  $dr^{lc}$ & 99.1 & 17.4 & 204 & 21 & 0.4 & 564 & 18 & 1.5 & 4 & 9 & 1.6 & 7 & 8 & 0.9 & 68576 & 18 & 1000 & 982 \\
  $means$ & 88.1 & 5.8 & 2 & 7 & 1.1 & 1926 & 21 & 1.5 & 2 & 7 & 1.9 & 27 & 17 & -0.0 & 706 & 7 & 1000 & 1000 \\
  $radmatchx^{lc}$ & 59.8 & 6.4 & 12 & 13 & 0.2 & 214 & 14 & 4.2 & 184 & 19 & 4.2 & 183 & 19 & -2.5 & 193223 & 19 & 1000 & 1000 \\
  $pairmatch^{lc}$ & 59.1 & 6.4 & 12 & 14 & 0.2 & 267 & 16 & 4.4 & 197 & 20 & 4.4 & 196 & 20 & -2.7 & 207889 & 20 & 1000 & 1000 \\
  $radmatch^{lc}$ & 54.5 & 6.6 & 14 & 15 & 0.3 & 411 & 17 & 4.9 & 236 & 21 & 5.0 & 235 & 21 & -3.3 & 249088 & 21 & 1000 & 1000 \\
\bottomrule
\end{tabular}
\begin{tablenotes}
\item \scriptsize Notes: `cov' and `int' refer to coverage rate and length of confidence interval, `bias' denotes the absolute bias from the true treatment effect, `sd' is the standard
deviation of the estimator, `rmse' stands for root mean squared error, and `seb' indicates the median bias of the estimated bootstrap standard error. `diff' indicates the difference to the left best performer in percent (\%) and `r' is the rank. `nsimp' shows the number of simulations with point estimations smaller than $|10^{10}|$ and `nsimse' denotes the number of simulations used to compute `cov', `int', and `seb'. Conditional on `nsimp', simulations with standard errors greater than 150 times the (true) standard error are not considered (i.e.simulations with non-converging bootstrap standard errors are dropped).
\end{tablenotes}
\end{threeparttable}
\end{sidewaystable}

\begin{sidewaystable}[ht]
\centering
\begin{threeparttable}
\caption{DGP 23}
\label{DGP 23}
\small
\begin{tabular}{rrrrrrrrrrrrrrrrrrr}
\toprule
& cov & int & d & r & bias & d & r & sd & d & r & rmse & d & r & seb & d & r & nsimp & nsimse \\
\midrule
  $reg$ & 99.2 & 630.7 & 38 & 2 & 2.2 & 27 & 3 & 33.3 & 56 & 8 & 33.3 & 48 & 7 & -6.5 & 874 & 9 & 1000 & 987 \\
  $reg^{kernel}$ & 99.3 & 779.1 & 70 & 8 & 7.3 & 315 & 11 & 21.4 & 0 & 1 & 22.6 & 0 & 2 & 2.8 & 318 & 4 & 999 & 937 \\
  $randforest$ & 99.3 & 962.9 & 110 & 11 & 3.6 & 103 & 7 & 34.2 & 60 & 10 & 34.4 & 53 & 9 & -6.8 & 916 & 10 & 999 & 995 \\
  $tsls$ & 99.3 & 658.8 & 44 & 4 & 2.1 & 22 & 2 & 22.4 & 5 & 3 & 22.5 & 0 & 1 & 3.7 & 457 & 5 & 999 & 995 \\
  $radmatchx^{cbps}$ & 99.4 & 816.2 & 78 & 9 & 9.4 & 436 & 17 & 37.5 & 75 & 12 & 38.7 & 72 & 11 & -2.8 & 316 & 3 & 1000 & 989 \\
  $pairmatch^{x}$ & 99.4 & 1593.3 & 248 & 13 & 3.4 & 96 & 5 & 49.9 & 134 & 14 & 50.0 & 122 & 14 & -5.0 & 650 & 8 & 997 & 997 \\
  $ipw^{probit}$ & 99.5 & 631.9 & 38 & 3 & 6.2 & 251 & 10 & 25.7 & 20 & 4 & 26.4 & 17 & 4 & 0.7 & 0 & 1 & 1000 & 986 \\
  $dr^{cbps}$ & 99.5 & 673.9 & 47 & 5 & 3.5 & 98 & 6 & 30.1 & 41 & 7 & 30.3 & 35 & 6 & -1.1 & 67 & 2 & 1000 & 989 \\
  $ipw^{cbps}$ & 99.5 & 749.2 & 63 & 7 & 7.6 & 330 & 12 & 22.2 & 4 & 2 & 23.5 & 4 & 3 & 4.1 & 519 & 6 & 999 & 991 \\
  $dr^{probit}$ & 99.5 & 702.8 & 53 & 6 & 3.3 & 90 & 4 & 33.5 & 57 & 9 & 33.6 & 49 & 8 & -4.5 & 577 & 7 & 1000 & 992 \\
  $radmatchx^{probit}$ & 99.5 & 1011.1 & 121 & 12 & 9.5 & 443 & 18 & 48.8 & 128 & 13 & 49.7 & 121 & 13 & -13.2 & 1888 & 12 & 999 & 995 \\
  $means$ & 99.8 & 851.7 & 86 & 10 & 23.1 & 1216 & 21 & 27.0 & 26 & 5 & 35.5 & 58 & 10 & 8.0 & 1107 & 11 & 1000 & 984 \\
  $pairmatch^{probit}$ & 99.8 & 2577.8 & 462 & 16 & 8.8 & 399 & 15 & 67.5 & 216 & 16 & 68.0 & 202 & 16 & 62.4 & 9256 & 14 & 994 & 993 \\
  $radmatch^{cbps}$ & 99.8 & 1999.3 & 336 & 15 & 5.6 & 221 & 9 & 68.8 & 222 & 17 & 69.0 & 206 & 17 & 60.6 & 8986 & 13 & 995 & 995 \\
  $radmatch^{probit}$ & 99.9 & 2644.4 & 477 & 17 & 9.3 & 431 & 16 & 76.9 & 260 & 18 & 77.4 & 244 & 18 & 63.5 & 9423 & 15 & 992 & 989 \\
  $pairmatch^{cbps}$ & 99.9 & 1884.9 & 311 & 14 & 7.7 & 337 & 13 & 56.4 & 164 & 15 & 56.9 & 153 & 15 & 72.0 & 10704 & 16 & 996 & 991 \\
  $ipw^{lc}$ & 100.0 & 3689.9 & 705 & 19 & 13.1 & 648 & 20 & 36.7 & 72 & 11 & 38.9 & 73 & 12 & 417.0 & 62469 & 20 & 999 & 974 \\
  $dr^{lc}$ & 100.0 & 5157.3 & 1025 & 20 & 8.0 & 354 & 14 & 28.0 & 31 & 6 & 29.1 & 29 & 5 & 505.1 & 75681 & 21 & 999 & 929 \\
  $pairmatch^{lc}$ & 80.7 & 3456.3 & 654 & 18 & 4.4 & 149 & 8 & 135.9 & 535 & 19 & 135.9 & 503 & 19 & -73.2 & 10883 & 17 & 984 & 984 \\
  $radmatch^{lc}$ & 80.5 & 12658.4 & 2661 & 21 & 1.8 & 0 & 1 & 158.8 & 643 & 21 & 158.7 & 605 & 21 & -91.2 & 13581 & 18 & 977 & 977 \\
  $radmatchx^{lc}$ & 52.7 & 458.4 & 0 & 1 & 9.7 & 454 & 19 & 152.8 & 614 & 20 & 153.0 & 580 & 20 & -136.1 & 20315 & 19 & 977 & 977 \\
\bottomrule
\end{tabular}
\begin{tablenotes}
\item \scriptsize Notes: `cov' and `int' refer to coverage rate and length of confidence interval, `bias' denotes the absolute bias from the true treatment effect, `sd' is the standard
deviation of the estimator, `rmse' stands for root mean squared error, and `seb' indicates the median bias of the estimated bootstrap standard error. `diff' indicates the difference to the left best performer in percent (\%) and `r' is the rank. `nsimp' shows the number of simulations with point estimations smaller than $|10^{10}|$ and `nsimse' denotes the number of simulations used to compute `cov', `int', and `seb'. Conditional on `nsimp', simulations with standard errors greater than 150 times the (true) standard error are not considered (i.e.simulations with non-converging bootstrap standard errors are dropped).
\end{tablenotes}
\end{threeparttable}
\end{sidewaystable}

\begin{sidewaystable}[ht]
\centering
\begin{threeparttable}
\caption{DGP 24}
\label{DGP 24}
\small
\begin{tabular}{rrrrrrrrrrrrrrrrrrr}
\toprule
& cov & int & d & r & bias & d & r & sd & d & r & rmse & d & r & seb & d & r & nsimp & nsimse \\
\midrule
  $dr^{probit}$ & 95.0 & 6.2 & 7 & 9 & 0.4 & 85 & 5 & 1.6 & 5 & 8 & 1.6 & 4 & 4 & 0.0 & 168 & 6 & 1000 & 1000 \\
  $dr^{cbps}$ & 95.0 & 6.1 & 7 & 8 & 0.4 & 86 & 6 & 1.5 & 5 & 7 & 1.6 & 4 & 3 & 0.0 & 126 & 3 & 1000 & 1000 \\
  $reg$ & 95.0 & 6.1 & 6 & 7 & 0.3 & 45 & 4 & 1.5 & 4 & 6 & 1.6 & 2 & 2 & 0.0 & 0 & 1 & 1000 & 1000 \\
  $tsls$ & 95.0 & 6.0 & 4 & 3 & 0.3 & 20 & 2 & 1.5 & 3 & 2 & 1.5 & 0 & 1 & 0.0 & 159 & 5 & 1000 & 1000 \\
  $ipw^{probit}$ & 94.3 & 6.0 & 5 & 6 & 0.6 & 167 & 10 & 1.5 & 3 & 5 & 1.6 & 6 & 6 & 0.0 & 155 & 4 & 1000 & 1000 \\
  $pairmatch^{cbps}$ & 93.9 & 8.4 & 47 & 16 & 0.7 & 226 & 14 & 2.2 & 49 & 15 & 2.3 & 50 & 14 & -0.1 & 2126 & 10 & 1000 & 1000 \\
  $pairmatch^{probit}$ & 93.8 & 8.5 & 48 & 17 & 0.6 & 193 & 11 & 2.2 & 51 & 16 & 2.3 & 51 & 15 & -0.1 & 2237 & 11 & 1000 & 1000 \\
  $randforest$ & 93.8 & 6.0 & 5 & 5 & 0.3 & 44 & 3 & 1.6 & 7 & 10 & 1.6 & 5 & 5 & -0.1 & 1126 & 8 & 1000 & 1000 \\
  $pairmatch^{x}$ & 93.1 & 6.8 & 18 & 13 & 0.4 & 90 & 7 & 1.8 & 24 & 12 & 1.9 & 22 & 10 & -0.1 & 2460 & 12 & 1000 & 1000 \\
  $radmatch^{probit}$ & 93.0 & 8.7 & 52 & 19 & 0.6 & 199 & 12 & 2.4 & 63 & 18 & 2.5 & 62 & 18 & -0.2 & 5237 & 16 & 1000 & 1000 \\
  $radmatch^{cbps}$ & 92.9 & 8.6 & 49 & 18 & 0.7 & 220 & 13 & 2.4 & 60 & 17 & 2.5 & 60 & 16 & -0.2 & 4868 & 15 & 1000 & 1000 \\
  $ipw^{cbps}$ & 92.4 & 6.0 & 5 & 4 & 0.7 & 243 & 15 & 1.5 & 3 & 3 & 1.7 & 10 & 7 & 0.0 & 21 & 2 & 1000 & 1000 \\
  $reg^{kernel}$ & 91.8 & 6.0 & 4 & 2 & 0.8 & 258 & 17 & 1.6 & 8 & 11 & 1.8 & 15 & 9 & -0.1 & 1661 & 9 & 1000 & 930 \\
  $radmatchx^{cbps}$ & 90.7 & 7.1 & 24 & 14 & 0.8 & 285 & 19 & 2.0 & 32 & 13 & 2.1 & 38 & 12 & -0.1 & 3187 & 14 & 1000 & 1000 \\
  $radmatchx^{probit}$ & 90.3 & 7.2 & 25 & 15 & 0.8 & 283 & 18 & 2.0 & 33 & 14 & 2.1 & 38 & 13 & -0.1 & 3178 & 13 & 1000 & 1000 \\
  $ipw^{lc}$ & 100.0 & 84.1 & 1365 & 20 & 1.1 & 426 & 20 & 1.5 & 3 & 4 & 1.9 & 23 & 11 & 8.6 & 198955 & 20 & 1000 & 963 \\
  $dr^{lc}$ & 100.0 & 437.7 & 7521 & 21 & 0.7 & 254 & 16 & 1.6 & 7 & 9 & 1.8 & 14 & 8 & 106.9 & 2464904 & 21 & 1000 & 356 \\
  $means$ & 70.9 & 5.7 & 0 & 1 & 2.0 & 834 & 21 & 1.5 & 0 & 1 & 2.5 & 61 & 17 & -0.0 & 273 & 7 & 1000 & 1000 \\
  $radmatchx^{lc}$ & 45.2 & 6.4 & 11 & 11 & 0.5 & 114 & 8 & 5.3 & 261 & 19 & 5.3 & 248 & 19 & -3.7 & 85238 & 17 & 1000 & 1000 \\
  $radmatch^{lc}$ & 41.7 & 6.5 & 12 & 12 & 0.5 & 149 & 9 & 5.9 & 297 & 20 & 5.9 & 283 & 20 & -4.2 & 97140 & 18 & 1000 & 1000 \\
  $pairmatch^{lc}$ & 38.0 & 6.3 & 9 & 10 & 0.2 & 0 & 1 & 6.3 & 326 & 21 & 6.3 & 310 & 21 & -4.7 & 108256 & 19 & 1000 & 1000 \\
\bottomrule
\end{tabular}
\begin{tablenotes}
\item \scriptsize Notes: `cov' and `int' refer to coverage rate and length of confidence interval, `bias' denotes the absolute bias from the true treatment effect, `sd' is the standard
deviation of the estimator, `rmse' stands for root mean squared error, and `seb' indicates the median bias of the estimated bootstrap standard error. `diff' indicates the difference to the left best performer in percent (\%) and `r' is the rank. `nsimp' shows the number of simulations with point estimations smaller than $|10^{10}|$ and `nsimse' denotes the number of simulations used to compute `cov', `int', and `seb'. Conditional on `nsimp', simulations with standard errors greater than 150 times the (true) standard error are not considered (i.e.simulations with non-converging bootstrap standard errors are dropped).
\end{tablenotes}
\end{threeparttable}
\end{sidewaystable}

\begin{sidewaystable}[ht]
\centering
\begin{threeparttable}
\caption{DGP 25}
\label{DGP 25}
\small
\begin{tabular}{rrrrrrrrrrrrrrrrrrr}
\toprule
& cov & int & d & r & bias & d & r & sd & d & r & rmse & d & r & seb & d & r & nsimp & nsimse \\
\midrule
  $dr^{cbps}$ & 94.9 & 0.1 & 1 & 4 & 0.0 & 39 & 6 & 0.0 & 0 & 4 & 0.0 & 0 & 2 & 0.0 & 218 & 4 & 1000 & 1000 \\
  $reg$ & 94.9 & 0.1 & 1 & 2 & 0.0 & 36 & 5 & 0.0 & 0 & 3 & 0.0 & 0 & 1 & 0.0 & 204 & 2 & 1000 & 1000 \\
  $ipw^{probit}$ & 94.8 & 0.1 & 2 & 6 & 0.0 & 42 & 9 & 0.0 & 0 & 2 & 0.0 & 0 & 5 & 0.0 & 243 & 6 & 1000 & 1000 \\
  $dr^{probit}$ & 94.8 & 0.1 & 2 & 5 & 0.0 & 39 & 7 & 0.0 & 0 & 5 & 0.0 & 0 & 4 & 0.0 & 217 & 3 & 1000 & 1000 \\
  $radmatch^{cbps}$ & 94.8 & 0.2 & 15 & 17 & 0.0 & 53 & 11 & 0.0 & 15 & 14 & 0.0 & 15 & 12 & 0.0 & 0 & 1 & 1000 & 1000 \\
  $tsls$ & 94.8 & 0.1 & 1 & 3 & 0.0 & 40 & 8 & 0.0 & 0 & 1 & 0.0 & 0 & 3 & 0.0 & 245 & 7 & 1000 & 1000 \\
  $pairmatch^{cbps}$ & 95.3 & 0.2 & 19 & 19 & 0.0 & 70 & 14 & 0.0 & 15 & 13 & 0.0 & 15 & 13 & 0.0 & 749 & 14 & 1000 & 1000 \\
  $ipw^{cbps}$ & 94.7 & 0.1 & 2 & 8 & 0.0 & 56 & 12 & 0.0 & 1 & 7 & 0.0 & 1 & 6 & 0.0 & 233 & 5 & 1000 & 1000 \\
  $pairmatch^{probit}$ & 94.2 & 0.2 & 18 & 18 & 0.0 & 71 & 15 & 0.0 & 16 & 15 & 0.0 & 17 & 16 & 0.0 & 372 & 12 & 1000 & 1000 \\
  $reg^{kernel}$ & 94.2 & 0.1 & 3 & 9 & 0.0 & 130 & 17 & 0.0 & 1 & 10 & 0.0 & 5 & 9 & 0.0 & 257 & 9 & 1000 & 1000 \\
  $randforest$ & 93.9 & 0.1 & 0 & 1 & 0.0 & 0 & 1 & 0.0 & 3 & 11 & 0.0 & 2 & 7 & -0.0 & 327 & 10 & 1000 & 1000 \\
  $radmatch^{probit}$ & 93.7 & 0.2 & 15 & 16 & 0.0 & 57 & 13 & 0.0 & 18 & 17 & 0.0 & 17 & 18 & -0.0 & 330 & 11 & 1000 & 1000 \\
  $radmatchx^{cbps}$ & 93.4 & 0.2 & 11 & 12 & 0.0 & 2 & 2 & 0.0 & 17 & 16 & 0.0 & 16 & 15 & -0.0 & 1009 & 15 & 1000 & 1000 \\
  $pairmatch^{x}$ & 93.2 & 0.2 & 9 & 10 & 0.0 & 29 & 4 & 0.0 & 14 & 12 & 0.0 & 14 & 11 & -0.0 & 708 & 13 & 1000 & 1000 \\
  $radmatchx^{probit}$ & 92.7 & 0.2 & 11 & 11 & 0.0 & 10 & 3 & 0.0 & 18 & 18 & 0.0 & 17 & 17 & -0.0 & 1157 & 16 & 1000 & 1000 \\
  $means$ & 91.8 & 0.1 & 2 & 7 & 0.0 & 317 & 20 & 0.0 & 0 & 6 & 0.0 & 16 & 14 & 0.0 & 250 & 8 & 1000 & 1000 \\
  $ipw^{lc}$ & 98.5 & 0.3 & 88 & 20 & 0.0 & 200 & 18 & 0.0 & 1 & 8 & 0.0 & 8 & 10 & 0.0 & 8756 & 17 & 1000 & 998 \\
  $dr^{lc}$ & 99.3 & 0.4 & 200 & 21 & 0.0 & 119 & 16 & 0.0 & 1 & 9 & 0.0 & 4 & 8 & 0.0 & 11874 & 18 & 1000 & 994 \\
  $radmatchx^{lc}$ & 57.4 & 0.2 & 12 & 13 & 0.0 & 51 & 10 & 0.1 & 178 & 19 & 0.1 & 174 & 19 & -0.1 & 29556 & 19 & 1000 & 1000 \\
  $pairmatch^{lc}$ & 55.6 & 0.2 & 12 & 14 & 0.0 & 278 & 19 & 0.1 & 192 & 20 & 0.1 & 191 & 20 & -0.1 & 31833 & 20 & 1000 & 1000 \\
  $radmatch^{lc}$ & 49.9 & 0.2 & 15 & 15 & 0.0 & 359 & 21 & 0.1 & 234 & 21 & 0.1 & 234 & 21 & -0.1 & 38792 & 21 & 1000 & 1000 \\
\bottomrule
\end{tabular}
\begin{tablenotes}
\item \scriptsize Notes: `cov' and `int' refer to coverage rate and length of confidence interval, `bias' denotes the absolute bias from the true treatment effect, `sd' is the standard
deviation of the estimator, `rmse' stands for root mean squared error, and `seb' indicates the median bias of the estimated bootstrap standard error. `diff' indicates the difference to the left best performer in percent (\%) and `r' is the rank. `nsimp' shows the number of simulations with point estimations smaller than $|10^{10}|$ and `nsimse' denotes the number of simulations used to compute `cov', `int', and `seb'. Conditional on `nsimp', simulations with standard errors greater than 150 times the (true) standard error are not considered (i.e.simulations with non-converging bootstrap standard errors are dropped).
\end{tablenotes}
\end{threeparttable}
\end{sidewaystable}

\begin{sidewaystable}[ht]
\centering
\begin{threeparttable}
\caption{DGP 26}
\label{DGP 26}
\small
\begin{tabular}{rrrrrrrrrrrrrrrrrrr}
\toprule
& cov & int & d & r & bias & d & r & sd & d & r & rmse & d & r & seb & d & r & nsimp & nsimse \\
\midrule
  $dr^{probit}$ & 95.1 & 0.3 & 2 & 7 & 0.0 & 183 & 8 & 0.1 & 1 & 4 & 0.1 & 0 & 3 & 0.0 & 4295 & 9 & 1000 & 1000 \\
  $ipw^{cbps}$ & 94.9 & 0.3 & 2 & 8 & 0.0 & 246 & 14 & 0.1 & 2 & 7 & 0.1 & 1 & 6 & 0.0 & 2758 & 4 & 1000 & 1000 \\
  $ipw^{probit}$ & 95.2 & 0.3 & 2 & 4 & 0.0 & 195 & 12 & 0.1 & 1 & 2 & 0.1 & 0 & 1 & 0.0 & 3959 & 8 & 1000 & 1000 \\
  $reg^{kernel}$ & 95.2 & 0.3 & 4 & 9 & 0.0 & 574 & 19 & 0.1 & 4 & 9 & 0.1 & 5 & 8 & 0.0 & 0 & 1 & 1000 & 1000 \\
  $dr^{cbps}$ & 95.3 & 0.3 & 2 & 6 & 0.0 & 185 & 9 & 0.1 & 1 & 5 & 0.1 & 0 & 4 & 0.0 & 3935 & 7 & 1000 & 1000 \\
  $reg$ & 95.3 & 0.3 & 2 & 5 & 0.0 & 166 & 5 & 0.1 & 1 & 6 & 0.1 & 0 & 5 & 0.0 & 3422 & 6 & 1000 & 1000 \\
  $tsls$ & 95.4 & 0.3 & 2 & 3 & 0.0 & 187 & 10 & 0.1 & 1 & 3 & 0.1 & 0 & 2 & 0.0 & 3219 & 5 & 1000 & 1000 \\
  $pairmatch^{probit}$ & 95.6 & 0.3 & 19 & 18 & 0.0 & 194 & 11 & 0.1 & 17 & 12 & 0.1 & 15 & 12 & 0.0 & 8736 & 10 & 1000 & 1000 \\
  $pairmatch^{cbps}$ & 94.4 & 0.3 & 19 & 19 & 0.0 & 238 & 13 & 0.1 & 19 & 16 & 0.1 & 18 & 16 & 0.0 & 2739 & 3 & 1000 & 1000 \\
  $radmatch^{probit}$ & 94.2 & 0.3 & 15 & 16 & 0.0 & 182 & 7 & 0.1 & 20 & 17 & 0.1 & 18 & 17 & -0.0 & 10262 & 12 & 1000 & 1000 \\
  $randforest$ & 94.2 & 0.3 & 1 & 2 & 0.0 & 63 & 3 & 0.1 & 6 & 11 & 0.1 & 4 & 7 & -0.0 & 11327 & 13 & 1000 & 1000 \\
  $radmatch^{cbps}$ & 93.8 & 0.3 & 16 & 17 & 0.0 & 246 & 15 & 0.1 & 20 & 18 & 0.1 & 19 & 18 & -0.0 & 9989 & 11 & 1000 & 1000 \\
  $pairmatch^{x}$ & 93.8 & 0.3 & 12 & 12 & 0.0 & 73 & 4 & 0.1 & 19 & 14 & 0.1 & 17 & 15 & -0.0 & 18137 & 15 & 1000 & 1000 \\
  $radmatchx^{cbps}$ & 93.6 & 0.3 & 12 & 13 & 0.0 & 31 & 2 & 0.1 & 18 & 13 & 0.1 & 16 & 13 & -0.0 & 14822 & 14 & 1000 & 1000 \\
  $radmatchx^{probit}$ & 92.3 & 0.3 & 12 & 14 & 0.0 & 0 & 1 & 0.1 & 19 & 15 & 0.1 & 17 & 14 & -0.0 & 18501 & 16 & 1000 & 1000 \\
  $means$ & 92.1 & 0.3 & 0 & 1 & 0.0 & 1284 & 21 & 0.1 & 0 & 1 & 0.1 & 10 & 11 & 0.0 & 1017 & 2 & 1000 & 1000 \\
  $ipw^{lc}$ & 98.4 & 0.5 & 87 & 20 & 0.0 & 812 & 20 & 0.1 & 4 & 8 & 0.1 & 7 & 10 & 0.0 & 123392 & 17 & 1000 & 998 \\
  $dr^{lc}$ & 99.1 & 0.9 & 213 & 21 & 0.0 & 519 & 17 & 0.1 & 5 & 10 & 0.1 & 5 & 9 & 0.0 & 174324 & 18 & 1000 & 992 \\
  $radmatchx^{lc}$ & 56.9 & 0.3 & 11 & 11 & 0.0 & 168 & 6 & 0.2 & 192 & 19 & 0.2 & 188 & 19 & -0.1 & 494466 & 19 & 1000 & 1000 \\
  $pairmatch^{lc}$ & 55.1 & 0.3 & 10 & 10 & 0.0 & 477 & 16 & 0.2 & 207 & 20 & 0.2 & 203 & 20 & -0.1 & 537894 & 20 & 1000 & 1000 \\
  $radmatch^{lc}$ & 50.3 & 0.3 & 13 & 15 & 0.0 & 553 & 18 & 0.2 & 248 & 21 & 0.2 & 243 & 21 & -0.2 & 642631 & 21 & 1000 & 1000 \\
\bottomrule
\end{tabular}
\begin{tablenotes}
\item \scriptsize Notes: `cov' and `int' refer to coverage rate and length of confidence interval, `bias' denotes the absolute bias from the true treatment effect, `sd' is the standard
deviation of the estimator, `rmse' stands for root mean squared error, and `seb' indicates the median bias of the estimated bootstrap standard error. `diff' indicates the difference to the left best performer in percent (\%) and `r' is the rank. `nsimp' shows the number of simulations with point estimations smaller than $|10^{10}|$ and `nsimse' denotes the number of simulations used to compute `cov', `int', and `seb'. Conditional on `nsimp', simulations with standard errors greater than 150 times the (true) standard error are not considered (i.e.simulations with non-converging bootstrap standard errors are dropped).
\end{tablenotes}
\end{threeparttable}
\end{sidewaystable}

\begin{sidewaystable}[ht]
\centering
\begin{threeparttable}
\caption{DGP 27}
\label{DGP 27}
\small
\begin{tabular}{rrrrrrrrrrrrrrrrrrr}
\toprule
& cov & int & d & r & bias & d & r & sd & d & r & rmse & d & r & seb & d & r & nsimp & nsimse \\
\midrule
  $pairmatch^{probit}$ & 94.9 & 0.2 & 46 & 17 & 0.0 & 1589 & 10 & 0.1 & 51 & 16 & 0.1 & 43 & 16 & -0.0 & 2825 & 10 & 1000 & 1000 \\
  $pairmatch^{cbps}$ & 95.2 & 0.2 & 44 & 16 & 0.0 & 1314 & 8 & 0.1 & 48 & 15 & 0.1 & 40 & 15 & -0.0 & 2414 & 9 & 1000 & 1000 \\
  $radmatch^{cbps}$ & 94.7 & 0.2 & 46 & 18 & 0.0 & 1214 & 7 & 0.1 & 58 & 17 & 0.1 & 49 & 17 & -0.0 & 6220 & 14 & 1000 & 1000 \\
  $ipw^{probit}$ & 94.6 & 0.2 & 5 & 7 & 0.0 & 1539 & 9 & 0.0 & 7 & 8 & 0.0 & 1 & 3 & -0.0 & 450 & 2 & 1000 & 1000 \\
  $dr^{cbps}$ & 94.5 & 0.2 & 6 & 8 & 0.0 & 262 & 4 & 0.0 & 8 & 9 & 0.0 & 2 & 5 & -0.0 & 593 & 5 & 1000 & 1000 \\
  $dr^{probit}$ & 94.4 & 0.2 & 7 & 9 & 0.0 & 228 & 3 & 0.0 & 8 & 10 & 0.0 & 2 & 6 & -0.0 & 546 & 4 & 1000 & 1000 \\
  $ipw^{cbps}$ & 94.4 & 0.2 & 5 & 5 & 0.0 & 2489 & 12 & 0.0 & 6 & 4 & 0.0 & 2 & 4 & -0.0 & 646 & 7 & 1000 & 1000 \\
  $tsls$ & 94.3 & 0.2 & 5 & 4 & 0.0 & 213 & 2 & 0.0 & 6 & 3 & 0.0 & 0 & 1 & -0.0 & 510 & 3 & 1000 & 1000 \\
  $reg$ & 94.1 & 0.2 & 5 & 6 & 0.0 & 0 & 1 & 0.0 & 7 & 6 & 0.0 & 1 & 2 & -0.0 & 605 & 6 & 1000 & 1000 \\
  $radmatch^{probit}$ & 93.8 & 0.2 & 50 & 19 & 0.0 & 1789 & 11 & 0.1 & 61 & 18 & 0.1 & 53 & 18 & -0.0 & 6201 & 13 & 1000 & 1000 \\
  $reg^{kernel}$ & 93.6 & 0.1 & 3 & 2 & 0.0 & 3225 & 17 & 0.0 & 6 & 5 & 0.0 & 3 & 8 & -0.0 & 1956 & 8 & 1000 & 1000 \\
  $pairmatch^{x}$ & 93.6 & 0.2 & 16 & 13 & 0.0 & 472 & 5 & 0.0 & 27 & 12 & 0.0 & 19 & 12 & -0.0 & 5922 & 12 & 1000 & 1000 \\
  $randforest$ & 93.3 & 0.2 & 4 & 3 & 0.0 & 548 & 6 & 0.0 & 10 & 11 & 0.0 & 4 & 9 & -0.0 & 3420 & 11 & 1000 & 1000 \\
  $radmatchx^{cbps}$ & 92.7 & 0.2 & 22 & 14 & 0.0 & 2560 & 13 & 0.0 & 35 & 13 & 0.1 & 28 & 13 & -0.0 & 6832 & 15 & 1000 & 1000 \\
  $radmatchx^{probit}$ & 91.9 & 0.2 & 23 & 15 & 0.0 & 2701 & 14 & 0.1 & 37 & 14 & 0.1 & 31 & 14 & -0.0 & 7737 & 16 & 1000 & 1000 \\
  $means$ & 90.1 & 0.1 & 0 & 1 & 0.0 & 9363 & 21 & 0.0 & 0 & 1 & 0.0 & 12 & 11 & 0.0 & 0 & 1 & 1000 & 1000 \\
  $ipw^{lc}$ & 100.0 & 2.2 & 1417 & 20 & 0.0 & 5070 & 19 & 0.0 & 5 & 2 & 0.0 & 5 & 10 & 0.2 & 347953 & 20 & 1000 & 975 \\
  $dr^{lc}$ & 100.0 & 11.4 & 7736 & 21 & 0.0 & 2955 & 16 & 0.0 & 7 & 7 & 0.0 & 2 & 7 & 2.9 & 4488107 & 21 & 1000 & 379 \\
  $radmatchx^{lc}$ & 47.9 & 0.2 & 11 & 11 & 0.0 & 5437 & 20 & 0.1 & 231 & 19 & 0.1 & 215 & 19 & -0.1 & 126911 & 17 & 1000 & 1000 \\
  $radmatch^{lc}$ & 40.9 & 0.2 & 12 & 12 & 0.0 & 4744 & 18 & 0.1 & 286 & 20 & 0.1 & 265 & 20 & -0.1 & 157481 & 18 & 1000 & 1000 \\
  $pairmatch^{lc}$ & 31.2 & 0.2 & 9 & 10 & 0.0 & 2871 & 15 & 0.2 & 330 & 21 & 0.2 & 306 & 21 & -0.1 & 184802 & 19 & 1000 & 1000 \\
\bottomrule
\end{tabular}
\begin{tablenotes}
\item \scriptsize Notes: `cov' and `int' refer to coverage rate and length of confidence interval, `bias' denotes the absolute bias from the true treatment effect, `sd' is the standard
deviation of the estimator, `rmse' stands for root mean squared error, and `seb' indicates the median bias of the estimated bootstrap standard error. `diff' indicates the difference to the left best performer in percent (\%) and `r' is the rank. `nsimp' shows the number of simulations with point estimations smaller than $|10^{10}|$ and `nsimse' denotes the number of simulations used to compute `cov', `int', and `seb'. Conditional on `nsimp', simulations with standard errors greater than 150 times the (true) standard error are not considered (i.e.simulations with non-converging bootstrap standard errors are dropped).
\end{tablenotes}
\end{threeparttable}
\end{sidewaystable}

\begin{sidewaystable}[ht]
\centering
\begin{threeparttable}
\caption{DGP 28}
\label{DGP 28}
\small
\begin{tabular}{rrrrrrrrrrrrrrrrrrr}
\toprule
& cov & int & d & r & bias & d & r & sd & d & r & rmse & d & r & seb & d & r & nsimp & nsimse \\
\midrule
$pairmatch^{probit}$ & 99.5 & 28.0 & 100 & 14 & 0.1 & 204 & 6 & 1.0 & 69 & 10 & 1.0 & 66 & 10 & 0.5 & 4647 & 12 & 1000 & 986 \\
  $radmatch^{probit}$ & 99.5 & 28.6 & 104 & 15 & 0.1 & 116 & 4 & 1.5 & 146 & 14 & 1.5 & 140 & 14 & -0.2 & 1880 & 10 & 1000 & 987 \\
  $radmatchx^{probit}$ & 99.5 & 25.4 & 81 & 12 & 0.1 & 85 & 3 & 1.8 & 202 & 16 & 1.8 & 194 & 16 & -0.9 & 7925 & 14 & 1000 & 995 \\
  $radmatchx^{cbps}$ & 99.5 & 23.8 & 70 & 10 & 0.2 & 326 & 15 & 2.0 & 242 & 17 & 2.1 & 233 & 17 & -1.1 & 10169 & 15 & 1000 & 995 \\
  $means$ & 99.8 & 17.3 & 23 & 6 & 0.3 & 576 & 19 & 0.6 & 0 & 1 & 0.7 & 7 & 2 & 0.1 & 1145 & 7 & 1000 & 984 \\
  $dr^{cbps}$ & 99.8 & 15.5 & 11 & 2 & 0.1 & 208 & 8 & 0.7 & 20 & 6 & 0.7 & 18 & 6 & -0.1 & 477 & 4 & 1000 & 986 \\
  $pairmatch^{cbps}$ & 99.8 & 27.2 & 93 & 13 & 0.2 & 274 & 12 & 1.2 & 107 & 12 & 1.2 & 103 & 12 & 0.4 & 3382 & 11 & 1000 & 986 \\
  $dr^{probit}$ & 99.8 & 15.8 & 13 & 3 & 0.1 & 206 & 7 & 0.7 & 17 & 5 & 0.7 & 16 & 5 & -0.1 & 431 & 3 & 1000 & 989 \\
  $reg$ & 99.8 & 17.9 & 28 & 8 & 0.1 & 213 & 9 & 0.8 & 30 & 8 & 0.8 & 28 & 8 & -0.1 & 1127 & 6 & 1000 & 990 \\
  $ipw^{probit}$ & 99.8 & 16.5 & 17 & 5 & 0.1 & 227 & 10 & 0.8 & 28 & 7 & 0.8 & 26 & 7 & -0.1 & 707 & 5 & 1000 & 991 \\
  $tsls$ & 99.8 & 17.6 & 25 & 7 & 0.1 & 236 & 11 & 0.9 & 46 & 9 & 0.9 & 44 & 9 & -0.2 & 1847 & 8 & 1000 & 992 \\
  $pairmatch^{x}$ & 99.8 & 24.2 & 72 & 11 & 0.1 & 169 & 5 & 1.2 & 98 & 11 & 1.2 & 93 & 11 & -0.2 & 1867 & 9 & 1000 & 995 \\
  $ipw^{cbps}$ & 99.8 & 20.4 & 46 & 9 & 0.2 & 278 & 13 & 1.3 & 122 & 13 & 1.3 & 117 & 13 & -0.6 & 5679 & 13 & 1000 & 997 \\
  $radmatch^{cbps}$ & 99.8 & 29.0 & 107 & 16 & 0.3 & 529 & 18 & 2.9 & 376 & 18 & 2.9 & 364 & 18 & -1.5 & 13472 & 16 & 1000 & 997 \\
  $randforest$ & 99.9 & 14.0 & 0 & 1 & 0.1 & 82 & 2 & 0.6 & 2 & 2 & 0.6 & 0 & 1 & -0.0 & 355 & 2 & 1000 & 982 \\
  $reg^{kernel}$ & 99.9 & 16.3 & 16 & 4 & 0.2 & 332 & 16 & 0.7 & 12 & 4 & 0.7 & 13 & 4 & -0.0 & 0 & 1 & 1000 & 996 \\
  $ipw^{lc}$ & 100.0 & 49.6 & 253 & 21 & 0.3 & 528 & 17 & 1.5 & 150 & 15 & 1.5 & 147 & 15 & 2.3 & 21516 & 17 & 1000 & 980 \\
  $dr^{lc}$ & 100.0 & 38.6 & 175 & 19 & 0.2 & 313 & 14 & 0.7 & 12 & 3 & 0.7 & 12 & 3 & 3.7 & 33558 & 18 & 1000 & 976 \\
  $radmatch^{lc}$ & 80.6 & 35.3 & 151 & 18 & 0.4 & 925 & 21 & 17.5 & 2814 & 21 & 17.5 & 2734 & 21 & -16.1 & 148496 & 21 & 1000 & 1000 \\
  $pairmatch^{lc}$ & 80.5 & 49.1 & 250 & 20 & 0.0 & 0 & 1 & 8.3 & 1284 & 19 & 8.3 & 1246 & 19 & -7.2 & 66053 & 19 & 1000 & 997 \\
  $radmatchx^{lc}$ & 69.0 & 30.0 & 114 & 17 & 0.4 & 905 & 20 & 13.7 & 2187 & 20 & 13.7 & 2125 & 20 & -13.2 & 121296 & 20 & 1000 & 1000 \\
\bottomrule
\end{tabular}
\begin{tablenotes}
\item \scriptsize Notes: `cov' and `int' refer to coverage rate and length of confidence interval, `bias' denotes the absolute bias from the true treatment effect, `sd' is the standard
deviation of the estimator, `rmse' stands for root mean squared error, and `seb' indicates the median bias of the estimated bootstrap standard error. `diff' indicates the difference to the left best performer in percent (\%) and `r' is the rank. `nsimp' shows the number of simulations with point estimations smaller than $|10^{10}|$ and `nsimse' denotes the number of simulations used to compute `cov', `int', and `seb'. Conditional on `nsimp', simulations with standard errors greater than 150 times the (true) standard error are not considered (i.e.simulations with non-converging bootstrap standard errors are dropped).
\end{tablenotes}
\end{threeparttable}
\end{sidewaystable}

\begin{sidewaystable}[ht]
\centering
\begin{threeparttable}
\caption{DGP 29}
\label{DGP 29}
\small
\begin{tabular}{rrrrrrrrrrrrrrrrrrr}
\toprule
& cov & int & d & r & bias & d & r & sd & d & r & rmse & d & r & seb & d & r & nsimp & nsimse \\
\midrule
  $tsls$ & 95.0 & 0.3 & 13 & 5 & 0.0 & 29 & 2 & 0.1 & 8 & 3 & 0.1 & 0 & 1 & 0.0 & 616 & 9 & 1000 & 1000 \\
  $pairmatch^{cbps}$ & 95.1 & 0.4 & 58 & 16 & 0.0 & 97 & 9 & 0.1 & 59 & 15 & 0.1 & 47 & 15 & -0.0 & 528 & 6 & 1000 & 1000 \\
  $ipw^{probit}$ & 95.2 & 0.3 & 14 & 8 & 0.0 & 116 & 12 & 0.1 & 10 & 6 & 0.1 & 5 & 3 & 0.0 & 526 & 5 & 1000 & 1000 \\
  $pairmatch^{probit}$ & 94.8 & 0.4 & 60 & 18 & 0.0 & 100 & 10 & 0.1 & 65 & 16 & 0.1 & 53 & 16 & -0.0 & 1372 & 11 & 1000 & 1000 \\
  $randforest$ & 94.7 & 0.3 & 13 & 6 & 0.0 & 53 & 4 & 0.1 & 13 & 11 & 0.1 & 5 & 6 & -0.0 & 291 & 3 & 1000 & 1000 \\
  $dr^{cbps}$ & 95.6 & 0.3 & 17 & 11 & 0.0 & 54 & 7 & 0.1 & 13 & 9 & 0.1 & 5 & 4 & 0.0 & 538 & 7 & 1000 & 1000 \\
  $dr^{probit}$ & 95.7 & 0.3 & 17 & 12 & 0.0 & 53 & 5 & 0.1 & 13 & 10 & 0.1 & 5 & 5 & 0.0 & 561 & 8 & 1000 & 1000 \\
  $ipw^{cbps}$ & 94.3 & 0.3 & 13 & 7 & 0.0 & 158 & 13 & 0.1 & 10 & 5 & 0.1 & 6 & 7 & 0.0 & 352 & 4 & 1000 & 1000 \\
  $reg$ & 95.9 & 0.3 & 15 & 10 & 0.0 & 30 & 3 & 0.1 & 11 & 7 & 0.1 & 2 & 2 & 0.0 & 657 & 10 & 1000 & 1000 \\
  $radmatch^{cbps}$ & 94.0 & 0.4 & 60 & 17 & 0.0 & 95 & 8 & 0.1 & 68 & 17 & 0.1 & 56 & 17 & -0.0 & 2149 & 13 & 1000 & 1000 \\
  $radmatch^{probit}$ & 93.3 & 0.4 & 64 & 19 & 0.0 & 103 & 11 & 0.1 & 78 & 18 & 0.1 & 65 & 18 & -0.0 & 3506 & 16 & 1000 & 1000 \\
  $reg^{kernel}$ & 93.0 & 0.3 & 10 & 3 & 0.0 & 201 & 18 & 0.1 & 9 & 4 & 0.1 & 8 & 8 & -0.0 & 0 & 1 & 1000 & 1000 \\
  $pairmatch^{x}$ & 92.7 & 0.3 & 30 & 13 & 0.0 & 54 & 6 & 0.1 & 39 & 12 & 0.1 & 28 & 12 & -0.0 & 2156 & 14 & 1000 & 1000 \\
  $radmatchx^{probit}$ & 92.4 & 0.3 & 36 & 15 & 0.0 & 197 & 17 & 0.1 & 46 & 14 & 0.1 & 40 & 14 & -0.0 & 2374 & 15 & 1000 & 1000 \\
  $radmatchx^{cbps}$ & 92.4 & 0.3 & 35 & 14 & 0.0 & 194 & 16 & 0.1 & 44 & 13 & 0.1 & 38 & 13 & -0.0 & 2073 & 12 & 1000 & 1000 \\
  $ipw^{lc}$ & 100.0 & 4.3 & 1722 & 20 & 0.0 & 260 & 19 & 0.1 & 6 & 2 & 0.1 & 9 & 9 & 0.4 & 166746 & 20 & 1000 & 962 \\
  $dr^{lc}$ & 100.0 & 23.2 & 9682 & 21 & 0.0 & 179 & 14 & 0.1 & 12 & 8 & 0.1 & 9 & 10 & 6.1 & 2258920 & 21 & 1000 & 459 \\
  $means$ & 86.6 & 0.2 & 0 & 1 & 0.1 & 434 & 21 & 0.1 & 0 & 1 & 0.1 & 18 & 11 & -0.0 & 53 & 2 & 1000 & 1000 \\
  $radmatchx^{lc}$ & 44.5 & 0.3 & 15 & 9 & 0.0 & 0 & 1 & 0.2 & 268 & 19 & 0.2 & 237 & 19 & -0.2 & 57603 & 17 & 1000 & 1000 \\
  $radmatch^{lc}$ & 37.9 & 0.3 & 12 & 4 & 0.0 & 298 & 20 & 0.3 & 338 & 20 & 0.3 & 304 & 20 & -0.2 & 73855 & 18 & 1000 & 1000 \\
  $pairmatch^{lc}$ & 31.2 & 0.3 & 9 & 2 & 0.0 & 183 & 15 & 0.3 & 397 & 21 & 0.3 & 356 & 21 & -0.2 & 87944 & 19 & 1000 & 1000 \\
\bottomrule
\end{tabular}
\begin{tablenotes}
\item \scriptsize Notes: `cov' and `int' refer to coverage rate and length of confidence interval, `bias' denotes the absolute bias from the true treatment effect, `sd' is the standard
deviation of the estimator, `rmse' stands for root mean squared error, and `seb' indicates the median bias of the estimated bootstrap standard error. `diff' indicates the difference to the left best performer in percent (\%) and `r' is the rank. `nsimp' shows the number of simulations with point estimations smaller than $|10^{10}|$ and `nsimse' denotes the number of simulations used to compute `cov', `int', and `seb'. Conditional on `nsimp', simulations with standard errors greater than 150 times the (true) standard error are not considered (i.e.simulations with non-converging bootstrap standard errors are dropped).
\end{tablenotes}
\end{threeparttable}
\end{sidewaystable}

\begin{sidewaystable}[ht]
\centering
\begin{threeparttable}
\caption{DGP 30}
\label{DGP 30}
\small
\begin{tabular}{rrrrrrrrrrrrrrrrrrr}
\toprule
& cov & int & d & r & bias & d & r & sd & d & r & rmse & d & r & seb & d & r & nsimp & nsimse \\
\midrule
  $ipw^{probit}$ & 95.0 & 0.2 & 1 & 7 & 0.0 & 199 & 12 & 0.0 & 0 & 3 & 0.0 & 0 & 3 & 0.0 & 402 & 8 & 1000 & 1000 \\
  $dr^{probit}$ & 95.0 & 0.2 & 1 & 6 & 0.0 & 186 & 7 & 0.0 & 0 & 1 & 0.0 & 0 & 1 & 0.0 & 424 & 9 & 1000 & 1000 \\
  $ipw^{cbps}$ & 95.0 & 0.2 & 1 & 8 & 0.0 & 253 & 14 & 0.0 & 1 & 7 & 0.0 & 1 & 6 & 0.0 & 285 & 3 & 1000 & 1000 \\
  $reg$ & 95.0 & 0.2 & 0 & 4 & 0.0 & 169 & 6 & 0.0 & 0 & 4 & 0.0 & 0 & 4 & 0.0 & 294 & 4 & 1000 & 1000 \\
  $reg^{kernel}$ & 95.1 & 0.2 & 2 & 9 & 0.0 & 582 & 18 & 0.0 & 3 & 8 & 0.0 & 5 & 8 & -0.0 & 0 & 1 & 1000 & 1000 \\
  $tsls$ & 95.1 & 0.2 & 0 & 3 & 0.0 & 192 & 10 & 0.0 & 0 & 5 & 0.0 & 0 & 5 & 0.0 & 297 & 5 & 1000 & 1000 \\
  $dr^{cbps}$ & 95.2 & 0.2 & 0 & 5 & 0.0 & 188 & 9 & 0.0 & 0 & 2 & 0.0 & 0 & 2 & 0.0 & 400 & 7 & 1000 & 1000 \\
  $pairmatch^{probit}$ & 95.4 & 0.2 & 17 & 18 & 0.0 & 198 & 11 & 0.0 & 15 & 12 & 0.0 & 15 & 12 & 0.0 & 1079 & 10 & 1000 & 1000 \\
  $pairmatch^{cbps}$ & 94.3 & 0.2 & 18 & 19 & 0.0 & 247 & 13 & 0.0 & 18 & 16 & 0.0 & 18 & 16 & 0.0 & 310 & 6 & 1000 & 1000 \\
  $randforest$ & 94.0 & 0.2 & 0 & 1 & 0.0 & 66 & 3 & 0.0 & 5 & 11 & 0.0 & 5 & 7 & -0.0 & 1440 & 12 & 1000 & 1000 \\
  $radmatch^{probit}$ & 93.8 & 0.2 & 14 & 16 & 0.0 & 187 & 8 & 0.0 & 18 & 17 & 0.0 & 18 & 17 & -0.0 & 1189 & 11 & 1000 & 1000 \\
  $pairmatch^{x}$ & 93.7 & 0.2 & 9 & 10 & 0.0 & 70 & 4 & 0.0 & 17 & 14 & 0.0 & 16 & 15 & -0.0 & 2521 & 16 & 1000 & 1000 \\
  $radmatch^{cbps}$ & 93.6 & 0.2 & 14 & 17 & 0.0 & 253 & 15 & 0.0 & 19 & 18 & 0.0 & 19 & 18 & -0.0 & 1474 & 13 & 1000 & 1000 \\
  $radmatchx^{cbps}$ & 93.6 & 0.2 & 10 & 12 & 0.0 & 33 & 2 & 0.0 & 16 & 13 & 0.0 & 15 & 13 & -0.0 & 2028 & 14 & 1000 & 1000 \\
  $radmatchx^{probit}$ & 92.1 & 0.2 & 10 & 11 & 0.0 & -0 & 1 & 0.0 & 17 & 15 & 0.0 & 16 & 14 & -0.0 & 2448 & 15 & 1000 & 1000 \\
  $means$ & 91.8 & 0.2 & 0 & 2 & 0.0 & 1338 & 21 & 0.0 & 1 & 6 & 0.0 & 12 & 11 & 0.0 & 58 & 2 & 1000 & 1000 \\
  $ipw^{lc}$ & 98.4 & 0.3 & 86 & 20 & 0.0 & 838 & 20 & 0.0 & 4 & 9 & 0.0 & 8 & 10 & 0.0 & 17441 & 17 & 1000 & 998 \\
  $dr^{lc}$ & 99.1 & 0.5 & 202 & 21 & 0.0 & 527 & 17 & 0.0 & 4 & 10 & 0.0 & 5 & 9 & 0.0 & 24495 & 18 & 1000 & 995 \\
  $pairmatch^{lc}$ & 56.4 & 0.2 & 10 & 13 & 0.0 & 510 & 16 & 0.1 & 197 & 20 & 0.1 & 196 & 20 & -0.1 & 73037 & 20 & 1000 & 1000 \\
  $radmatchx^{lc}$ & 56.4 & 0.2 & 11 & 14 & 0.0 & 151 & 5 & 0.1 & 189 & 19 & 0.1 & 188 & 19 & -0.1 & 69923 & 19 & 1000 & 1000 \\
  $radmatch^{lc}$ & 51.4 & 0.2 & 13 & 15 & 0.0 & 603 & 19 & 0.1 & 241 & 21 & 0.1 & 240 & 21 & -0.1 & 89408 & 21 & 1000 & 1000 \\
\bottomrule
\end{tabular}
\begin{tablenotes}
\item \scriptsize Notes: `cov' and `int' refer to coverage rate and length of confidence interval, `bias' denotes the absolute bias from the true treatment effect, `sd' is the standard
deviation of the estimator, `rmse' stands for root mean squared error, and `seb' indicates the median bias of the estimated bootstrap standard error. `diff' indicates the difference to the left best performer in percent (\%) and `r' is the rank. `nsimp' shows the number of simulations with point estimations smaller than $|10^{10}|$ and `nsimse' denotes the number of simulations used to compute `cov', `int', and `seb'. Conditional on `nsimp', simulations with standard errors greater than 150 times the (true) standard error are not considered (i.e.simulations with non-converging bootstrap standard errors are dropped).
\end{tablenotes}
\end{threeparttable}
\end{sidewaystable}

\begin{sidewaystable}[ht]
\centering
\begin{threeparttable}
\caption{DGP 31}
\label{DGP 31}
\small
\begin{tabular}{rrrrrrrrrrrrrrrrrrr}
\toprule
& cov & int & d & r & bias & d & r & sd & d & r & rmse & d & r & seb & d & r & nsimp & nsimse \\
\midrule
  $radmatch^{probit}$ & 99.1 & 78.8 & 808 & 19 & 0.7 & 174045 & 20 & 22.3 & 4374 & 21 & 22.3 & 4346 & 21 & -19.4 & 64165 & 21 & 1000 & 1000 \\
  $pairmatch^{x}$ & 99.2 & 32.6 & 276 & 13 & 0.1 & 33402 & 12 & 4.4 & 788 & 14 & 4.4 & 782 & 14 & -3.4 & 11225 & 13 & 1000 & 998 \\
  $tsls$ & 99.4 & 12.7 & 46 & 2 & 0.0 & 3790 & 4 & 0.7 & 49 & 6 & 0.7 & 48 & 6 & -0.2 & 455 & 8 & 1000 & 994 \\
  $pairmatch^{probit}$ & 99.4 & 46.6 & 437 & 15 & 0.2 & 59091 & 16 & 5.7 & 1044 & 15 & 5.7 & 1037 & 15 & -2.9 & 9652 & 12 & 1000 & 997 \\
  $pairmatch^{cbps}$ & 99.4 & 51.5 & 493 & 16 & 0.3 & 87678 & 18 & 8.5 & 1611 & 16 & 8.5 & 1600 & 16 & -5.8 & 19108 & 14 & 1000 & 999 \\
  $reg$ & 99.6 & 13.6 & 57 & 4 & 0.0 & 179 & 2 & 0.6 & 23 & 4 & 0.6 & 23 & 4 & -0.0 & 0 & 1 & 1000 & 987 \\
  $randforest$ & 99.6 & 22.0 & 153 & 12 & 0.4 & 95136 & 19 & 12.7 & 2444 & 17 & 12.7 & 2427 & 17 & -12.1 & 40009 & 17 & 1000 & 998 \\
  $radmatch^{cbps}$ & 99.6 & 45.2 & 420 & 14 & 0.0 & 0 & 1 & 2.6 & 423 & 11 & 2.6 & 419 & 11 & 0.1 & 336 & 6 & 1000 & 1000 \\
  $radmatchx^{cbps}$ & 99.7 & 15.9 & 83 & 8 & 0.1 & 28188 & 11 & 0.8 & 69 & 8 & 0.8 & 69 & 8 & -0.1 & 127 & 3 & 1000 & 989 \\
  $radmatchx^{probit}$ & 99.7 & 19.8 & 129 & 11 & 0.1 & 24626 & 10 & 1.2 & 140 & 10 & 1.2 & 139 & 10 & -0.4 & 1360 & 10 & 1000 & 995 \\
  $dr^{probit}$ & 99.8 & 14.9 & 71 & 7 & 0.0 & 3430 & 3 & 0.6 & 24 & 5 & 0.6 & 23 & 5 & 0.0 & 54 & 2 & 1000 & 987 \\
  $dr^{cbps}$ & 99.8 & 15.9 & 83 & 9 & 0.0 & 4059 & 5 & 0.6 & 17 & 3 & 0.6 & 16 & 3 & 0.1 & 141 & 4 & 1000 & 992 \\
  $ipw^{lc}$ & 99.9 & 71.5 & 723 & 18 & 0.1 & 35043 & 13 & 1.0 & 102 & 9 & 1.0 & 102 & 9 & 8.4 & 27770 & 15 & 1000 & 963 \\
  $ipw^{probit}$ & 99.9 & 13.1 & 51 & 3 & 0.1 & 15128 & 7 & 0.5 & 0 & 1 & 0.5 & 0 & 1 & 0.1 & 197 & 5 & 1000 & 985 \\
  $dr^{lc}$ & 99.9 & 114.7 & 1221 & 20 & 0.2 & 53434 & 15 & 3.5 & 597 & 13 & 3.5 & 594 & 13 & 8.8 & 29025 & 16 & 1000 & 986 \\
  $ipw^{cbps}$ & 99.9 & 14.3 & 65 & 6 & 0.1 & 16178 & 9 & 0.8 & 59 & 7 & 0.8 & 58 & 7 & -0.2 & 574 & 9 & 1000 & 987 \\
  $reg^{kernel}$ & 99.9 & 18.7 & 116 & 10 & 0.2 & 46471 & 14 & 2.7 & 433 & 12 & 2.7 & 431 & 12 & -2.1 & 7012 & 11 & 1000 & 999 \\
  $means$ & 100.0 & 13.8 & 60 & 5 & 0.3 & 66342 & 17 & 0.5 & 4 & 2 & 0.6 & 16 & 2 & 0.2 & 409 & 7 & 1000 & 980 \\
  $radmatch^{lc}$ & 77.7 & 200.6 & 2211 & 21 & 0.1 & 15469 & 8 & 16.9 & 3298 & 18 & 16.9 & 3275 & 18 & -15.8 & 52222 & 18 & 998 & 998 \\
  $pairmatch^{lc}$ & 76.3 & 65.6 & 656 & 17 & 0.0 & 8160 & 6 & 18.9 & 3697 & 19 & 18.9 & 3671 & 19 & -17.9 & 59160 & 19 & 998 & 998 \\
  $radmatchx^{lc}$ & 48.3 & 8.7 & 0 & 1 & 0.9 & 235248 & 21 & 19.7 & 3846 & 20 & 19.7 & 3823 & 20 & -19.3 & 63980 & 20 & 1000 & 1000 \\
\bottomrule
\end{tabular}
\begin{tablenotes}
\item \scriptsize Notes: `cov' and `int' refer to coverage rate and length of confidence interval, `bias' denotes the absolute bias from the true treatment effect, `sd' is the standard
deviation of the estimator, `rmse' stands for root mean squared error, and `seb' indicates the median bias of the estimated bootstrap standard error. `diff' indicates the difference to the left best performer in percent (\%) and `r' is the rank. `nsimp' shows the number of simulations with point estimations smaller than $|10^{10}|$ and `nsimse' denotes the number of simulations used to compute `cov', `int', and `seb'. Conditional on `nsimp', simulations with standard errors greater than 150 times the (true) standard error are not considered (i.e.simulations with non-converging bootstrap standard errors are dropped).
\end{tablenotes}
\end{threeparttable}
\end{sidewaystable}

\begin{sidewaystable}[ht]
\centering
\begin{threeparttable}
\caption{DGP 32}
\label{DGP 32}
\small
\begin{tabular}{rrrrrrrrrrrrrrrrrrr}
\toprule
& cov & int & d & r & bias & d & r & sd & d & r & rmse & d & r & seb & d & r & nsimp & nsimse \\
\midrule
  $dr^{probit}$ & 95.0 & 0.2 & 8 & 9 & 0.0 & 80 & 5 & 0.0 & 7 & 10 & 0.0 & 3 & 4 & 0.0 & 35205 & 5 & 1000 & 1000 \\
  $reg$ & 95.0 & 0.2 & 6 & 6 & 0.0 & 43 & 2 & 0.0 & 6 & 7 & 0.0 & 1 & 2 & 0.0 & 25950 & 2 & 1000 & 1000 \\
  $tsls$ & 95.1 & 0.2 & 6 & 4 & 0.0 & 45 & 3 & 0.0 & 4 & 2 & 0.0 & 0 & 1 & 0.0 & 56207 & 7 & 1000 & 1000 \\
  $dr^{cbps}$ & 94.9 & 0.2 & 7 & 8 & 0.0 & 82 & 6 & 0.0 & 7 & 9 & 0.0 & 3 & 3 & 0.0 & 29219 & 3 & 1000 & 1000 \\
  $randforest$ & 94.5 & 0.2 & 6 & 5 & 0.0 & 74 & 4 & 0.0 & 10 & 11 & 0.0 & 5 & 6 & -0.0 & 163440 & 10 & 1000 & 1000 \\
  $ipw^{probit}$ & 94.0 & 0.2 & 6 & 7 & 0.0 & 163 & 10 & 0.0 & 5 & 6 & 0.0 & 4 & 5 & 0.0 & 55555 & 6 & 1000 & 1000 \\
  $pairmatch^{cbps}$ & 93.7 & 0.2 & 45 & 16 & 0.0 & 209 & 13 & 0.1 & 47 & 15 & 0.1 & 44 & 15 & -0.0 & 74738 & 8 & 1000 & 1000 \\
  $ipw^{cbps}$ & 93.6 & 0.2 & 6 & 3 & 0.0 & 228 & 14 & 0.0 & 5 & 4 & 0.0 & 6 & 7 & 0.0 & 33086 & 4 & 1000 & 1000 \\
  $pairmatch^{probit}$ & 93.0 & 0.2 & 47 & 17 & 0.0 & 152 & 8 & 0.1 & 52 & 16 & 0.1 & 47 & 16 & -0.0 & 264367 & 11 & 1000 & 1000 \\
  $radmatchx^{cbps}$ & 92.9 & 0.2 & 24 & 14 & 0.0 & 271 & 17 & 0.1 & 31 & 13 & 0.1 & 31 & 13 & -0.0 & 344403 & 12 & 1000 & 1000 \\
  $radmatch^{cbps}$ & 92.8 & 0.2 & 47 & 18 & 0.0 & 193 & 12 & 0.1 & 56 & 17 & 0.1 & 52 & 17 & -0.0 & 371182 & 13 & 1000 & 1000 \\
  $radmatchx^{probit}$ & 92.6 & 0.2 & 25 & 15 & 0.0 & 268 & 16 & 0.1 & 33 & 14 & 0.1 & 32 & 14 & -0.0 & 374673 & 14 & 1000 & 1000 \\
  $pairmatch^{x}$ & 92.5 & 0.2 & 18 & 13 & 0.0 & 94 & 7 & 0.0 & 27 & 12 & 0.0 & 22 & 11 & -0.0 & 399714 & 15 & 1000 & 1000 \\
  $reg^{kernel}$ & 92.3 & 0.2 & 3 & 2 & 0.0 & 281 & 18 & 0.0 & 5 & 3 & 0.0 & 8 & 8 & -0.0 & 77962 & 9 & 1000 & 1000 \\
  $radmatch^{probit}$ & 92.0 & 0.2 & 51 & 19 & 0.0 & 153 & 9 & 0.1 & 63 & 18 & 0.1 & 57 & 18 & -0.0 & 542388 & 16 & 1000 & 1000 \\
  $ipw^{lc}$ & 100.0 & 2.2 & 1366 & 20 & 0.0 & 360 & 19 & 0.0 & 5 & 5 & 0.0 & 12 & 10 & 0.2 & 26897177 & 20 & 1000 & 961 \\
  $dr^{lc}$ & 100.0 & 11.6 & 7603 & 21 & 0.0 & 263 & 15 & 0.0 & 6 & 8 & 0.0 & 8 & 9 & 3.0 & 350936629 & 21 & 1000 & 366 \\
  $means$ & 87.6 & 0.2 & 0 & 1 & 0.0 & 598 & 21 & 0.0 & 0 & 1 & 0.0 & 22 & 12 & -0.0 & 0 & 1 & 1000 & 1000 \\
  $radmatchx^{lc}$ & 48.2 & 0.2 & 11 & 11 & 0.0 & 0 & 1 & 0.1 & 244 & 19 & 0.1 & 226 & 19 & -0.1 & 10331147 & 17 & 1000 & 1000 \\
  $radmatch^{lc}$ & 41.5 & 0.2 & 12 & 12 & 0.0 & 383 & 20 & 0.1 & 291 & 20 & 0.2 & 274 & 20 & -0.1 & 12349127 & 18 & 1000 & 1000 \\
  $pairmatch^{lc}$ & 35.2 & 0.2 & 9 & 10 & 0.0 & 189 & 11 & 0.2 & 324 & 21 & 0.2 & 303 & 21 & -0.1 & 13991649 & 19 & 1000 & 1000 \\
\bottomrule
\end{tabular}
\begin{tablenotes}
\item \scriptsize Notes: `cov' and `int' refer to coverage rate and length of confidence interval, `bias' denotes the absolute bias from the true treatment effect, `sd' is the standard
deviation of the estimator, `rmse' stands for root mean squared error, and `seb' indicates the median bias of the estimated bootstrap standard error. `diff' indicates the difference to the left best performer in percent (\%) and `r' is the rank. `nsimp' shows the number of simulations with point estimations smaller than $|10^{10}|$ and `nsimse' denotes the number of simulations used to compute `cov', `int', and `seb'. Conditional on `nsimp', simulations with standard errors greater than 150 times the (true) standard error are not considered (i.e.simulations with non-converging bootstrap standard errors are dropped).
\end{tablenotes}
\end{threeparttable}
\end{sidewaystable}


%

\end{document}